\documentclass[longauth]{aa}  
\usepackage{xcolor}
\usepackage{graphicx}
\usepackage{txfonts}
\usepackage{tikz,lipsum,lmodern}
\usepackage[most]{tcolorbox}
\usepackage{subcaption}
\usepackage{booktabs}
\usepackage[colorlinks=true,linkcolor=purple,citecolor=blue,filecolor=black,urlcolor=cyan,final=true]{hyperref}
\usepackage{lineno}

\makeatletter
\renewcommand*\aa@pageof{, page \thepage{} of \pageref*{LastPage}}
\makeatother

\begin{document}

   \title{The 17 April 2021 widespread solar energetic particle event} 

   \subtitle{}

    \author{N.~Dresing\inst{1}
            \and 
        L.~Rodr\'iguez-Garc\'ia\inst{2}
            \and
        I.~C.~Jebaraj\inst{1,3}
            \and
        A.~Warmuth\inst{4}
            \and 
        S.~Wallace\inst{5}
            \and
        L.~Balmaceda\inst{6}
            \and
        T.~Podladchikova\inst{7,31}
            \and
        R.~D.~Strauss\inst{8}
            \and
        A.~Kouloumvakos\inst{9}
            \and
        C.~Palmroos\inst{1}
            \and 
        V. Krupar\inst{6,34}
            \and
        J.~Gieseler\inst{1}
            \and
        Z.~Xu\inst{10}
            \and
        J.~G.~Mitchell\inst{6,5}
            \and
        C.~M.~S.~Cohen\inst{11}
            \and 
        G.~A.~de~Nolfo\inst{6}
            \and 
        E.~Palmerio\inst{12}
                \and
        F.~Carcaboso\inst{6,13}
            \and
        E.~K.~J.~Kilpua\inst{14}
            \and
        D.~Trotta\inst{15}
         \and
        U.~Auster\inst{16}
            \and
        E.~Asvestari\inst{14}
                \and
        D.~da~Silva\inst{6,17,18}
            \and
        W.~Dr\"oge\inst{19}
            \and
        T.~Getachew\inst{6,13}
            \and
        R.~G\'omez-Herrero\inst{2}
            \and
        M.~Grande\inst{20}
            \and
        D.~Heyner\inst{16}
            \and
        M.~Holmstr{\"o}m\inst{21}
            \and
        J.~Huovelin\inst{14}
            \and
        Y.~Kartavykh\inst{10}
            \and
        M.~Laurenza\inst{22}
            \and
        C.~O.~Lee\inst{23}
            \and
        G.~Mason\inst{9}
            \and
        M.~Maksimovic\inst{24}
            \and
        J.~Mieth\inst{16}
            \and 
        G.~Murakami\inst{25}
            \and
        P.~Oleynik\inst{1}
            \and
        M.~Pinto\inst{26,27}
            \and
        M.~Pulupa\inst{23}
            \and
        I.~Richter\inst{16}
            \and
        J.~Rodr\'iguez-Pacheco\inst{2}
            \and
        B.~S\'anchez-Cano\inst{28}
            \and 
        F.~Schuller\inst{4}    
            \and
        H.~Ueno\inst{29}
            \and
        R.~Vainio\inst{1}
            \and 
        A.~Vecchio\inst{30,24}
            \and
        A.~M.~Veronig\inst{31,32}
            \and
        N.~Wijsen\inst{6,33}
        }
        
   \institute{Department of Physics and Astronomy,
            University of Turku, Turku, Finland\\
            \email{nina.dresing@utu.fi}
        \and 
            Universidad de Alcalá, Space Research Group (SRG-UAH), Plaza de San Diego s/n, 28801 Alcalá de Henares, Madrid, Spain 
        \and
            Center for mathematical Plasma Astrophysics---CmPA, Department of Mathematics, KU Leuven, Leuven, Belgium
        \and
            Leibniz-Institut f\"ur Astrophysik Potsdam (AIP), Potsdam, Germany
        \and
            Department of Physics, George Washington University, Washington, DC, USA
        \and 
            Heliophysics Science Division, NASA Goddard Space Flight Center, Greenbelt, MD, USA
        \and    
            Skolkovo Institute of Science and Technology, Moscow, Russia
        \and
            Centre for Space Research, North-West University, Potchefstroom, South Africa
        \and
            The Johns Hopkins University Applied Physics Laboratory, Laurel, MD, USA
        \and 
            Institute of Experimental and Applied Physics, Kiel University, Kiel, Germany
        \and 
            California Institute of Technology, Pasadena, CA, USA
        \and
            Predictive Science Inc., San Diego, CA, USA
        \and
            Physics Department, The Catholic University of America, Washington, DC, USA
        \and
            Faculty of Science, University of Helsinki, Helsinki, Finland
        \and
           The Blackett Laboratory, Department of Physics, Imperial College London, London, UK
        \and 
           Institut für Geophysik und extraterrestrische Physik, Technische Universit{\"a}t Braunschweig, Braunschweig, Germany
        \and
           University of Maryland Baltimore County, Baltimore, MD, USA
        \and
           Laboratory for Atmospheric and Space Physics, University of Colorado at Boulder, Boulder, CO, USA
        \and
            Institute for Theoretical Physics and Astrophysics, University of W\"urzburg, W\"urzburg, Germany
        \and  
            Department of Physics, Aberystwyth University, Aberystwyth, UK
        \and
            Swedish Institute of Space Physics, Kiruna, Sweden
        \and
            Institute of Space Astrophysics and Planetology -- INAF, Roma, Italy
        \and
            Space Sciences Laboratory, University of California--Berkeley, Berkeley, CA, USA
        \and
            LESIA, CNRS, Observatoire de Paris, Universit{\'e} PSL, Sorbonne Universit{\'e}, Universit{\'e} de Paris, Meudon, France
        \and
            Institute of Space and Astronautical Science, Japan Aerospace Exploration Agency, Kanagawa, Japan
        \and
            European Space Research and Technology Centre, European Space Agency, Noordwijk, The Netherlands
        \and
            Laborat{\'o}rio de Instrumenta\c{c}\~{a}o e F{\'i}sica Experimental de Part{\'i}culas, Lisbon, Portugal
        \and 
            School of Physics and Astronomy, University of Leicester, Leicester, UK 
        \and
            Research and Development Directorate, Japan Aerospace Exploration Agency, Tsukuba, Japan
        \and
            Radboud Radio Lab -- Department of Astrophysics, Radboud University, Nijmegen, The Netherlands
        \and 
            Institute of Physics, University of Graz, Graz, Austria
        \and
            Kanzelh\"ohe Observatory for Solar and Environmental Research, University of Graz, Treffen, Austria
        \and 
            University of Maryland, College Park, MD, USA 
        \and 
            Goddard Planetary Heliophysics Institute, University of Maryland, Baltimore, MD, USA 
        }

   \date{}

 
  \abstract
   {A complex and long-lasting solar eruption on 17 April 2021 produced a widespread Solar Energetic Particle (SEP) event that was observed by five longitudinally well-separated observers in the inner heliosphere covering distances to the Sun from 0.42 to 1~au: BepiColombo, Parker Solar Probe, Solar Orbiter, STEREO~A, and near-Earth spacecraft. The event was the second widespread SEP event  dected in solar cycle 25 and produced relativistic electrons and protons. It was associated with a long-lasting solar hard X-ray flare showing multiple hard X-ray peaks over a duration of one hour. The event was further accompanied by a medium fast Coronal Mass Ejection (CME) with a speed of 880~km~s$^{-1}$ driving a shock, an EUV wave as well as long-lasting and complex radio burst activity showing four distinct type III~burst groups over a period of 40 minutes.}
   {We aim at understanding the reason for the  the wide spread of elevated SEP intensities in the inner heliosphere as well as identifying the underlying source regions of the  observed energetic electrons and protons.}
   {A comprehensive multi-spacecraft analysis of remote-sensing observations and in-situ measurements of the energetic particles and interplanetary context is applied to attribute the SEP observations at the different locations to the various potential source regions at the Sun. An ENLIL simulation is used to characterize the complex interplanetary state and its role for the energetic particle transport. 
   The magnetic connection between  each spacecraft and the Sun is determined using ballistic backmapping in combination with potential field source surface extrapolations in the lower corona. In combination with a reconstruction of the coronal shock front we then determine the times when the shock establishes magnetic connections with the different observers. 
   Radio observations are used to characterize the directivity of the four main injection episodes, which are then employed in a 2D SEP transport simulation to test the importance of these different injection episodes.}
   {A comprehensive timing analysis of the inferred solar injection times of the SEPs observed at  each spacecraft suggests different source processes being important for the electron and the proton event. Comparison  among the characteristics and timing of the potential  particle sources, such as the CME-driven shock or the flare, suggests a stronger shock contribution for the proton event and a more likely flare-related source of the electron event. }
   { In contrast to earlier studies on widespread SEP events, we find that in this event an important ingredient for the wide SEP spread was the wide longitudinal range of about 110$^{\circ}$ covered by distinct SEP injections, which is also supported by our SEP transport modeling.}


   \maketitle
  
%
\section{Introduction} \label{sec:intro}
Solar energetic particle (SEP) events are characterized by a rich and complex set of physical processes responsible for the acceleration and propagation of the particles. Since the early observations of \citet{Forbush1946}, an enormous amount of knowledge has been built around SEPs, highlighting their importance for understanding the behavior of the outer layers of the Sun's atmosphere, as well as addressing fundamental questions related to energetic particle propagation in astrophysical environments \citep[e.g.,][]{Reames2021}.
Multi-point observations of SEP events at different heliospheric locations provide an invaluable opportunity to study the production and transport of energetic particles, with several recent studies addressing the problem from a variety of perspectives \citep[e.g.,][]{Dresing2014, Gomez-Herrero2015, Klein2017, 2021Rodriguez-Garcia, Frassati2022}.

\begin{figure*}
    \includegraphics[width=0.48\textwidth]{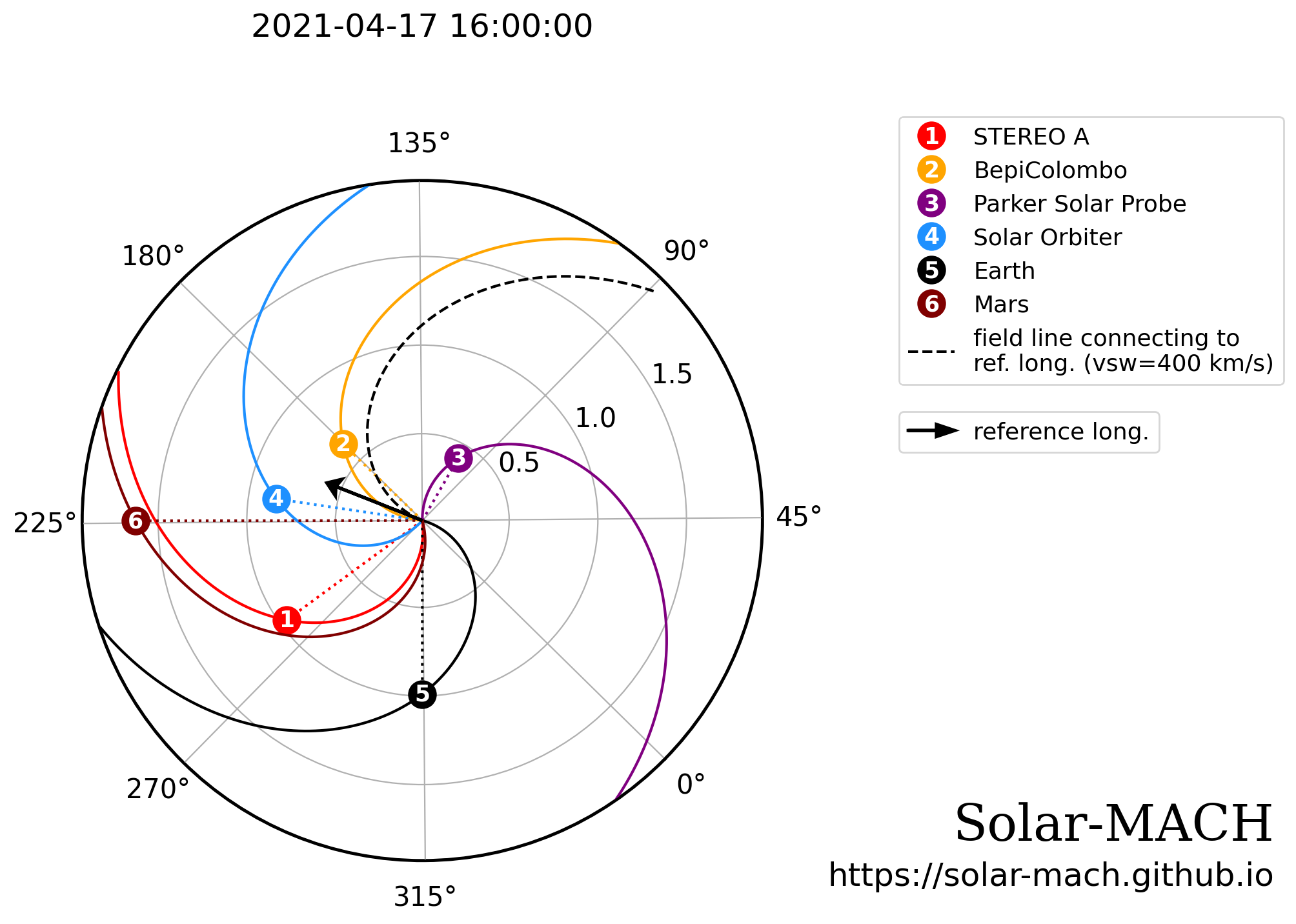}
   \includegraphics[width=0.5\textwidth]{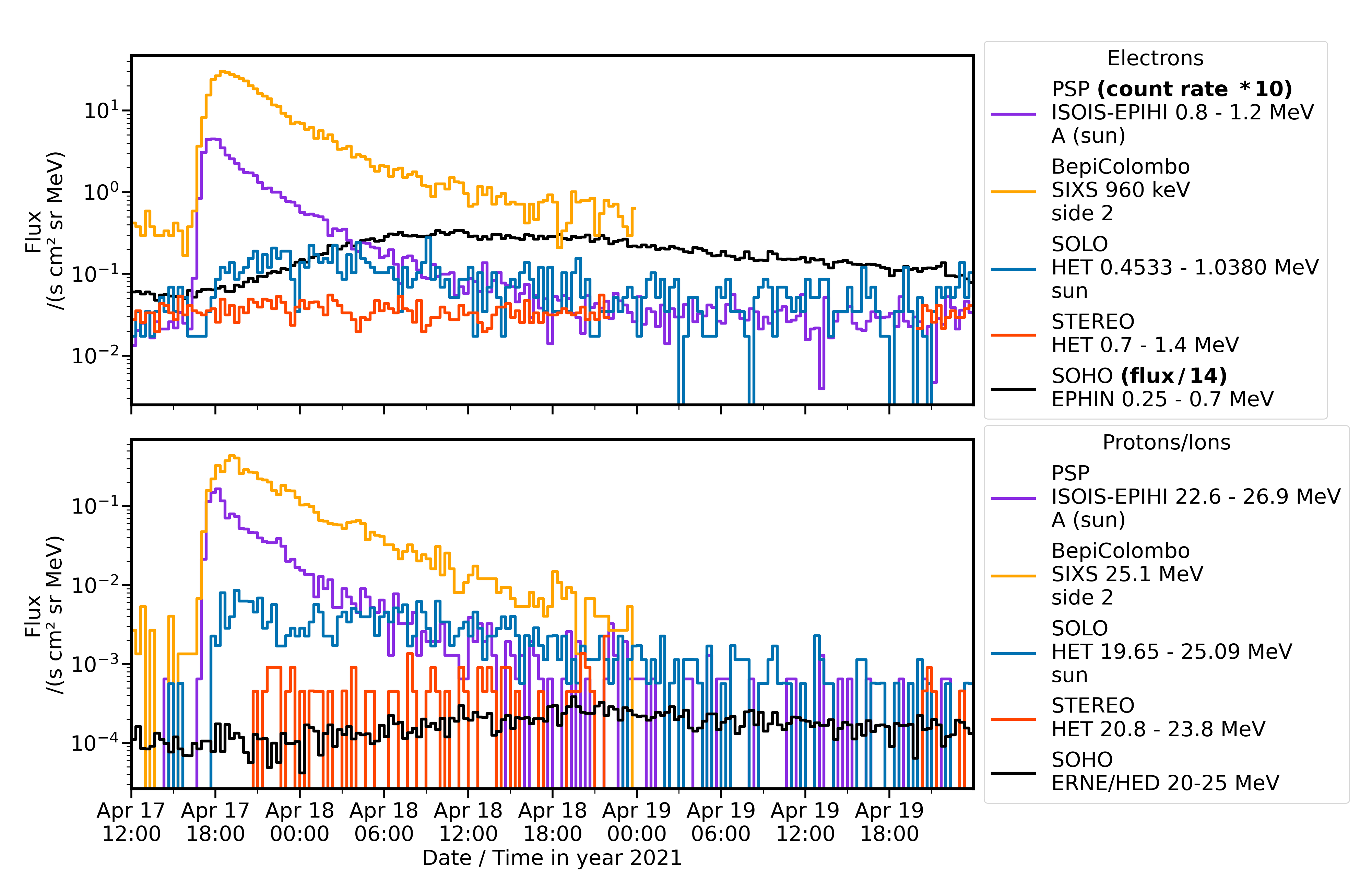}
      \caption{Longitudinal spacecraft constellation and magnetic connectivity  at 16:00 UT on 17 April 2021 (left) together with multi-spacecraft SEP observations (right). The upper panel shows $\sim$1~MeV  electron intensities and the lower  panel $\sim$25~MeV  proton intensities (respectively ions) observed by the spacecraft indicated by the same colors in the left figure. The observers' configuration plot  (left panel) has been produced using the Solar MAgnetic Connection HAUS \citep[Solar-MACH;][]{Gieseler2022} tool.
              }
         \label{fig:solar-mach_and_multi_sc_SEP}
   \end{figure*}

On 2021 April 17 a SEP event was observed by multiple spacecraft at well-separated locations in the inner heliosphere (within 1 au) but also by spacecraft in orbit about Mars (at 1.63 au from the Sun). The solar origin of the SEP event was temporary associated with a solar flare from behind the southeastern limb of Earth-facing Sun. This event can be considered the second widespread SEP event of solar cycle as it was detected over a longitude span of 210$^{\circ}$ \citep[with the first widespread SEP event of solar cycle 25 occurring on 2020 Nov 29 and analyzed by e.g.][]{Kollhoff2021, Kouloumvakos2022, Palmerio2022}. It is the first SEP event observed at five well-separated locations in the inner heliosphere (within 1~au) and also constrained by observations at Mars.
Figure~\ref{fig:solar-mach_and_multi_sc_SEP} (left) illustrates the observer locations in the heliographic equatorial plane together with nominal Parker field lines connecting each observer with the Sun depicted in the center of the plot. The black arrow marks the longitude of the associated flare  (identified using Solar Orbiter STIX measurements as described in Sect. \ref{sec:flare}), and the dashed black spiral denotes the nominal magnetic field line connecting to this location. BepiColombo (yellow) was  the spacecraft with the best nominal connection to the flare site. Parker Solar Probe (purple) and Solar Orbiter (blue) were approximately equally  separated in longitude from the flaring region, but on different sides. STEREO~A (red) and Earth (black) were further separated to the west of the flare. Despite the large angular separation between all spacecraft, the SEP event was observed at these five locations as shown in Fig.~\ref{fig:solar-mach_and_multi_sc_SEP} (right), which makes it a widespread event \citep[e.g.,][]{Dresing2014}. The top panel shows $\sim$1~MeV  electron intensities and the bottom panel $\sim$25~MeV  proton intensities. As expected due to its closest magnetic connection, BepiColombo observed the highest intensities. Parker Solar Probe, being situated closest  at 0.42 au from the Sun to the Sun, observed significantly higher intensities than Solar Orbiter, although their total separation angles are comparable. STEREO~A observed a weak proton event but no  significant increase of MeV electron intensities. At Earth/L1 (SOHO),  the location with the poorest nominal magnetic connection with the flare site, the electron event  seems more intense and distinct when compared with STEREO~A. While Parker Solar Probe and Solar Orbiter were situated close to the ecliptic plane at slightly northern latitudes,  BepiColombo, STEREO~A and Earth were situated at southern latitudes (see Table~\ref{table:ADAPT-WSA connectivity},  which summarizes the observer locations and their magnetic connections to the Sun) with a maximum of $-7.2^{\circ}$ in the case of BepiColombo.
The 17 April 2021 event, therefore, shows not only a spatial asymmetry with respect to the flare longitude but also clear differences between the electron and proton distributions.

We investigate here the drivers for this wide SEP spread as well as the reasons for the observed asymmetries. The most common explanations for widespread events have been a large acceleration region, e.g.\ an extended coronal shock \citep[e.g.,][]{Rouillard2012, Gomez-Herrero2015, Lario2016, 2021Rodriguez-Garcia, Kouloumvakos2022}, and/or efficient perpendicular transport in the corona or interplanetary medium \citep[e.g.,][]{Dresing2012, Droege2016}. 

The paper is organized as follows. After discussing the instrumentation used in this study in Sect.~\ref{sec:instrumentation}, we begin with a detailed analysis of the magnetic connectivity between the different observer locations and the Sun (Sect.~\ref{sec:connectivity}). Section~\ref{sec:flare} discusses the complex and long-lasting flare of the event, Sect.~\ref{sec:CME} describes the analysis of the associated coronal mass ejection (CME) and CME-driven shock. Sect.~\ref{sec:EUV_obs} presents observation and analysis of the associated extreme ultra-violet (EUV) wave. Observations and a reconstruction of the coronal CME-driven shock are presented in Sect.~\ref{sec:shock_obs}. Sect.~\ref{sec:radio_obs} provides an analysis of the various type~II and III radio bursts observed during the event.
In Sect.~\ref{sec:in_situ_obs}, we study the interplanetary context, involving also simulations of the state of the interplanetary medium using 3D magnetohydrodynamic (MHD) simulations, and present an overview of the multi-spacecraft SEP observations. 
A more detailed analysis of SEP onset times, velocity dispersion, and pitch-angle distributions is presented in Sect.~\ref{sec:sep_obs}.
In Sect.~\ref{sec:timing}, we relate the timing of SEP arrivals with solar counterpart observations to infer the parent source regions of the SEPs. 
In Sect.~\ref{Sec:modelling}, interplanetary transport modeling results are presented assuming two different scenarios: the first being the standard scenario of a single SEP injection into interplanetary space from a single source region, while a second scenario assumes multiple SEP injections from different particle sources at different times.
Sections~\ref{sec:discussion} and \ref{sec:summary} provide the discussion and conclusions of the study presented here, respectively.

\section{Instrumentation} \label{sec:instrumentation}

{\it BepiColombo}

Several data sets from the cruise phase of BepiColombo \citep{Benkhoff2021} en route to Mercury are used in this study, such as  from the Solar Intensity X-Ray and Particle Spectrometer \citep[SIXS;][]{Huovelin2020} on board the Mercury Planetary Orbiter (MPO, the European spacecraft involved in the BepiColombo mission). SIXS provides measurements of high-energy electrons and protons with the SIXS-P particle detector. This instrument consists of five orthogonal 150~$\mu$m thick Si PIN detectors, also called `Sides', and a $5\times 5\times 6.3$~mm\textsuperscript{3} CsI(Tl) scintillator with photodiode read-out. It detects electrons in the range 50~keV to 3~MeV and protons in the range 1 to 30~MeV with a total nominal geometric factor of about 0.19~cm\textsuperscript{2} sr. We note that Sides 0 and 4 are partially and totally obstructed by the spacecraft cruise shield, respectively.

We also use data from the BepiColombo Environment Radiation Monitor \citep[BERM;][]{Pinto2021} on board MPO, which is a particle detector that consists of a single silicon stack telescope with a small particle entrance of 0.5~mm\textsuperscript{2} and a 50~$\mu$m beryllium cutoff window. Particles are arranged into 5 channels for electrons ($\sim$0.15--10~MeV), 8 channels for protons (1.5--100~MeV), and 5 channels for heavy ions (1--50~MeV$\cdot$mg\textsuperscript{-1}$\cdot$cm\textsuperscript{-2}). BERM is mounted behind the radiator panel and faces the anti-sunward direction. 

In addition, data from the Solar Particle Monitor (SPM) on board the Japanese s/c Mercury Magnetospheric Orbiter \citep[MMO, also known as Mio;][]{Murakami2020} are employed. SPM is a particle detector that forms part of the housekeeping suite, and it consists of two silicon photodiodes (SPM1 and SPM2), each one with an effective area of 10~mm $\times$ 10~mm and a depletion layer thickness of 0.3~mm. Each sensor has four different deposited energy channels, which cover the energy ranges 70--1170 keV and 50--200 MeV respectively. A calibration of the sensors is currently being performed with Monte Carlo simulations based on Geant4 \citep{Agostinelli2003}. 

Finally, we also use data from the BepiColombo MPO magnetometer \citep[MPO-MAG;][]{Heyner2021}, which is composed of two tri-axial fluxgate magnetometers placed on a 2.9~m boom. MPO-MAG measures magnetic field up to 128~Hz in a $\pm$2048~nT range. 
\\
\\
{\it Parker Solar Probe}

Energetic particles at Parker Solar Probe \citep[PSP;][]{Fox2016} are measured by the Integrated Science Investigation of the Sun \citep[IS$\odot$IS;][]{McComas2016} suite. Low-energy  electrons and ions ($\sim$20~keV to 20~MeV/nucleon over $2\pi$ stereoradians) are covered by the Energetic Particle Instrument-Low \citep[EPI-Lo;][]{Hill2017}, consisting of 80 time-of-flight apertures. The high-energy particles are measured with the Energetic Particle Instrument-High \citep[EPI-Hi;][]{Wiedenbeck2017}, consisting of three telescopes of stacked solid-state detectors, using the standard $dE/dx$ versus residual energy technique to measure ions from $\sim$1 to $>$100 MeV/nuc and electrons in the range $\sim$ 0.5--6~MeV. The  first two Low Energy Telescopes (LETs) of EPI-Hi consist of a double-ended detector, providing oppositely viewing apertures (LETA and LETB) and one single-ended detector (LETC) with a viewing axis perpendicular to that of LETA. The third telescope (High Energy Telescope; HET) covers the highest energies and is double-ended with two apertures (HETA and HETB) providing roughly sunward and anti-sunward viewing directions along the nominal Parker spiral. 
 
Observations of the magnetic field are obtained from the fluxgate magnetometer part of the FIELDS \citep{Bale2016} suite, and solar wind measurements are provided by the Solar Probe Cup (SPC) instrument part of the Solar Wind Electrons Alphas and Protons \citep[SWEAP;][]{Kasper2016} investigation.       
 
Radio observations are provided by the Radio Frequency Spectrometer \citep[RFS;][]{Pulupa2017} part of FIELDS, which is a dual-channel digital spectrometer designed for both remote-sensing of radio waves and in-situ measurements of electrostatic fluctuations between $10$~kHz and $19.2$~MHz. Here, we use the RFS data when input channels were set to the two pairs of crossed dipoles. Besides the radio flux density, it also allows us to retrieve the degree of circular polarization \citep{Pulupa20}. 
\\
\\
{\it Solar Orbiter}

Data from several instruments on board Solar Orbiter \citep{Muller2020} are used. The Spectrometer/Telescope for Imaging X-rays \citep[STIX;][]{Krucker2020} provides imaging spectroscopy in the X-ray range (4-150~keV). It has a full-disk field of view (FOV) and sub-second time resolution. The Radio and Plasma Waves \citep[RPW;][]{Maksimovic20b,Maksimovic2021,Vecchio2021} instrument on Solar Orbiter consists of several subsystems including the Thermal Noise and High Frequency Receiver (TNR-HFR or THR) with a dual channel sweeping receiver in the range from 4~kHz up to 16~MHz.
In particular, THR provides measurements of the plasma quasi-thermal noise (QTN) in the range 4~kHz -- 1~MHz. When the QTN signal is quite strong, the spectral peak at the electron plasma frequency can be identified from which the in-situ absolute electron density can be derived \citep{Meyer-Vernet2017, Khotyaintsev2021}. 
The properties of energetic particles as measured by Solar Orbiter are studied using the Electron Proton Telescope (EPT) and the High Energy Telescope (HET) of the Energetic Particle Detector \citep[EPD;][]{Rodriguez-Pacheco2020} instrument suite.  Both sensors consist of two double-ended telescopes. EPT and measures ions and electrons in the energy ranges 20~keV -- 15~MeV and 20--400~keV, respectively, and HET relativistic electrons between 0.3 and 30 MeV and protons between 7 and 107 MeV. 

The Solar Orbiter Magnetometer \citep[MAG;][]{Horbury2020} is a fluxgate vector magnetometer, yielding in-situ measurements of the interplanetary magnetic field with 16~vectors/s (normal mode) and up to 128~vectors/sec (burst mode). 

The lower-energy, thermal, and suprathermal particles are measured by the Solar Wind Analyzer \citep[SWA;][]{Owen2020} suite. In this work, measurements from the SWA Proton and Alphas Sensor (PAS), sampling 3D velocity distribution functions of protons and alpha particles in the 0.2--20~keV energy range with a 4~s time cadence, are used to address the in-situ plasma moments, such as the solar wind's bulk flow speed and density.
\\
\\
{\it STEREO~A}

Observations from several instruments on board the Solar Terrestrial Relations Observatory \citep[STEREO;][]{Kaiser2008} are used in this study. As the STEREO~B spacecraft is inactive since October 2014 due to multiple hardware anomalies, only data from instruments on board STEREO~A are available for the period under consideration. 

The STEREO/WAVES \citep[S/WAVES;][]{2008SSRv..136..487B} instrument provides comprehensive measurements of all components of the electric field fluctuations between 2.5~kHz and 16~MHz. It allows us to locate sources, and calculate the polarization state (including apparent source sizes) of radio emissions in a heliocentric distance range from 4\,$R_{\odot}$ to 1~au, while the flux density can be measured even down to 2\,$R_{\odot}$ \citep{2014SoPh..289.4633K}. Unfortunately, direction-finding data were not available for this event.

Interplanetary magnetic field measurements are provided by the Magnetic Field Experiment \citep[MFE;][]{Acuna2008}, part of the In situ Measurements of Particles And CME Transients \citep[IMPACT;][]{Luhmann2008} instrument suite. MFE is a triaxial fluxgate magnetometer mounted on a telescopic boom at a distance of $\sim$3~m from the spacecraft body, reaching a maximum cadence of 32~vectors/s.

Energetic particle observations with 1-minute cadence are provided by several instruments, part of the IMPACT investigation. The Solar Electron Proton Telescope \citep[SEPT;][]{Muller-Mellin2008} consists of dual double-ended magnet/foil particle telescopes measuring 30--400~keV electrons and 60--7000~keV ions. Two separate units provide anisotropy information in four different looking directions: Sun, Asun (pointing sunward and anti-sunward along the nominal Parker spiral, respectively), North, and South (pointing towards the North and South ecliptic poles, respectively). Since July 2015, after the solar conjunction, the spacecraft was rolled $180^{\circ}$ about the spacecraft--Sun line and these nominal pointing directions changed, with the Sun and Asun telescopes pointing perpendicular to the nominal Parker spiral direction, North pointing southward and South pointing northward. The Low Energy Telescope \citep[LET;][]{Mewaldt2008} measures protons from $\sim$2 to $\sim$13~MeV and heavier ions from $\sim$2 to $>$40~MeV/nuc (species-dependent energy range). The field of view is divided into 16 different sectors, providing directional information. The High Energy Telescope \citep[HET;][]{vonRosenvinge2008} provides the highest energy measurements, including 0.7--4~MeV electrons and 13--100~MeV protons.

Physical properties of the solar wind plasma are obtained by the Plasma and Suprathermal Ion Composition \citep[PLASTIC;][]{Galvin2008} instrument, in particular by the Solar Wind Sector (SWS), sampling solar wind proton bulk parameters.

Remote-sensing observations  from STEREO~A are provided by several instruments that are part of the Sun Earth Connection Coronal and Heliospheric Investigation \citep[SECCHI;][]{Howard2008}. This instrument suite includes an Extreme UltraViolet Imager \citep[EUVI;][]{Wuelser2004}, two coronagraphs (COR1 and COR2) imaging the corona from 1.4 up to 15\,$R_{\odot}$, and two Heliospheric Imager \citep[HI;][]{Eyles2009} cameras (HI1 and HI2).
\\
\\
{\it Near-Earth spacecraft}

From  the Wind spacecraft \citep{Ogilvie1997}, we use the Magnetic Field Investigation \citep[MFI;][]{Lepping1995} instrument, measuring at a cadence of 11~vectors/s. The Solar Wind Experiment \citep[SWE;][]{Ogilvie1995} and Three-Dimensional Plasma and Energetic Particle Investigation \citep[3DP;][]{Lin1995} instruments provide energetic particle measurements  from which we use electron observations in the range of $\sim$40-600 keV. Finally, the Wind/WAVES \citep[WAVES;][]{1995SSRv...71..231B} instrument measures the electric field from 0.3~Hz up to 13~MHz using three dipolar antennas.

From the Solar and Heliospheric Observatory \citep[SOHO;][]{Domingo1995}, we use energetic proton measurements of the Energetic and Relativistic Nuclei and Electron \citep[ERNE;][]{Torsti1995} covering energies of a few to a hundred MeV, and energetic electron measurements in the MeV range by the Electron Proton Helium Instrument (EPHIN), part of the Comprehensive Suprathermal and Energetic Particle Analyser \citep[COSTEP;][]{Muller-Mellin1995} suite, and coronagraph observations by the Large Angle and Spectrometric Coronagraph \citep[LASCO;][]{Brueckner1995}.

Extreme ultraviolet (EUV) solar images were obtained by the Atmospheric Imaging Assembly \citep[AIA;][]{Lemen2012} on board the Solar Dynamics Observatory \citep[SDO;][]{Pesnell2012}.
\\
\\
{\it Mars}

Observations at Mars have been obtained from two spacecraft in orbit around the planet. Energetic particle data come from the Solar Energetic Particle \citep[SEP;][]{Larson2015} instrument on board the Mars Atmosphere and Volatile Evolution \citep[MAVEN;][]{Jakosky2015} mission. MAVEN/SEP is a solid-state telescopic detector with two identical sensors (SEP1 and SEP2), each containing two oppositely arranged double-ended telescopes (A and B) and measuring ions in the energy range $\sim$20--6000~keV and electrons in the range of $\sim$20--200~keV. SEP1 and SEP2 are body-mounted onto the MAVEN spacecraft to provide orthogonal look directions, with each telescope providing a  42$^{\circ}$ $\times$ 31$^{\circ}$ FOV coverage.

Solar wind density and velocity come from the Analyzer of Space Plasmas and Energetic Atoms \citep[ASPERA-3;][]{Barabash2006} on board Mars Express \citep[MEX;][]{chicarro2004}, and in particular from the Ion Mass Analyser (IMA) sensor that measures ions in the energy range 10~eV/q -- 30~keV/q. 

\section{Spacecraft constellation and magnetic connectivity} \label{sec:connectivity}

For all locations at which this event was observed, with the exception of Mars, we derive the instantaneous magnetic connectivity to the solar surface at 16:00~UT on 17 April 2021 using two different coronal models. The first model is a standard potential-field source surface \citep[PFSS;][]{Schatten1969,Altschuler1969,WangSheeley1992} model out to 2.5~\(R_\odot\), and the second is the Wang--Sheeley--Arge \citep[WSA;][]{ArgePizzo2000,Arge2003a,Arge2004,McGregor2008} model. The latter makes use of the Schatten Current Sheet (SCS) model to extend the PFSS solution in this work to 5~\(R_\odot\), providing a more realistic magnetic field topology of the upper corona. This height is appropriate for applications of deriving spacecraft connectivity since WSA is magnetostatic and is designed to be most accurate in low beta regimes, whereas when coupled with an MHD model as its inner boundary condition one would want to derive the coronal field to between 20--30~\(R_\odot\) to ensure that the solar wind is supersonic and super alfv\'enic \citep{Arge2004}.  In both the WSA and traditional PFSS approach, the observed solar wind speed at the time of the event is used to backmap the event to the model-derived coronal field, assuming a Parker spiral. In the case of BepiColombo, which did not measure the solar wind speed during cruise phase, a nominal value of 400~km~s$^{-1}$ is used, which coincides with the simulated solar wind given by the ENLIL model \citep{Odstrcil2004}, as discussed in Sect.~\ref{sec:in_situ_obs}. In this section, we present the results from both models, discuss any differences between the solutions, and elaborate on any uncertainties associated with magnetic connectivity. 


Both the PFSS and WSA coronal solutions are obtained using an identical Air Force Data Assimilative Photospheric Flux Transport \citep[ADAPT;][]{ADAPT_Arge2009,Arge2010,Arge2013,Hickmann2015} time-dependent photospheric field map, derived using data from Global Oscillation Network Group \citep[GONG;][]{Harvey1996} magnetograms. ADAPT uses flux-transport modeling \citep{WordenHarvey2000} to account for solar time-dependent phenomena (e.g., differential rotation, meridional, and supergranulation flows) when observational data are not available. This is especially useful for studying events that originate on the solar far-side (i.e., the solar hemisphere not visible from Earth). Since ADAPT is an ensemble model, it provides 12 possible states (i.e., realizations) of the solar surface magnetic field, ideally representing the best estimate of the range of possible global photospheric flux distribution solutions at any given moment in time. When coupled with the WSA model, ADAPT-WSA derives an ensemble of 12 realizations representing the global coronal field and spacecraft magnetic connectivity to 1~\(R_\odot\) (i.e., the solar surface) for a given moment in time, providing an estimate of the uncertainty in the ADAPT-WSA solution. The best realization is then determined by comparing the model-derived and the in-situ observed radial interplanetary magnetic field (IMF) and solar wind speed. 

\begin{figure}
   \centering
   \includegraphics[width=0.45\textwidth]{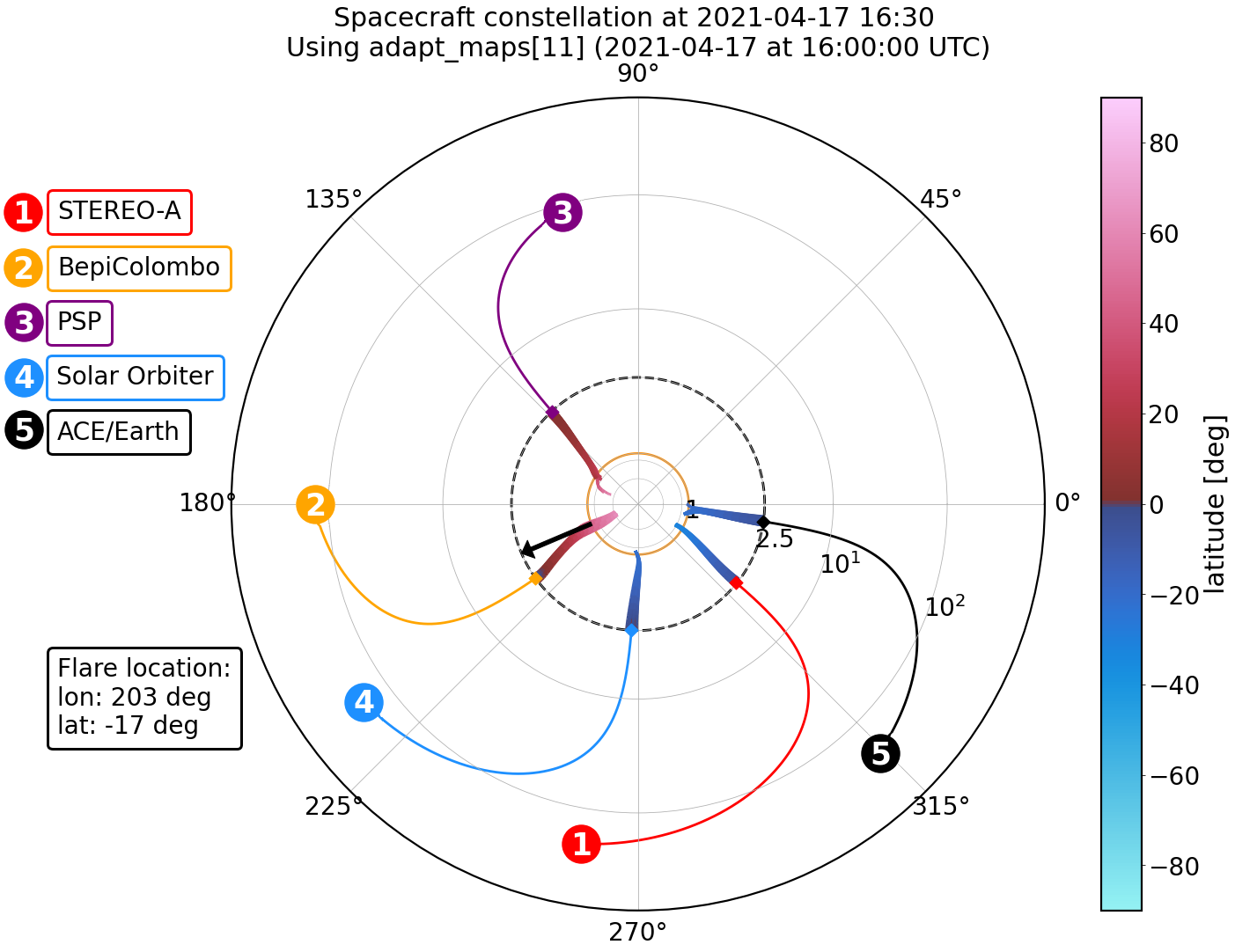}
      \caption{ Semi-logarithmic representation of the spacecraft constellation in Carrington coordinate system. The plot is linear inside the dashed black circle, which marks the distance of the potential field source surface at 2.5 ~\(R_\odot\) in this case), and the orange circle marks the Sun.  Above 2.5 ~\(R_\odot\), the plot is logarithmic in distance. Color-coded solid circles mark the various spacecraft of the constellation, and the lines connected to them represent the nominal Parker spiral solutions  computed considering their heliocentric distances and the observed solar wind speeds. Inside the dashed black line, the magnetic connection is extrapolated with a PFSS solution, where the color of the lines corresponds to heliospheric latitude. The black arrow corresponds to the flare location.
              }
         \label{fig:solar_mach_pfss}
   \end{figure}

Figure~\ref{fig:solar_mach_pfss} shows the instantaneous connectivity derived with the PFSS coronal field solution, with the corresponding footpoint connectivity listed in Table~\ref{table:ADAPT-WSA connectivity} columns (6)--(7). The plot shows the Sun in the center, the source surface (dashed circle), which is the outer boundary of potential-field models, and the spacecraft constellation in the heliospheric Carrington coordinate system, where the unit of distance is the solar radius. The magnetic connectivity to different spacecraft is estimated as a nominal Parker spiral connecting to the source surface, from which magnetic field lines are tracked downwards to the photosphere using a PFSS extrapolation. Table~\ref{table:ADAPT-WSA connectivity} shows the magnetic connection points from the various spacecraft to the photosphere and the observed solar wind speed that is used to calculate the Parker spiral. Note that the scale in the plot is logarithmic above the source surface and linear below.

\begin{figure*}  
   \centering
	\includegraphics[width=\textwidth]{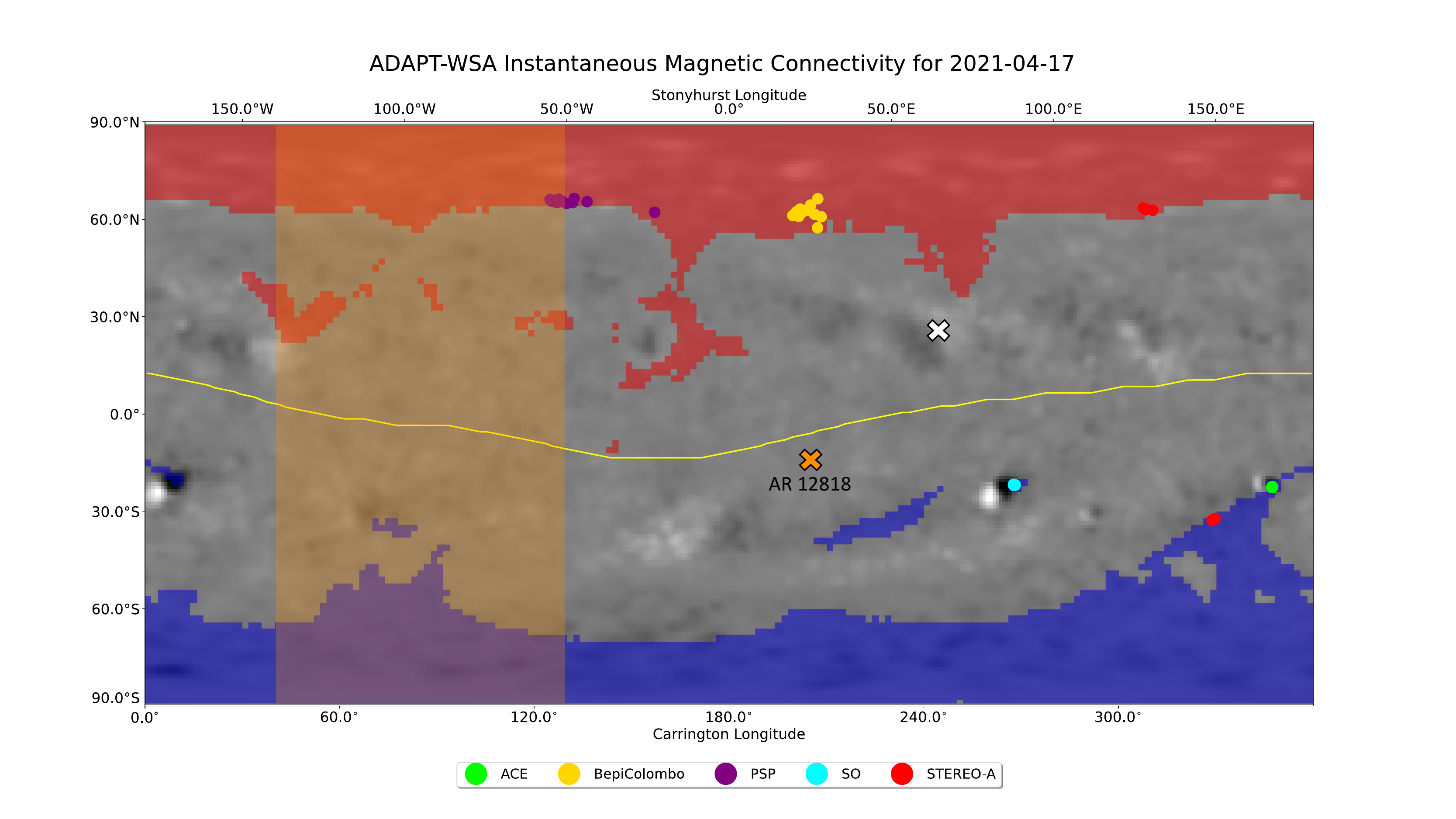}
	\caption{ADAPT-WSA derived instantaneous magnetic connectivity on 17 April 2021 at 16:00:00~UT for five of the spacecraft observing the SEP event, overlain onto the corresponding ADAPT-GONG map used to derive the coronal field, and the WSA-derived coronal holes (red/positive, blue/negative). The footpoint connectivity for each spacecraft is labeled by 12 colored points, one for each ADAPT-WSA realization, and the WSA-derived HCS is overplotted in yellow. The locations of two ARs that emerge on the solar far-side are labeled with an ``$\times$'', yet are not incorporated into the ADAPT map until several days after this event. AR~12818 associated with a solar flare (discussed in Sect.~\ref{sec:remote_obs}) is labeled with an orange ``$\times$''. Orange shading marks the portion of the Sun not observed by remote imagers on board STEREO~A, Solar Orbiter, or spacecraft near Earth.} 
	\label{ADAPT-WSA Connectivity}
\end{figure*}

Figure~\ref{ADAPT-WSA Connectivity} shows the WSA-derived instantaneous magnetic connectivity and thus the model-estimated magnetic footpoint for each of the five spacecraft on 17 April 2021 at 16:00~UT for all 12 realizations of the ADAPT-WSA output. The 12 realizations often produce similar results causing overlapping footpoints, which are overlaid onto the WSA-derived coronal holes shaded in red (positive/outwardly-directed field) and blue (negative/inwardly-directed field). Since this plot is a summary of the connectivity for all 12 realizations, the coronal hole shading represents any grid cell derived as open by any of the 12 realizations. The heliospheric current sheet (HCS) is overplotted in yellow and is nearly parallel with the solar equatorial plane. The average and standard deviation over all realizations of footpoint connectivity are calculated and shown in Table~\ref{table:ADAPT-WSA connectivity} columns (9)--(10), with the exception of STEREO~A. For this spacecraft, the values in Table~\ref{table:ADAPT-WSA connectivity} represent eight of the 12 ADAPT-WSA realizations which derived the magnetic footpoint at the southern polar coronal hole boundary. The other four realizations derive the source region of this event at the northern polar coronal hole boundary, with an average and standard deviation of 308.8$^{\circ}$$\pm$1.1$^{\circ}$ Carrington longitude, 63.1$^{\circ}$$\pm$0.3$^{\circ}$ heliographic latitude. This is common when the spacecraft is near the HCS (discussed in more detail below). It is important to note that in several instances on this table, the standard deviation of the footpoint connectivity falls within the 2.0$^{\circ}$ resolution limit of the WSA model. The standard deviation is only included to show the precision and range of variance among the 12 realizations. 

For all spacecraft except STEREO~A, the 12 ADAPT-WSA realizations produce very similar results for the model-determined magnetic footpoints. The largest standard deviation that was calculated was 8.4$^{\circ}$ for Parker Solar Probe's longitudinal footpoint connectivity (purple). This is likely because Parker Solar Probe observed this event on the solar far-side as seen from Earth, where there are no observations of the photospheric field to update our solution. All other spacecraft have nominal standard deviations in footpoint latitude and Carrington longitude, giving us more confidence in our results.

When comparing the results from the PFSS model vs. WSA, both the model-derived polarities (Table~\ref{table:ADAPT-WSA connectivity} column (8)) and the footpoint connectivity agree overall, with the exception of the magnetic footpoint derived for Parker Solar Probe. The PFSS model derives the source region of the Parker Solar Probe-observed event on the boundary of the northern polar coronal hole extension (at 28.6$^\circ$ latitude, 146.7$^\circ$ Carrington longitude), whereas the ADAPT-WSA derived source region is at the boundary of the northern polar coronal hole (at 65.3$^\circ$ latitude, 131.1$^\circ$ Carrington longitude). Differences in the two model solutions could arise for a few reasons, a primary one being that Parker Solar Probe observed this event on the solar far-side where we do not have recent observations of the photospheric field to drive coronal models. Nevertheless, both models derived the footpoint locations of this event for all five spacecraft as originating from the boundaries of coronal holes, with each spacecraft situated within 5$^{\circ}$ of the HCS. When the solar wind originates from the HCS at locations where it is nearly parallel to the solar equatorial plane, there is increased uncertainty in the backmapped locations of the magnetic footpoints for observers at 1~\(R_\odot\) when using any coronal model. This is because a difference of a few degrees from the HCS (i.e., 1\,--\,2 model grid cells) could connect the spacecraft to either side of the streamer belt. It is also common in this scenario for the spacecraft-measured polarity to fluctuate between inward and outward connectivity as the spacecraft never becomes sufficiently separated from the HCS. However, for this event, both models accurately derive the solar wind magnetic field polarity, which is measured by each of the five spacecraft in situ (Table~\ref{table:ADAPT-WSA connectivity} column (8)), giving us more confidence in our results. 

Selecting the best ADAPT input map to drive both models is particularly challenging for this event because the spacecraft were widely separated in longitude, whereas this type of modeling produces the most accurate results for spacecraft connected to the most recently added photospheric field observations (i.e., in this case Earth and STEREO~A). Additionally, Parker Solar Probe and BepiColombo observed this event on the solar far-side. Solar Orbiter was also located on the far-side; however, the spacecraft was connected to the near-side (i.e., the solar hemisphere visible from Earth) at 1~R$_\sun$. Lastly, there were two far-side active regions (ARs) that rotated onto the near-side on 19 and 22 April (labeled with an ``$\times$'' in  Fig.~\ref{ADAPT-WSA Connectivity}). Although they are not visible in the ADAPT map from 17 April, the locations of these ARs are labeled with crosses in Fig.~\ref{ADAPT-WSA Connectivity}. New far-side AR emergence is problematic for all coronal models \citep{Wallace2022}. To account for this evolution, we test various input maps from 17\,--\,23 April into both the PFSS model and WSA. We find that the connectivity for each spacecraft does not change drastically by using any particular map in this date range. Therefore, we select a map from the time closest to the SEP event, 17 April 2021 at 16:00:00~UT. It is important to note that the two far-side ARs fall inside the visible hemispheres observed by Solar Orbiter and STEREO~A during the time of the SEP event, making it possible to identify if any of these ARs are associated with a solar flare. One of the far-side ARs located at $-$19.09$^\circ$ latitude, 204.73$^\circ$ Carrington longitude was associated with a solar flare (discussed in detail in Sect.~\ref{sec:remote_obs}). Additionally, we can be confident that none of these far-side ARs produced flares within the longitudinal sector in which no remote observations of the solar corona are available (i.e., from $\sim$45\,--\,125$^\circ$ Carrington longitude, shaded in orange in Fig.~\ref{ADAPT-WSA Connectivity}) comprising the location of Parker Solar Probe, situated at 104$^\circ$ Carrington longitude.  

\begin{table*}
\caption{Magnetic connectivity between spacecraft and the Sun. Columns (1)--(4) present the respective observer and its location in Carrington coordinates (with the first row providing the flare location). Column (5) lists the measured solar wind speed, (6)--(7) and (9)--(10) provide the backmapped magnetic footpoints of the observer at the solar surface using the simple PFSS and the ADAPT-WSA models, respectively. Column (8) presents the magnetic field polarity observed (O) in situ and modeled (M) by ADAPT-WSA  and PFSS.}
\label{table:ADAPT-WSA connectivity}      
\centering          
\begin{tabular}{l cccc|cc|ccc}     
\toprule
(1) & (2) & (3) & (4) & (5) & (6) & (7) & (8) & (9) & (10) \\
  & & & & & \multicolumn{2}{c}{PFSS$^{(a)}$} &  \multicolumn{3}{c}{ADAPT-WSA$^{(a)}$} \\ 
  Spacecraft & r & Long.$^{(b)}$ & Lat.$^{(b)}$ & V$_{obs}$ & Long.$^{(b)}$ & Lat.$^{(b)}$ & Polarity  & Long.$^{(b, c)}$ & Lat.$^{(b, c)}$ \\
    &  (au) & ($^{\circ}$) & ($^{\circ}$) & (km s$^{-1}$)  & ($^{\circ}$) & ($^{\circ}$) & (O, M) & ($^{\circ}$) & ($^{\circ}$) \\
  \hline
  Flare & --- & 203 & $-$17 & --- & --- & --- & --- & --- & --- \\
  BepiColombo & 0.63 & 180.3 & $-$7.2 & 400$^{(d)}$ & 202.5 & 62.4 & (+, +) & 204.1$\pm$2.8 & 62.2$\pm$2.1   \\
  PSP  & 0.42 & 104.3 & 3.8 & 328  & 146.7 & 28.6 & (+, +) & 131.1$\pm$8.4 & 65.3$\pm$1.1 \\
  Solar Orbiter  & 0.84 & 215.8 & 0.4 & 375 & 266.7 & $-$21.5 & (--, --) & 267.9$\pm$0.2 & $-$21.8$\pm$0.1  \\
  STA  & 0.97 & 260.5 & $-$7.2 & 385 & 329.1 & $-$32.7 & (--, --) & 329.3$\pm$0.3 & $-$32.5$\pm$0.2  \\
  L1  & 1.00 & 314.2 & $-$5.4 & 601 & 347.4 & $-$21.8 & (--, --) & 347.4$\pm$0.1 & $-$22.5$\pm$0.1 \\
\bottomrule
\end{tabular}
\tablefoot{
\tablefoottext{a}{PFSS and ADAPT-WSA footpoints at 1~R$_\sun$;}
\tablefoottext{b}{Longitude and latitude values are given in the Carrington coordinate system;}
\tablefoottext{c}{Average and standard deviation values calculated from all 12 realizations of the ADAPT-WSA.}
\tablefoottext{d} This is not an observational value but a nominal value consistent with ENLIL simulations
}
\end{table*}

\begin{table}
\caption{Separation angles between location of the flare and spacecraft magnetic footpoints based on ADAPT-WSA values}
\label{table:sep_angles}      
\centering          
\begin{tabular}{l rrr}     
\hline   
  Spacecraft & Lon. sep. & Lat. sep. & Total sep. \\
   / Location& ($^{\circ}$) & ($^{\circ}$) &  ($^{\circ}$) \\
  \hline
  BepiColombo & 1.1 & 79.2&  79.2\\
  PSP & -71.9 & 82.3 &  98.1\\
  Solar Orbiter & 64.9 & 4.8 &  61.0 \\
  STA & 126.3 & 15.5 & 108.7 \\
  L1 & 144.4 & 5.5 &  127.3 \\
\hline                    
\end{tabular}
\end{table}

\section{Remote-sensing observations of the solar corona} \label{sec:remote_obs}

\subsection{Observations of the associated flare}\label{sec:flare}

\begin{figure}  
   \centering
	\includegraphics[width=0.5\textwidth]{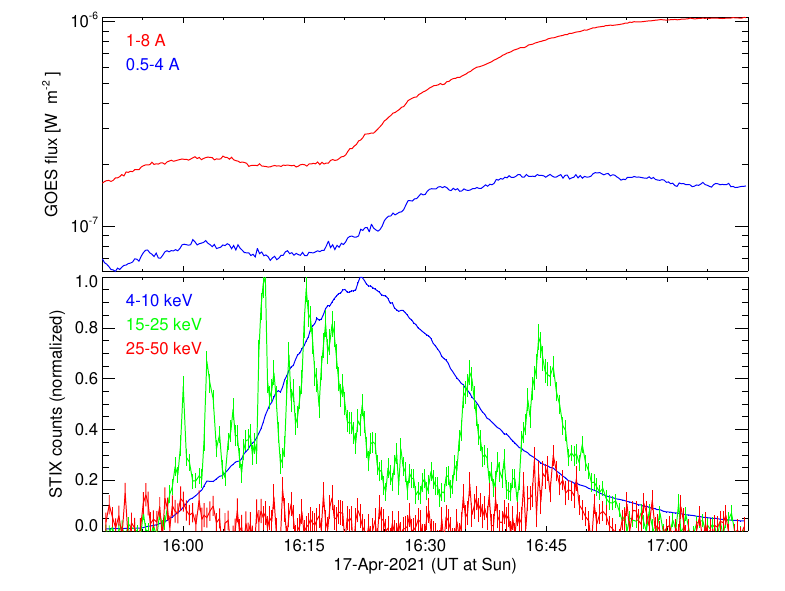}
	\caption{X-ray observations of the associated flare. Top: Soft X-ray fluxes as recorded by GOES-16. Bottom: Normalized background-subtracted STIX count rates integrated over two different energy bands. Note the gradual evolution of the thermal emission at 4-10~keV (blue) as opposed to the multiple non-thermal spikes seen at 15-25~keV (green) and at 25-50~keV (red; multiplied by 0.3 for clarity). For both GOES and STIX, times have been shifted so that they refer to UT at the Sun.} 
	\label{fig_stix_lc}
\end{figure}

The SEP event was associated with a solar flare occurring in the active region that was assigned the NOAA AR number 12818 when it rotated onto the Earth-facing hemisphere three days later. While the flare was clearly visible in the field of view of the STEREO~A EUVI instrument, it was initially occulted as seen from from Earth, but later phases of the eruption could be seen above the limb. Starting around 15:45~UT, the eruption of a flux rope was observed in EUV with SDO/AIA above the northeastern limb. Note that because we use observational assets at different locations, we have shifted all times pertaining to flare observations to UT at the Sun. From 16:03~UT on, the cusp-shaped top of a flaring arcade became visible in the 131~\AA\ channel. At wavelengths corresponding to lower temperatures, flaring loops appeared only after 18:15~UT, consistent with a considerable occultation angle. The GOES soft X-ray flux started to increase at 16:15~UT and peaked at 17:10~UT as a GOES class of B9.7 (see top panel in Fig.~\ref{fig_stix_lc}).

In contrast to Earth-based assets, the whole flare was visible from Solar Orbiter, and observed in hard X-rays (HXR) with STIX. We show STIX count rates integrated over two energy bins in the bottom panel of Fig.~\ref{fig_stix_lc}. The counts are background-subtracted and normalized to the peak count rate in the two ranges. The thermal HXR emission at 4--10~keV (generated by the hot plasma) increased from 15:55~UT onward, peaked at 16:22~UT, and decayed to pre-event background levels only at around 19:00~UT, thus indicating a long-duration event. Again, these times refer to when events have happened on the Sun. At Solar Orbiter, they were observed 7~min later. Based on a statistical comparison of STIX and GOES/SXR X-ray fluxes for flares that were fully visible for both instruments, the true GOES class can be estimated to $\sim$C5\footnote{See the STIX website for a description of the method: \url{https://datacenter.stix.i4ds.net/wiki/index.php?title=GOES_Flux_vs_STIX_counts}. The discrepancy between the B9.7 obtained from actual GOES observations and the class estimate from STIX is mainly due to occultation of the majority of hot flare plasma as seen from Earth.} This is also shown by the fact that the GOES fluxes peak more than half an hour after the STIX thermal count rate, since GOES sees significant emission only when larger loops due to magnetic reconnection become filled by hot plasma later in the event.

Above 15~keV, the STIX light curves show the spiky behavior typical for the non-thermal HXR emission generated by accelerated electrons that precipitate into the chromosphere. At least 13 non-thermal spikes are identified. While the non-thermal emission phase is usually restricted to a few minutes in typical C-class flares \citep[cf.][]{Veronig2002}, this flare shows non-thermal emission over 50~min. In particular, there are two major peaks in the late phase that in contrast to the other show emission above 25~keV, indicative of a comparatively harder spectrum. In Appendix~\ref{flare_ap}, we provide a full spectral analysis of the event using the STIX data.

\begin{figure*}  
   \centering
	\includegraphics[width=\textwidth]{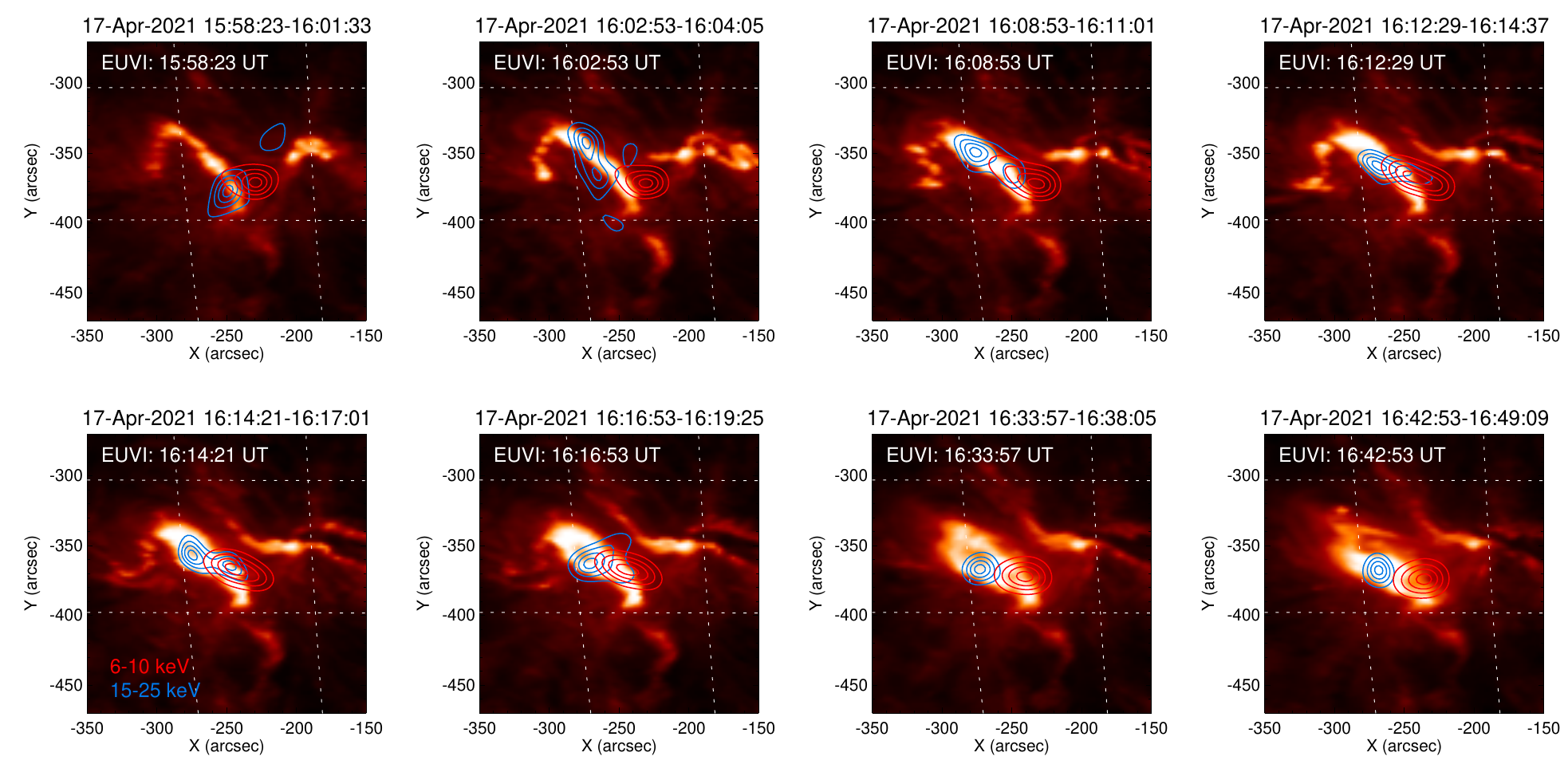}
	\caption{Flare evolution as seen in a series of STIX HXR images overlaid on STEREO~A EUVI 304~\AA\ images that have been rotated so that they correspond to the viewpoint of Solar Orbiter. Depicted are the coronal thermal source (red contours) and the chromospheric non-thermal footpoints (blue contours) reconstructed with the Expectation Maximization algorithm. The integration times (UT at the Sun) correspond to the eight non-thermal HXR peaks with the highest number of counts above 15~keV. Additionally, the observation times of the EUVI images are shown. For reference, a longitude-latitude grid (in Stonyhurst coordinates) with a spacing of 5$^{\circ}$ is overplotted.} 
	\label{fig_stix_imgs}
\end{figure*}

HXR images can be reconstructed from pixelated STIX science data. Figure~\ref{fig_stix_imgs} shows the HXR sources for the eight HXR peaks that had the largest number of non-thermal counts overplotted on STEREO~A EUVI 304~\AA\ images that have been rotated so that they correspond to the viewpoint of Solar Orbiter. In the EUVI frames, we mainly see the chromospheric flare ribbons, thus such a reprojection that assumes that all features are lying on the solar surface does not introduce significant projection artefacts. Red contours show the coronal thermal source (6--10~keV), and the blue contours show the chromospheric non-thermal footpoints (15--25~keV). All images are reconstructed with the Expectation Maximization algorithm \citep{Massa2019}. Normally, the precise source locations are provided by the STIX Aspect System \citep[][]{Warmuth2020}. However, Solar Orbiter was at a heliocentric distance of 0.84~au, which is too far from the Sun to provide a reliable pointing solution. We therefore apply the average image displacement obtained from other events from the cruise phase where aspect information was available as implemented in the STIX imaging software. This method yields a mean position uncertainty of about $\pm$10$\arcsec$. The flare position (plotted in heliocentric Cartesian coordinates in Fig.~\ref{fig_stix_imgs}) is at the heliographic coordinates of E111S18 (203$^{\circ}$ Carrington longitude). As seen from Earth, this corresponds to an occultation angle of 20$^{\circ}$.

The coronal thermal source undergoes little evolution little evolution throughout the flare. One might expect to observe a pair of non-thermal sources consistent with the footpoints of the magnetic loops containing the hot plasma \citep[cf.][]{Fletcher2011}. However, most HXR peaks only show a single footpoint at the eastern edge of the thermal source. The issue here is that while the individual non-thermal peaks are all very clearly defined, the total number of counts above 15~keV per peak is quite low, on the order of 1\,000--2\,000 counts. This is marginal for imaging, particularly in case there is more than one source present. Nevertheless, we find that all non-thermal peaks are originating from the same active region, and there is no evidence of a second remote source. While the presence of such a secondary source cannot be ruled out, it would have to be weaker than the main source by a factor of 5--10. We conclude that the footpoint brightness was very asymmetric in this event, with the eastern footpoint clearly dominating. This is consistent with the flare ribbons seen at 304~\AA, where also the southwestern ribbon is the dominating one. The different non-thermal peaks are not associated with changing footpoint locations.

\subsection{CME observations}\label{sec:CME}
\label{CME_obs}
   \begin{figure}
   \centering
   \includegraphics[width=0.49\textwidth]{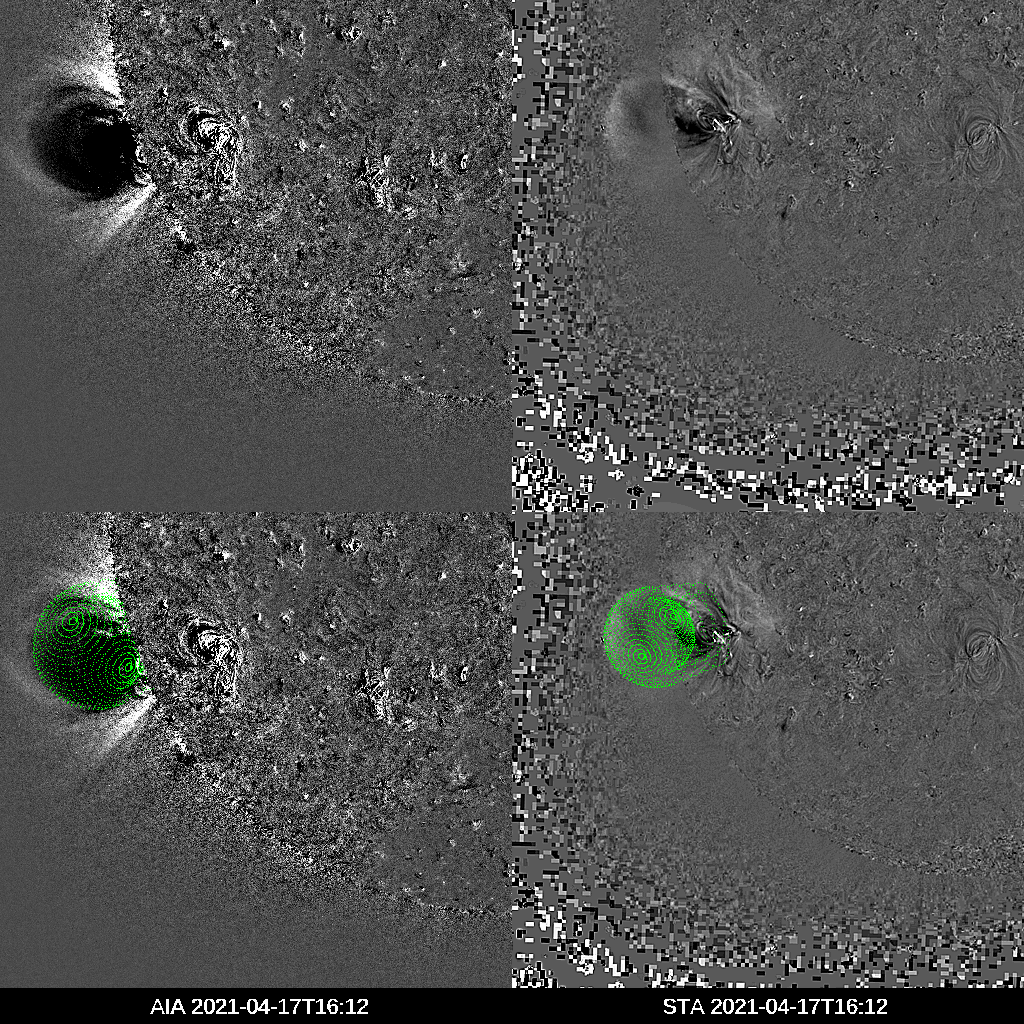}
      \caption{EUV observations by SDO/AIA (\textit{left}) and STEREO~A/EUVI (\textit{right}) at the same instant of time. The green mesh corresponding to the 3D reconstruction of the CME is overlaid to base-difference images shown in the \textit{upper panels}. }
         \label{fig:CME_AIA_EUI}
   \end{figure}
   \begin{figure*}
   \centering
   \includegraphics[width=0.9\textwidth]{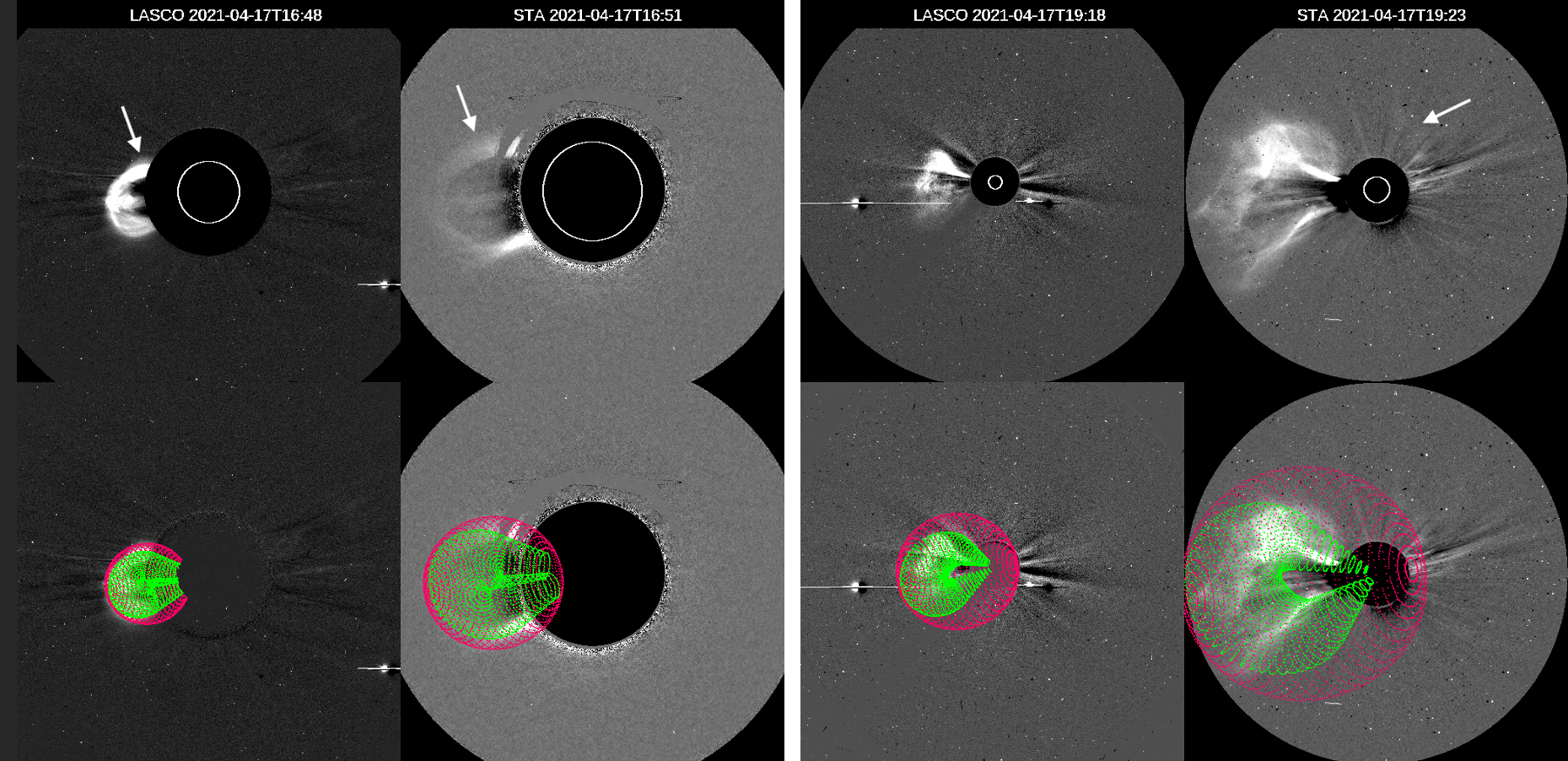}
      \caption{Base-difference images of the coronagraph observations by SOHO/LASCO/C2 and STEREO~A/COR1 (\textit{left}) and LASCO/C3 and STEREO~A/COR2 (\textit{right}) at different times. The green (red) mesh corresponding to the 3D reconstruction of the CME (CME-driven shock) is shown in the \textit{lower panels}. The white arrows indicate the signatures of the CME-driven shock.  }
         \label{fig:CME_COR2A-COR3}
   \end{figure*}

A CME erupted from the same active region as the associated flare, NOAA AR 12818, located in the southern solar hemisphere at  heliographic coordinates E111S18 (203$^{\circ}$ Carrington longitude) on the day of the event. The active region entered Earth's field of view on 20 April. The evolution of the CME was observed $\sim$30$^\circ$ from the eastern limb towards the central meridian as seen from STEREO~A. The bottom panels of Fig.~\ref{fig:CME_AIA_EUI} show EUV images taken by SDO/AIA (left) and STEREO~A/EUVI (right) at $\sim$16:10~UT (all times refer to observation times at the spacecraft). At this time, we observe the first clear indication of the eruption, when the CME exhibits prominent signatures of over-expansion \citep[e.g.][]{Patsourakos2009}, evidenced by the bubble-like appearance in EUV images. At the same time, the flare ribbons are activated along an arched path. Between 16:12~UT and 16:25~UT, the CME continued to expand as it reached the edge of the field-of-view of the EUV instruments. The left panel of Fig.~\ref{fig:CME_COR2A-COR3} shows the time ($\sim$16:50~UT) at which the CME was clearly visible by the coronagraph imagery: COR1 (second column) and LASCO/C2 data (first column). As shown in the images, the CME morphology in white light images at these heights is consistent with classic flux-rope characteristics, namely featuring the presence of a bright outer rim followed by a cavity \citep[e.g.][]{Vourlidas2013}.

The angular separation between STEREO~A and Earth was $\sim$53$^\circ$, which still enables a reliable 3D reconstruction of the CME \citep[e.g.][]{balmaceda2018, Verbeke2022}. For this purpose, we used the graduated cylindrical shell \citep[GCS;][]{Thernisien2006GCS} model to reproduce the CME appearance by fitting pairs of EUV (at distances below 1.5~$R_{\odot}$) and white-light (from $\sim$2.5 to 22~$R_{\odot}$) images. The model consists of a croissant-like structure fully described by six free parameters: three for location and orientation (latitude and longitude of the CME leading-edge, and tilt or inclination of the main axis of the CME with respect to the solar equator), and three for the geometry (height; aspect ratio, which sets the rate of expansion versus the height of the CME; and angular separation of the legs or half-angle). The sensitivity (deviations) in the parameters of the GCS analysis is given in Table~2 of \cite{Thernisien2009}. It is worth noting that these parameters are sensitive to image quality and human interpretation \citep{Verbeke2022}. The routine used for the reconstruction is \textit{rtcloudwidget.pro}, available as part of the \textit{scraytrace} package in the SolarSoft IDL library\footnote{\url{http://www.lmsal.com/solarsoft/}}.

The bottom panels of Fig.~\ref{fig:CME_AIA_EUI} and Fig.~\ref{fig:CME_COR2A-COR3} show the GCS fit analysis, where the green mesh represents the flux rope structure. The 3D reconstruction shows that the CME follows a non-radial path towards the solar equator in the early evolution with the latitude varying from $-14^{\circ}$ to $-9^\circ$ from 16:12 to 17:23 UT. The longitude and the tilt angle, meanwhile, do not show deviations, staying at fixed values of $-116^\circ$ and $-70^\circ$, respectively. The GCS parameters were chosen to better describe the portion of the CME oriented towards Solar Orbiter, as the croissant-like shape used for the fitting could not represent fully the CME due to its non-radial propagation and curved axis. The last term was introduced by \cite{Rodriguez-Garcia2022CME} to refer to flux ropes that may deviate from the nominal semi-circular (croissant-like) shape and have instead an undulating axis. 
The CME speed at the leading-edge estimated from the linear fit to the height--time measurements is 880~km~s$^{-1}$. The width of the CME is estimated based on \citet{Dumbovic2019}, where the semi-angular extent in the equatorial plane is expressed by ${R\textsubscript{maj}-{(R\textsubscript{maj}-R\textsubscript{min})} \times |tilt|/90}$. Then, the total angular extent of the CME is 46$^{\circ}$. The value of $R\textsubscript{maj}$ (face-on CME half-width) is calculated by adding $R\textsubscript{min}$ (edge-on CME half-width) to the half-angle, and $R\textsubscript{min}$ was calculated as the $\arcsin(aspect~ratio)$. The CME width deviation was derived from the mean half-angle error, estimated by \cite{Thernisien2009} as +13$^{\circ}$/$-7^{\circ}$.
Thus, at the latest time of the 3D reconstruction at 19:23 UT,  corresponding to a CME height of 15.5~$R_{\odot}$, the narrow CME ($\sim$46$^{\circ}$) is propagating in the direction E116S09 with a moderate speed ($\sim$880~km~s\textsuperscript{-1}).


\begin{table*}
\caption{First intersection between the coronal shock and magnetic field lines connecting to the spacecraft as determined with the ADAPT-WSA model (Sect.~\ref{sec:connectivity}). Times refer to observation times at 1~au.}  

\label{table:shock_connectivity}      
\centering          
\begin{tabular}{l cccc}      
\hline  
  \\ Spacecraft & r  & V$_{obs}$ & Estimated & Shock    
  \\  &  (au) & (km s$^{-1}$)  &  Intersection  & Height\textsuperscript{a}   \\
  & & & [UT] & [R$_\sun$] \\
  \hline
  BepiColombo & 0.63 & 400 & 16:30$\pm$3 min & 1.58  \\
  Parker Solar Probe  & 0.42 & 328  & 17:19$\pm$3 min & 1.45  \\
  Solar Orbiter  & 0.84 & 375 & 16:55$\pm$3 min & 1.07  \\
  STEREO~A  & 0.97 & 385 & 17:24$\pm$3 min & 1.04  \\
  L1  & 1.00 & 601 & 17:30$\pm$3 min & 1.01 \\
\hline                    
\end{tabular}

\footnotesize{ \textbf{Notes}. \textsuperscript{a} Height from the solar center of the intersection point between the reconstructed CME-driven shock and the field line connecting to the respective spacecraft.} 
\end{table*}


\subsection{EUV wave observations}
\label{sec:EUV_obs}
\begin{figure*}  
   \centering
	\includegraphics[width=\textwidth]{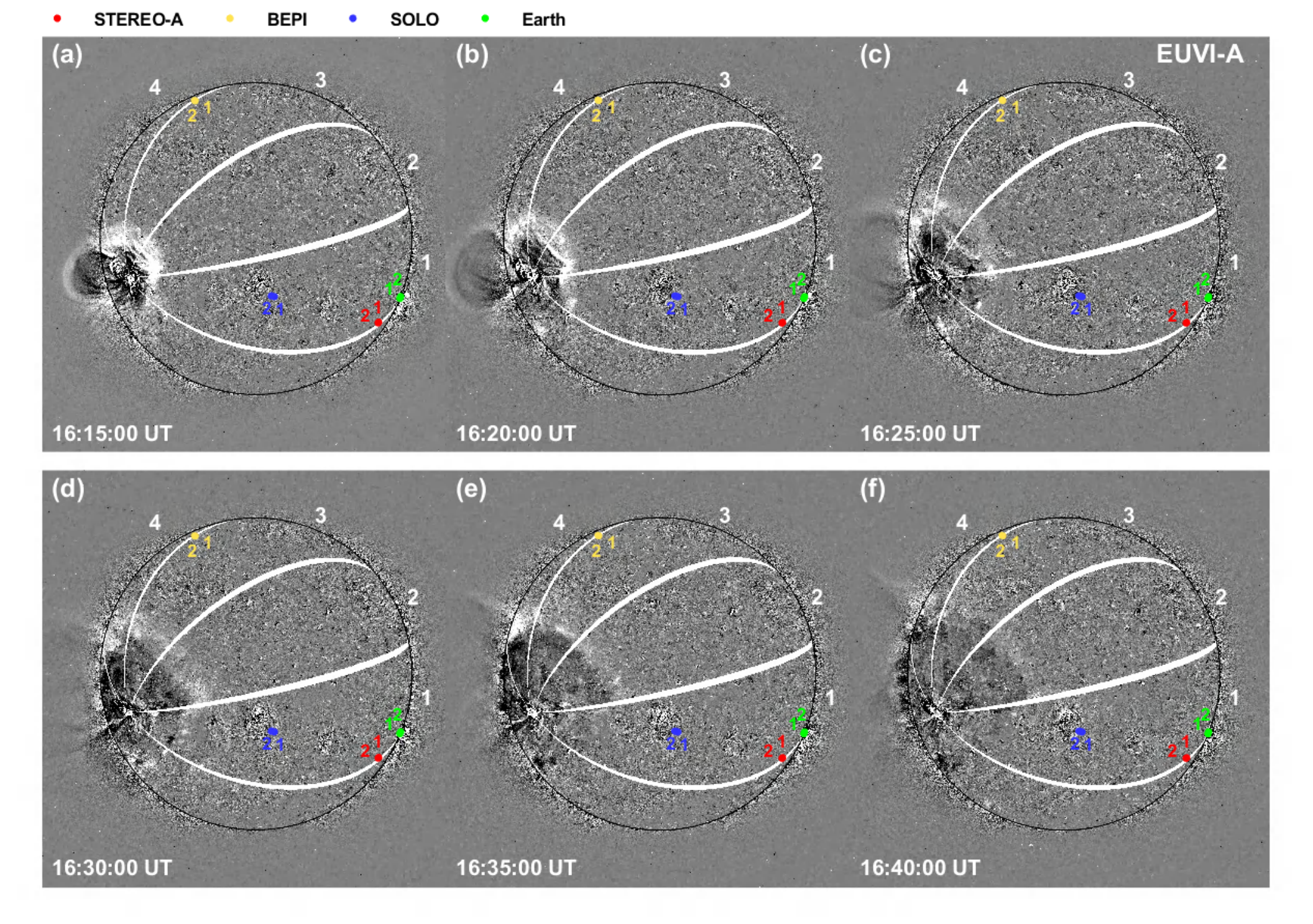}
	\caption{EUV wave overview as observed in STEREO~A/EUVI 195~{\AA} running-difference images from 16:15 to 16:40~UT. We follow the EUV wave in four angular sectors 1--4. A movie accompanying the figure is available online (movie1). Markers show magnetic footpoints derived for STEREO~A (red), BepiColombo (yellow), Solar Orbiter (blue), and ACE (green) spacecraft. The magnetic footpoints are determined using a combination of ballistic backmapping in the heliosphere and backmapping below the source surface using ADAPT-WSA to 1~Rs (points 1), a standard PFSS model to 1~$R_{\odot}$ (points 2), and to a height of 100~Mm above the photosphere (points 3). As seen in the accompanying video, the EUV wave reaches the BepiColombo footpoints at around 16:55~UT (point 3) and at around 17:00~UT (points 1 and 2). Times refer to the observation time at STEREO~A.} 
	\label{fig_wave_overview}
\end{figure*}

Figure~\ref{fig_wave_overview} together with the accompanying movie shows an overview of the EUV wave evolution in STEREO~A/EUVI 195~{\AA} running-difference images created with a lag of 150~s. The prominent signatures of the EUV wave, which exhibits a quasi-circular propagation away from the eruptive center over the solar disk, are already clearly seen around 16:10~UT and can be followed for about 40~min in STEREO~A. As follows from the derivation of EUV wave kinematics and perturbation characteristics in Appendix~\ref{EUV_wave_ap}, the EUV wave on the solar disk extends to about 680~Mm from the source region with a mean velocity of 223--327~km~s$^{-1}$. Above the solar limb, in the northern direction, the EUV wave can be followed to a distance of about 740~Mm, propagating with speeds of 260--450~km~s$^{-1}$ (for heights increasing from 1.05 to 1.15~$R_{\odot}$). At the same time, in the southern direction, the wave is seen only to about 350~Mm, propagating with speeds of 220--300~km~s$^{-1}$. As seen from the movie, accompanying Fig.~\ref{fig_wave_overview}, the EUV wave reaches the backmapped magnetic footpoints of BepiColombo (yellow) at around 16:55~UT (point 3) and at around 17:00~UT (points 1 and 2). Each of the points 1, 2 and 3 correspond to the spacecraft's magnetic field footpoints obtained using either ADAPT-WSA (point 1), the PFSS model at 1 R$_{\odot}$  (point 2) and PFSS at a height of 100 Mm above the photosphere (point 3). The footpoints of other spacecraft, which lie on the visible hemisphere as seen by STEREO~A, are displayed in other colors as described in the figure legend.

\subsection{CME-driven shock observations}
\label{sec:shock_obs}

In white-light images  the signatures of a shock wave formed in front of the expanding flux rope are faint. We use calibrated, excess-mass images (i.e. pre-event image subtracted) and display them in Fig.~\ref{fig:CME_COR2A-COR3}. By 16:30~UT, when the CME front is visible in COR1 FOV, the EUV wave is still visible on the surface. The CME exhibits a diffuse front ahead the brighter rim, more clearly seen at the north flank in both COR1 and LASCO-C2 images (marked with white arrows in the left panels of Fig.~\ref{fig:CME_COR2A-COR3}). This typical ``two-front'' morphology is generally interpreted as evidence of a CME-driven shock in white-light images \citep{ontiveros2009,Vourlidas2013}. The EUV wave is visible on the disk until 16:50~UT. By this time, the CME reaches the edge of COR1 FOV. At larger distances, namely COR2 FOV, a diffuse arched-shaped feature (white arrow in right panel of Fig.~\ref{fig:CME_COR2A-COR3}) is also seen propagating on the northwest quadrant. This feature is best visible between 18:23~UT and 20:23~UT in COR2 FOV and may result from the compression of a relatively weak shock wave against the underlying coronal structures. We use these features to estimate the angular extension of the shock. For this, a spherical surface \citep{kwon2017} is used to model the 3D appearance of the shock (represented by the red mesh in Fig.~\ref{fig:CME_COR2A-COR3}). 

From the 3D reconstruction, we estimate that the shock reaches a speed of $\sim$1500~km~s$^{-1}$ below 5~R\textsubscript{$\odot$} and is propagating on the direction between Solar Orbiter and BepiColombo, consistent with the direction estimated from the CME 3D modeling in Sect.~\ref{sec:CME}. Following \cite{kwon2017}, shown in their figure 2, we determine the angular width of the shock to be $\sim$180$^\circ$ at 19:23~UT, corresponding to a height of the shock nose of $\sim$16.3~R\textsubscript{$\odot$}. Table~\ref{table:shock_connectivity} shows the timing of the first intersection between the coronal shock reconstruction and the magnetic field lines  obtained with the ADAPT-WSA model connecting to the different spacecraft analyzed in this study. All times refer to observation times at 1~au.


\subsection{Radio observations}\label{sec:radio_obs}

   \begin{figure}
   \centering
   \includegraphics[width=0.49\textwidth]{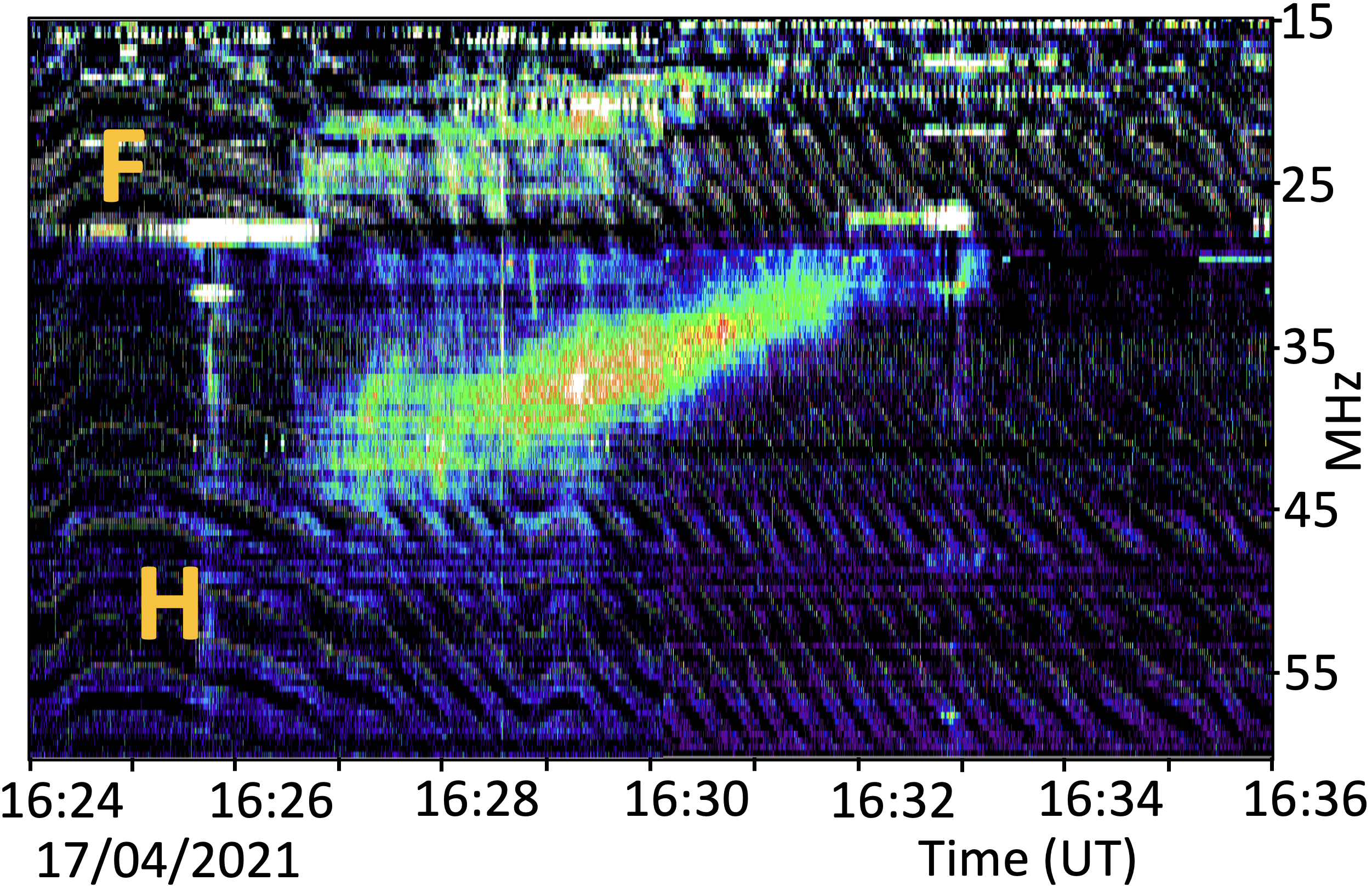}
      \caption{ Decametric type II radio burst observed by the Swiss-Landschlacht e-Callisto receiver. Fundamental and second harmonic lanes are marked by F and H, respectively.
              }
         \label{fig:ground_based_spectra}
   \end{figure}

The radio emission associated with the eruption on 17 April 2021 was observed by ground-based and space-borne instruments and includes both type II and type III radio bursts. Type II (TII hereon) bursts are related to acceleration of energetic electrons at shock waves \citep[][and references therein]{Krasnoselskikh85, Benz88}, while type III (TIII hereon) radio bursts are signatures of fast electron beams propagating via open (or quasi-open) magnetic field lines from the corona to interplanetary space \citep[][]{Zheleznyakov65, Jebaraj22}. In Fig.~\ref{fig:ground_based_spectra}, we present a dynamic radio spectrum using measurements from the ground-based e-Callisto instrument located at Swiss Landschlacht, providing observations in the 80~MHz to 10~MHz range. This spectrum shows a poorly-observed decametric TII radio burst starting at 16:26~UT and exhibiting both fundamental and second harmonic emission lanes. Harmonic emission is brighter compared to the fundamental due to the large angle between the source and Earth \citep[directivity of harmonic emission has a wider angle than the fundamental;][]{Zheleznyakov70a, Tkachenko21}. 

\begin{figure*}
   \centering
   \sidecaption
    \includegraphics[width=0.9\textwidth]{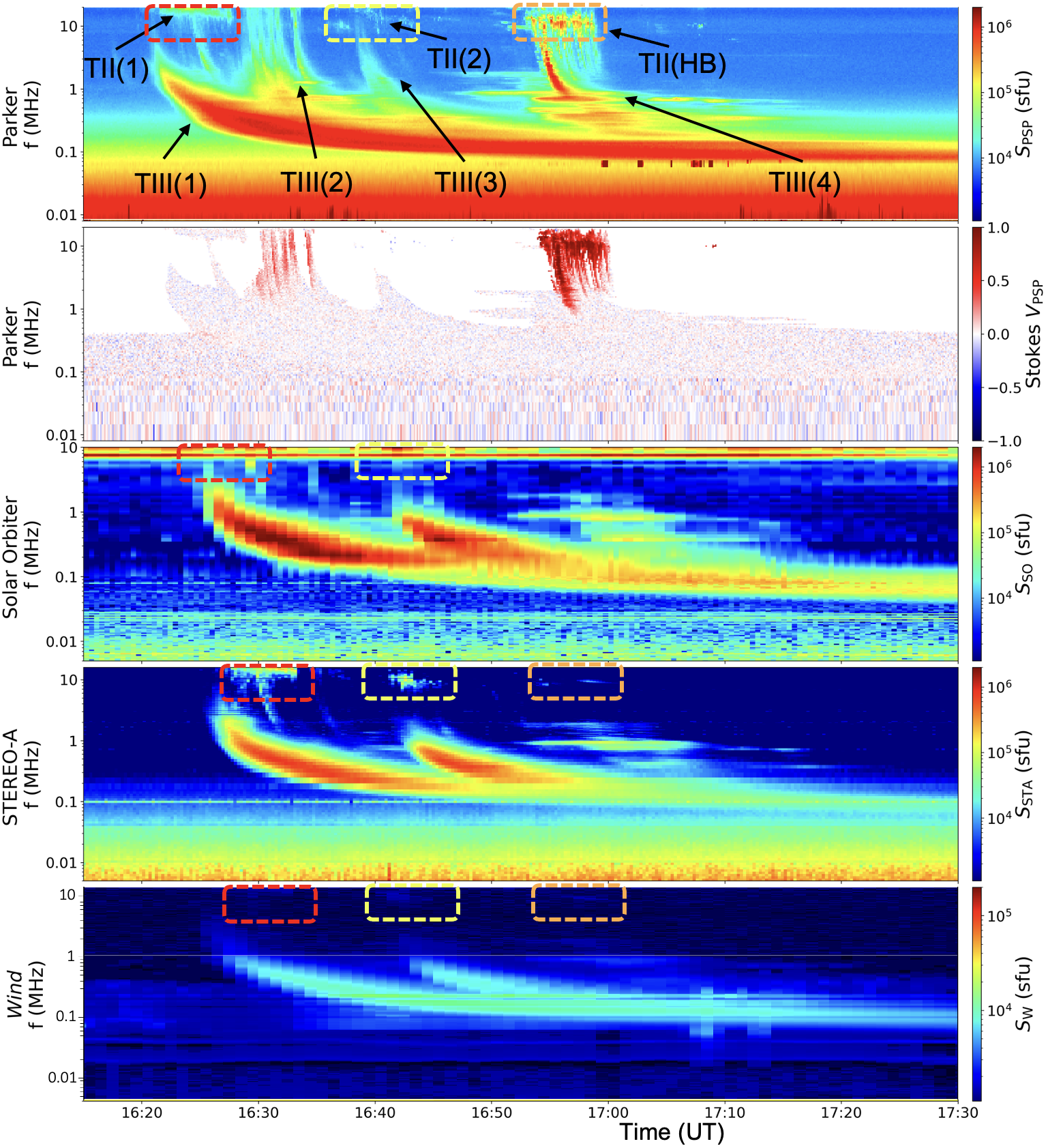}
      \caption{Radio spectrograms from all available space-borne observatories. Panels 1 \& 2 show Stokes I and the net polarization (Stokes V/I) from Parker Solar Probe. Panels 3 -- 5 show the Stokes I measurements from Solar Orbiter, STEREO~A, and Wind, respectively. The different bursts are indicated in panel 1. The TII bursts are marked in other panels by rectangular boxes of red (TII(1)), yellow (TII(2)), and orange (TII(HB)). 
              }
         \label{fig:radio_multi_sc}
   \end{figure*}

Figure~\ref{fig:radio_multi_sc} presents a combined dynamic radio spectrum of all available hecto-kilometer observations from all the observing spacecraft, namely, Parker Solar Probe, Solar Orbiter, STEREO~A, and Wind. The spectrum shows a number of different radio emissions including groups of TIII bursts and distinctly patchy TII emission. A list of the starting times of each radio burst as observed by different spacecraft is provided in Table~\ref{table:timing}. An interesting aspect is that most of the TIII radio burst groups were best or exclusively observed by PSP/FIELDS/RFS partly due to its enhanced resolution and sensitivity \citep{Pulupa2017}. But the radial distance of the spacecraft from the Sun and also the directivity of the emission at the source \citep{Thejappa12} also play a key role. 
\citet{Jebaraj20} have suggested that the intensity of a radio burst is higher in the direction of the source propagation. Therefore, the intensity of the radio emission at different observers depends on both the position of the observing spacecraft and the intrinsic directivity of the radio source. This explains the emission intensity at Parker Solar Probe, which was the closest spacecraft to the Sun during the flare--CME event. 
As we show in Fig.~\ref{fig:solar-mach_and_multi_sc_SEP} (left), the spacecraft were located at different longitudes and radial distances (Table~\ref{table:ADAPT-WSA connectivity}). In the following, we use the Parker Solar Probe spectra (Fig.~\ref{fig:radio_multi_sc} panel 1) to describe the spectral morphology of the TIII and TII bursts. The identification of the different TIII and TII bursts in Parker Solar Probe observations provide the foundation for the multi-spacecraft directivity analysis presented in Appendix~\ref{radio_ap}, where we combine the identification of the bursts with the cross-calibrated data from other spacecraft to locate the source in interplanetary space (Fig.~\ref{fig:radio_directions}).


The different groups of TIII radio bursts and multiple components of the TII bursts exhibit interesting characteristics as far as their spectral morphology is concerned. The first and third TIII groups (TIII(1) and TIII(3) hereon) are rather faint at the short-hectometer wavelengths and appear to be intense across all spacecraft observations. The second and fourth type III groups (TIII(2) and TIII(4) hereon) were observed almost only by Parker Solar Probe and consisted of a large number of individual TIII bursts that were better distinguishable in the short-hectometer wavelengths. This indicates that during the time when TIII(2) and TIII(4) were observed, there were multiple smaller episodes of electron acceleration and subsequent release into the open magnetic field lines in the direction of Parker Solar Probe. 

This is further corroborated by the polarization measurements made by PSP/FIELDS/RFS, which indicate that the energetic electron beams (the sources of type III bursts) were strongly directed towards Parker Solar Probe. Appendix~\ref{radio_ap} discusses the details of the polarization measurements extensively. The results indicate that TIII(2) and TIII(4) originated from a region of negative magnetic field polarity. The relatively high degree of polarization (Fig.~\ref{fig:radio_multi_sc} panel 2) of TIII(4) at its origin also hints at a region with high magnetic field strength (e.g., ramp of a quasi-perpendicular shock wave). 
As for the magnetic connectivity, Parker Solar Probe also observed Langmuir waves \citep[see;][]{Ginzburg58, Melrose85} on multiple occasions, close to the local plasma frequency. This indicates that the electron beams generating TIII(2) and TIII(4) were directly sampled by Parker Solar Probe.


The TII bursts associated with the event are distinctly patchy and complex in the hectometer wavelengths (see Fig.~\ref{fig:radio_multi_sc}). It is likely that the different TII components are associated with the same shock wave but at different regions. All TII bursts also appear bursty in terms of intensity variations (marked in Fig.~\ref{fig:radio_multi_sc}), suggesting an on-and-off emission process at the shock front \citep[][]{Mann95c}. On-and-off TII bursts are believed to be emitted from locations on the shock wave where the upstream plasma conditions induce rapid changes to its obliquity and other characteristics \citep[e.g.,][]{Schmidt14, Jebaraj21, Kouloumvakos21}. Due to their patchy nature, it is somewhat difficult to distinguish between them, but we identify two main TII radio components (TII(1) and TII(2), which are marked in Fig.~\ref{fig:radio_multi_sc}) observed in the short hectometer wavelengths (16--13 MHz) together with TIII(1) and TIII(3), and TIII(4), respectively. Overall, it seems these patchy TII bursts were observed from the start of the event and the beginning of TIII(1) and continued even after TIII(4). 

Furthermore, we note another interesting temporal and spectral phenomenon observed together with TIII(4), namely, the presence of TII herringbone-like features (TII(HB) hereon). The observation of such a feature may either indicate interaction of the shock wave and TIII(4), or that some of the electron beams generating TIII(4) may just be herringbones accelerated at the near-perpendicular shock front. Herringbone features with no clear backbone emission are often clear indicators of shock fronts with near-perpendicular geometry ($\theta_{Bn} = 87^{\circ} - 89.9^{\circ}$), which are able to accelerate electrons along either side of the magnetic field lines interacting with the said shock front \citep[][]{Mann2005}. We also note that observations beyond TIII(4) of patchy TII bursts may be associated with both TII(1) and TII(2).


A similar mechanism may also contribute to TIII(2), which was also observed uniquely by Parker Solar Probe. The polarization analysis of TIII(2) presented in Appendix~\ref{radio_ap} suggests that if there would have been a herringbone-like feature at the origin of these type III bursts, it would be observed in the decamater wavelengths. However, due to the lack of meter-decameter observations, it is not possible to make such a conclusion.  



Solar Orbiter was the second-closest radio observer radially and also the second-closest spacecraft to the flaring active region in terms of the magnetic connectivity (see Fig.~\ref{fig:solar-mach_and_multi_sc_SEP} and Fig.~\ref{fig:solar_mach_pfss}). The spacecraft observed mainly TIII(1) and TIII(3) and also TIII(4) at lower frequencies. Due to the limited survey-mode observations during the initial phase of the mission \citep[][]{Maksimovic20b}, the low resolution HFR observations from the Solar Orbiter/RPW instrument make it difficult to recognize the strongly patchy type II burst. However, the likely intensity variations (in the frequency range 16--5~MHz) of the type II bursts can be seen in Fig.~\ref{fig:radio_multi_sc} panel 3.

At the time of the event, STEREO~A was located almost diametrically opposite ($\sim$180$^\circ$) from Parker Solar Probe. It observed well the TIII(1) and TIII(3) and also partially the TIII(2). TIII(4) was observed faintly at lower frequencies at this location. This indicates that for STEREO~A the source region of the TIII(2) and TIII(4) may have been partially and fully occulted. A number of type II patches corresponding to the ones observed by Parker Solar Probe were also observed. TII(1) and TII(2), which are indicated by the red and yellow rectangles in Fig.~\ref{fig:radio_multi_sc}, were observed to be nearly as intense as in Parker Solar Probe. However, TII(HB) was considerably weaker (marked by the orange rectangle). Such variations in intensity may indicate that the source directivity was in the direction of the spacecraft, which observed the brighter emission. In this case, the faintness of TII(HB) in STEREO~A observations further supports that the source of the herringbones was likely located close to the line-of-sight of Parker Solar Probe and at the periphery of STEREO~A.   

Wind was the furthest spacecraft from the flare location and therefore only observed the low frequency parts of TIII(1) and TIII(3). Most other bursts were either too faint or not observed at all by the spacecraft. An interesting feature here is that Wind observed very faint signatures of both TII(1) and TII(2) as indicated by the rectangles in Fig.~\ref{fig:radio_multi_sc}. Their fluxes however were an order of magnitude smaller than the ones observed by STEREO~A. Considering that TII(1) was observed by ground based instrumentation, it is likely that the source of the emission was visible from Earth and therefore for Wind as well. 

The multi-vantage point observations also introduce the phenomena of time delay (light travel time to spacecraft). By combining the time delay and the intensity variations between different spacecraft, it is possible to locate the spatial position of the source at a given frequency. We present a detailed analysis of the radio source propagation in Appendix~\ref{radio_ap}. Figure~\ref{fig:radio_directions} shows the radio source locations of TIII(1) and TIII(3) estimated using a directivity model. The results of the analysis suggest that TIII(1) propagated in the longitude $-121.0^\circ\pm3.2^\circ$ (slightly east of the flare longitude), and the electron beam generating TIII(3) propagated in the longitude $-98.3^\circ\pm4.1^\circ$ (slightly west of the flare longitude). This fits the by-eye analysis of the radio bursts based on their visibility to each observer.




%


\section{Interplanetary context and SEP observations} \label{sec:in_situ_obs}
%

%
The heliospheric conditions through which particles  and CME-driven shocks propagate at the time of their release can significantly affect the SEP timing and intensity profiles \citep[e.g.][]{Laitinen2013, Dalla2020, Lario2022}. We use both multi-point solar wind and IMF observations and the WSA-ENLIL+Cone model \citep{Odstrcil2004} to provide a comprehensive understanding of the geometry, not only of the interplanetary structures and their possible influence in the propagation of the SEPs, but also of the shocks and their role in forming the  observed intensity--time profiles. 
In this section, we first describe the ENLIL simulation and then discuss in-situ plasma, magnetic field, and multi-spacecraft SEP observations. 

\subsection{The state of the interplanetary medium as derived with the ENLIL model}\label{subsec:ENLIL_model}
ENLIL is a global 3D MHD model\footnote{\url{https://ccmc.gsfc.nasa.gov/models/modelinfo.php?model=ENLIL\%20with\%20Cone\%20Model}} that provides a time-dependent background characterization of the heliosphere outside 21.5~R\textsubscript{$\odot$}. ENLIL uses time-dependent magnetograms as a background, into which spheroidal-shaped high-pressure structures without any internal magnetic field can be inserted to mimic observed CME-associated solar wind disturbances. ENLIL-modelled CMEs have an artificially higher thermal pressure to compensate for the lack of a strong magnetic field \citep[][and references therein]{Odstrcil2004}. To improve the characterization of the heliosphere, multi-point coronagraph observations are used to infer CME parameters, using the GCS model described in Sect.~\ref{sec:CME}. The inner boundary condition is given by the WSA V5.2 model, using inputs from the standard quick-reduce zero-point corrected magnetograms from GONG (GONGZ), available on the National Solar Observatory website\footnote{\url{ftp://gong2.nso.edu/QR/zqs/}}. In this case, the GONGZ magnetograms fit the in-situ solar wind speed and magnetic field polarity better (not shown). The reliability of the CME arrival predictions depends strongly on the initial CME input parameters, such as speed, direction, and width \citep{Lee2013,Mays2015,Kay2020}, but also on the errors that can arise in the ambient model parameters and on the accuracy of the solar wind background derived from the magnetograms \citep{Lee2013}. Based on \cite{Wold2018}, the mean absolute arrival-time prediction error is 10.4 $\pm$ 0.9 hours, with a tendency to an early prediction of $-4.0$ hours.

The magnetic connectivity at the onset time  of the SEP event is relevant to the understanding of the SEP observations, and considering the ENLIL-modelled varying solar wind conditions to calculate the IMF lines is an alternative to using the nominal Parker spirals. The preconditioning of the heliosphere and the interaction of the IP structures with the ambient solar wind that might be present at the SEP onset time can actively influence this connectivity \citep[][]{2012Masson,2021Palmerio,2022Lario}. Therefore, we choose an ENLIL simulation time from 10 to 24 April 2021 (i.e.\ from seven days before to seven days after the SEP event onset).  This interval encompasses possible previous CMEs as well as subsequent CMEs propagating through the structured solar wind streams up to a distance of 2.1 au. All these structures may influence the propagation of particles and CME-driven shocks arriving at the different spacecraft. For this purpose, the GCS 3D reconstruction process presented in Sect.~\ref{sec:CME} is also used for the other nine relevant CMEs erupting in the time range of 10--24 April. The CMEs details and model set-up parameters and the results of the simulations are available on the Community Coordinated Modeling Center (CCMC) website\footnote{\url{https://ccmc.gsfc.nasa.gov/database_SH/Laura_Rodriguez-Garcia_041322_SH_1.php}\label{footnote enlil run}}.

\begin{figure*}
   \centering
   \includegraphics[width=0.8\textwidth]{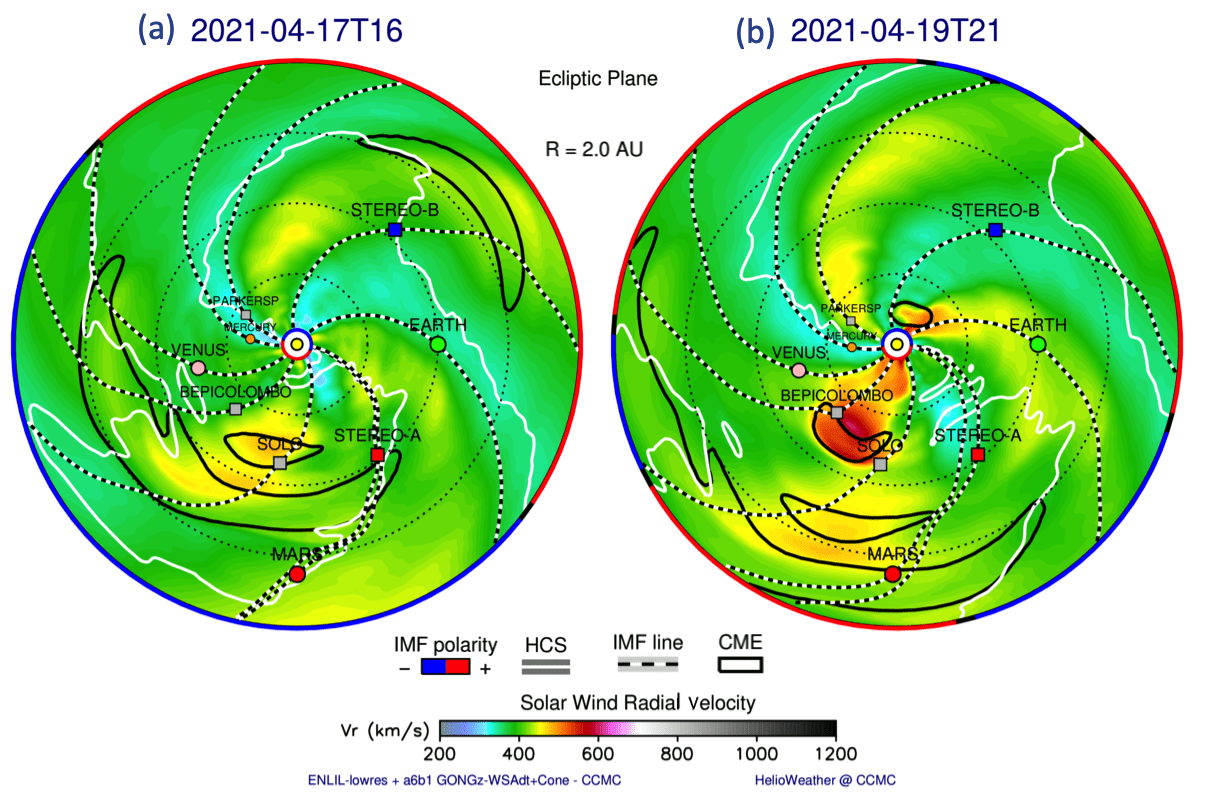}
      \caption{Radial velocity contour plot from the ENLIL simulation in the ecliptic plane. The black and white dashed lines represent the IMF lines, and the black contours track the ICME ejecta. The white lines correspond to the HCS, which separates the regions with opposite magnetic polarity, shown in blue (negative) or red (positive) on the outer edge of the simulation region. \textit{Left panel}: magnetic connectivity of the different spacecraft around the particle solar release time. \textit{Right panel}: SEP event-related ICME arrival to Solar Orbiter.}
         \label{fig:enlil}
   \end{figure*}

The left panel of Fig.~\ref{fig:enlil} shows a snapshot of the solar wind radial speed in the ENLIL simulation around the SEP onset time on 17 April 2021 at 16:00~UT. The black contours track the ICME ejecta. They are manifested in the simulation as coherent and outward propagating high-density regions. The pattern of slower ($\sim$300~km~s\textsuperscript{-1}) and a bit faster ($\sim$500~km~s\textsuperscript{-1}) solar wind streams is visible in the plot. The black and white dashed lines represent the IMF lines connecting the Sun with the various observer positions. The simulation shows several transient and corotating structures present near Solar Orbiter, Earth, STEREO~A, and Mars at the time of the onset of the particles that might modify the magnetic connectivity and SEP propagation conditions. There is a relatively small ICME reaching Solar Orbiter during the ongoing SEP event. According to ENLIL, this ICME does not extend to any other investigated spacecraft. Ahead of this ICME there is a clearly wider eruption covering about $140^{\circ}$ in longitude. Its western edge encloses STEREO~A at the time of the SEP injection from the Sun and it is between the Sun and Mars. None of the CMEs inserted into ENLIL impacts Earth, but the leading edge of a stream interaction region (SIR) is reaching the planet at the time of the onset of the  particle intensity increase seen at Earth.

The ENLIL simulation also shows that at the time of the  initial SEP injection from the Sun (left panel of  Fig.~\ref{fig:enlil}) the IP medium is  relatively undisturbed between the Sun and BepiColombo as well as Parker Solar Probe. We note that the wide ICME discussed previously crossed BepiColombo and Solar Orbiter, but this was before the SEPs were injected at the Sun.  Nevertheless, this ICME may still have an effect on the propagation conditions of SEPs. The simulated status of the heliosphere around the SEP onset time agrees overall with the in-situ plasma and magnetic field measurements as discussed further below. 
The right panel of Fig.~\ref{fig:enlil} shows the heliosphere two days later, on 19 April 2021 at 21:00~UT. The ICME that was associated to the SEP event has then reached BepiColombo and Solar Orbiter. The simulation suggests that the ICME nose propagates between these two spacecraft and both of them cross the structure near the flanks.

\begin{figure*}
  \includegraphics[width=1.0\textwidth]{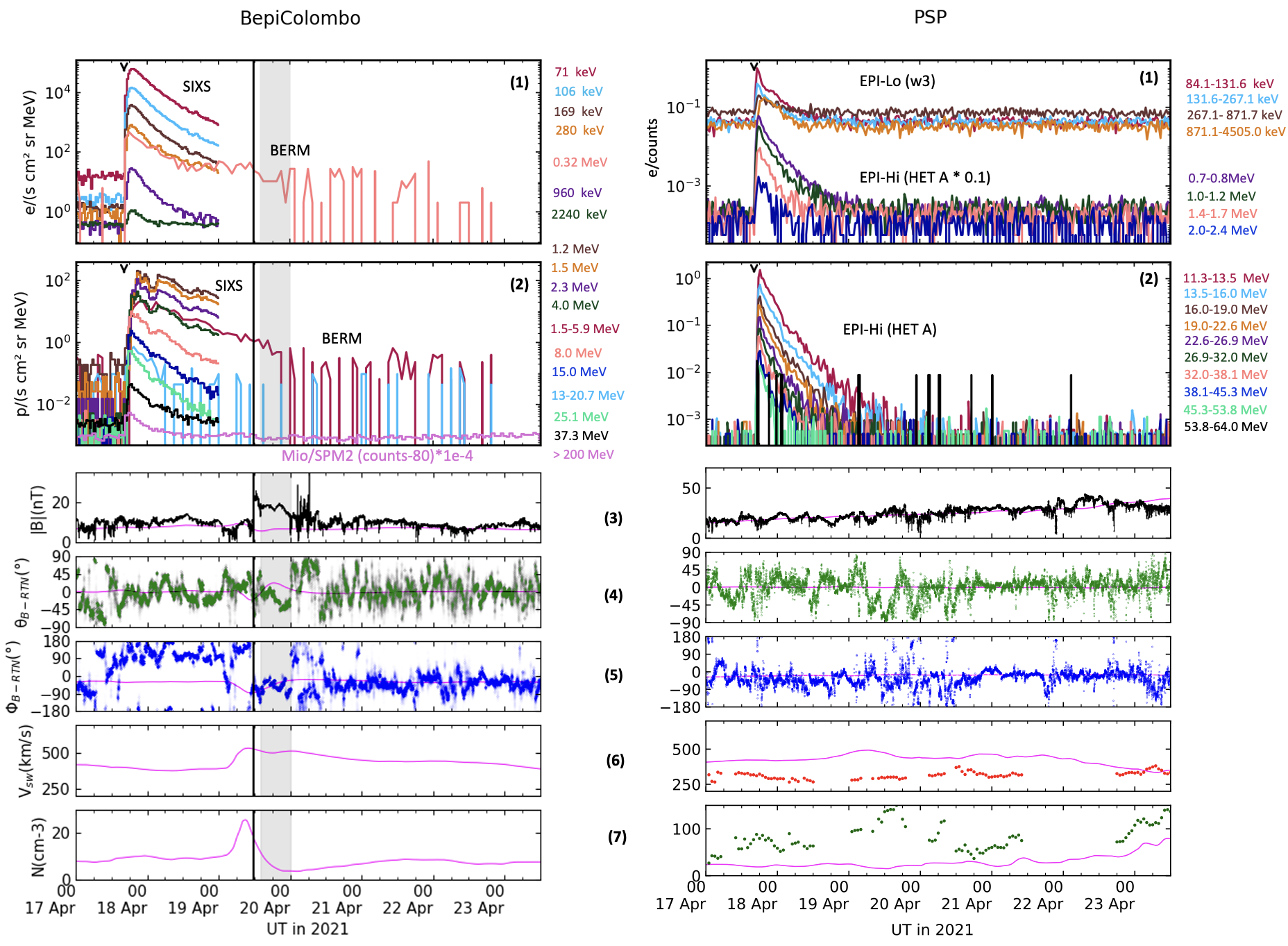}
      \caption{In-situ SEP time profiles as well as plasma and magnetic field observations by BepiColombo (left) and Parker Solar Probe (right). \textit{Top}: Energetic electron and proton temporal profiles observed from several energy channels and instruments. For SIXS, we use fluxes detected in side 2 of the detector. The flare eruption time is represented by the arrow on the upper x-axes. The vertical solid line and gray shaded area, respectively, indicate IP shock and ejecta transit observed by BepiColombo. \textit{Bottom}: In-situ plasma and magnetic field observations. The panels present, from top to bottom, the magnetic field magnitude, the magnetic field latitudinal and azimuthal angles, $\theta$\textsubscript{B-RTN} and $\phi$\textsubscript{B-RTN}, the solar wind speed, and the proton density, where RTN stands for radial-tangential-normal coordinates (IP structures as described in top panel). The pink lines represent the ENLIL simulation results.}
         \label{fig:IP_context_Bepi_PSP}
\end{figure*}

\begin{figure*}
  \sidecaption
  \includegraphics[width=1.0\textwidth]{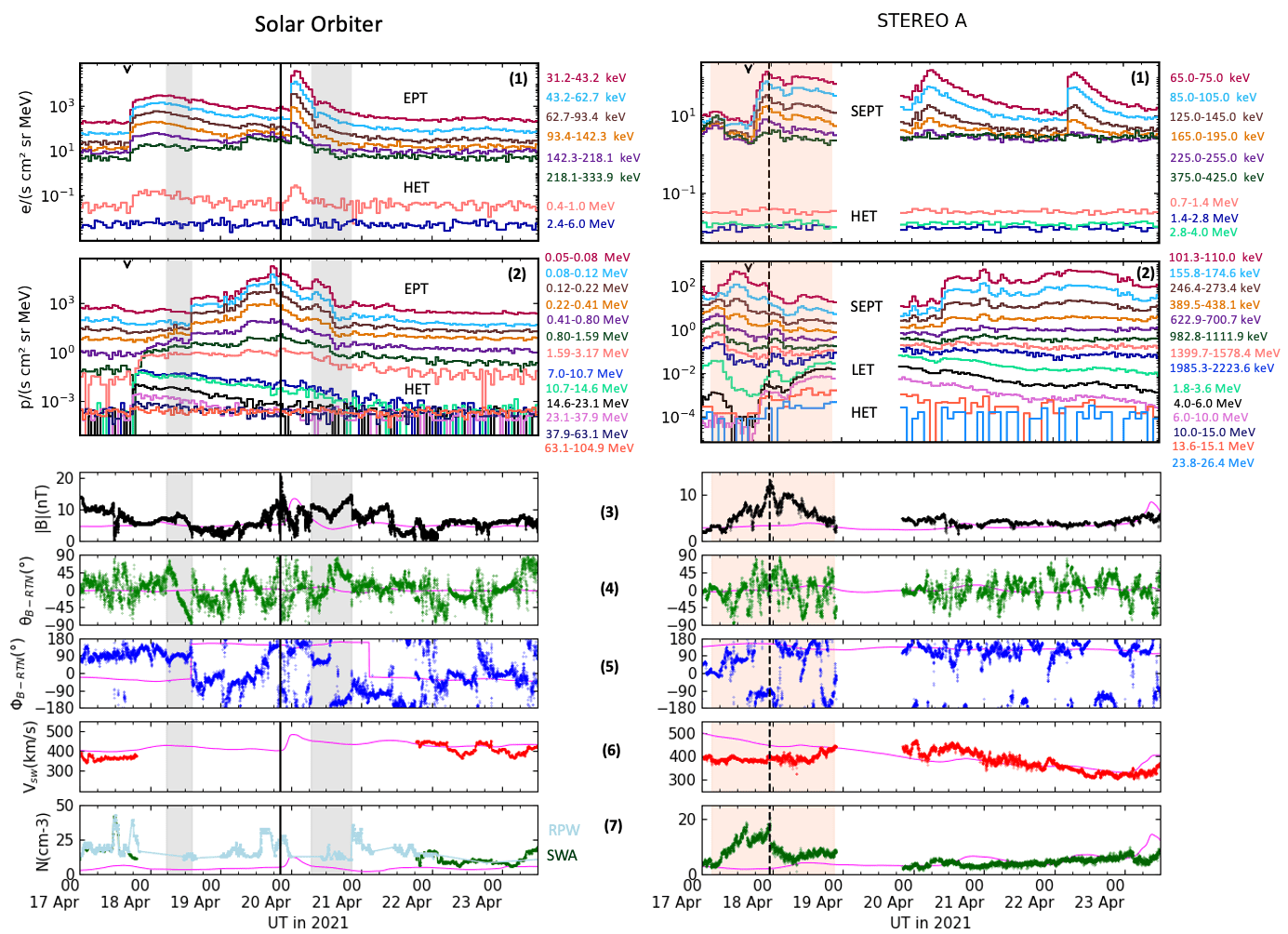}
      \caption{In-situ SEP time profiles as well as plasma and magnetic field observations by Solar Orbiter (left) and STEREO~A (right). \textit{Top}: Energetic electron and proton temporal profiles observed from several energy channels. We use the sunward looking sectors of Solar Orbiters' EPD/EPT and HET. For STEREO~A, as not all instruments provide sectored measurements, we use omni-directional data. The salmon shaded area indicates an SIR observed by STEREO~A, while the stream interface is shown as a dashed line. Flare time and rest of IP structures indicated as in Fig.~\ref{fig:IP_context_Bepi_PSP}. \textit{Bottom}: In-situ plasma and magnetic field observations. Solar wind densities for Solar Orbiter are obtained from RPW/QTN measurements.} Panels as in Fig.~\ref{fig:IP_context_Bepi_PSP}.
         \label{fig:IP_context_SOLO_STA}
  \end{figure*}

\begin{figure*}
  \includegraphics[width=1.0\textwidth]{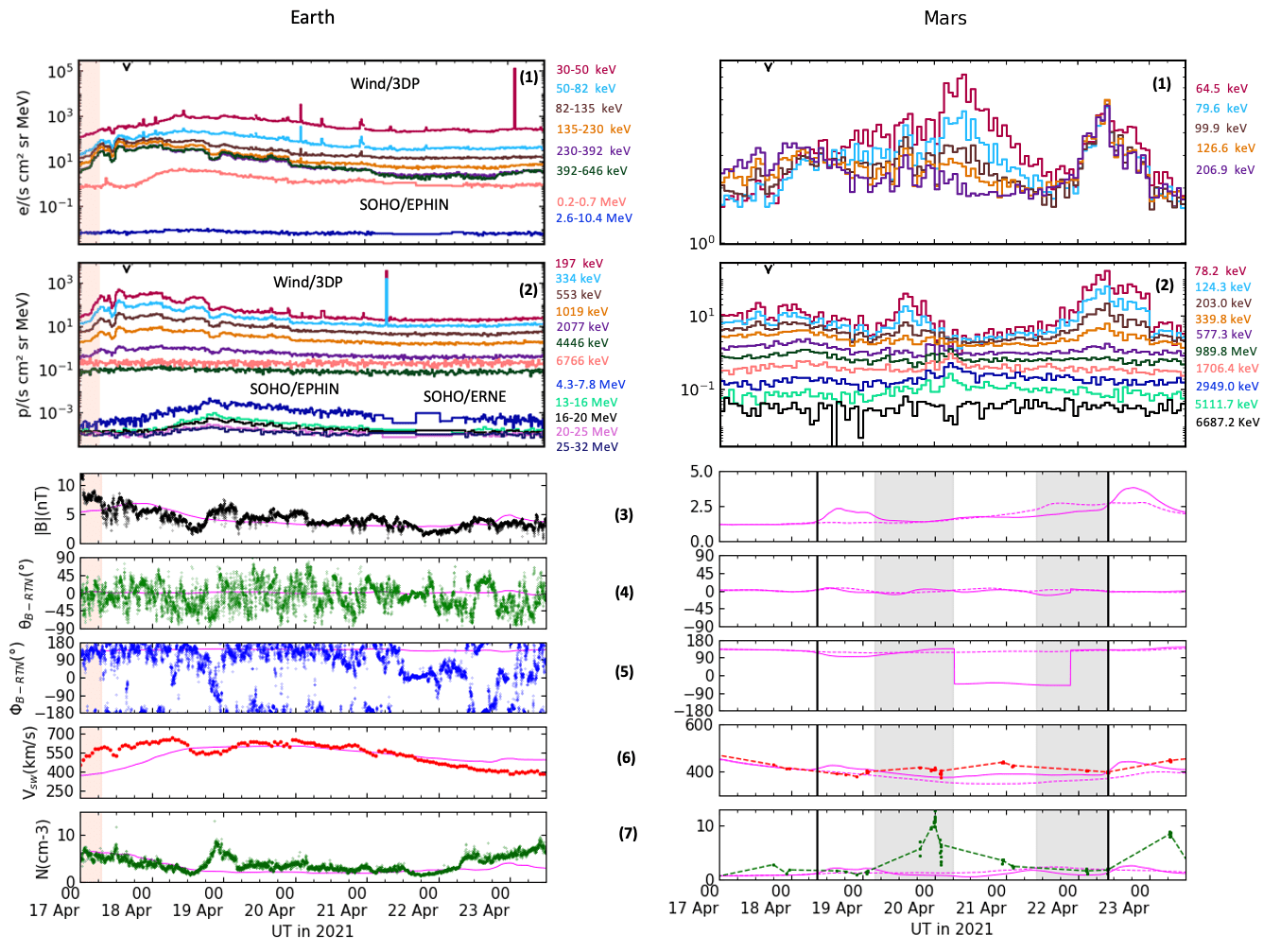}
      \caption{In-situ SEP time profiles as well as plasma and magnetic field observations by Earth (left) and Mars (right). \textit{Top}: Energetic electron and proton time profiles observed from several energy channels. Flare time and IP structures as in Fig.~\ref{fig:IP_context_Bepi_PSP}. \textit{Bottom}: In-situ plasma and magnetic field observations. Panels as in Fig.~\ref{fig:IP_context_Bepi_PSP}. The pink dashed lines are the ENLIL background solar wind with no CMEs included in the simulation.   
              }
         \label{fig:IP_context_Maven_L1}
  \end{figure*}

The  five bottom panels of Figs.~\ref{fig:IP_context_Bepi_PSP}, \ref{fig:IP_context_SOLO_STA}, and \ref{fig:IP_context_Maven_L1} present the in-situ plasma and magnetic field data over-plotted with the pink line showing the result of the ENLIL simulation from 17 April to mid 23 April.  The whole set of panels in these figures present, from top to bottom, energetic electron intensities at different energies (1), proton/ion intensities at different energies (2), the magnetic field magnitude (3), the magnetic field latitudinal (4) and azimuthal (5) angles in spacecraft centered radial-tangential-normal (RTN) coordinates, namely  $\theta$\textsubscript{B-RTN} and $\phi$\textsubscript{B-RTN}, the solar wind proton speed (6), and the solar wind proton density (7). As specified in the following section, ENLIL follows the general trend of the measured solar wind speed at the locations of Solar Orbiter and Mars, which were only separated by 9$^{\circ}$ in longitude during the SEP event, while at the remaining locations there are some differences with in-situ measurements. Although ENLIL reproduced the overall features of high-speed streams present in the heliosphere during the period of study, the differences between the modeled and measured solar wind profiles could be explained by complex  coronal holes that render the comparison between measurements and ENLIL results at Earth difficult, as well as preventing accurately resolving the glancing encounter of the SIR at STEREO A, as discussed below.

ENLIL successfully predicts the arrival of the several ICMEs observed in situ within the uncertainty of the model, as shown in the increase of the speed, density, or magnetic field in the pink profiles in Figs.~\ref{fig:IP_context_Bepi_PSP}, \ref{fig:IP_context_SOLO_STA}, and \ref{fig:IP_context_Maven_L1}. Due to the absence of the internal magnetic field in the simulated CMEs, the magnetic field magnitude increase is, however, lower than what was measured in situ. In particular, based on ENLIL simulations and measured in situ as discussed in Sect.~\ref{subsec:insitu_obs}, the ICME related to the SEP event is intercepted by BepiColombo and Solar Orbiter, while Mars might be only observing the associated IP shock, not the ICME ejecta. We relate the better simulation of the arrival time of this ICME at Solar Orbiter location in comparison with BepiColombo to the fact that we chose the CME parameters which better reproduced the  portion of the CME oriented towards Solar Orbiter as ENLIL input, as discussed in Sect.~\ref{sec:CME}. The minimum longitudinal extent of the ICME related to the SEP event is $\sim$45{$^{\circ}$}, as shown in the right panel of Fig.~\ref{fig:enlil}. This value is in agreement with the angular extent of the CME along the equatorial plane ($\sim$46{$^{\circ}$}) estimated from the GCS reconstruction presented in Sect.~\ref{sec:CME}.


  %
 
\subsection{Multi-spacecraft in-situ plasma, magnetic field and SEP observations in context with the ENLIL simulation}\label{subsec:insitu_obs}

As discussed in Sect.~\ref{subsec:ENLIL_model}, during the period of study there are several IP structures impacting the spacecraft under consideration, which may in turn affect the SEP particle profiles. In the following, we discuss the energetic particle observations and their relation with the interplanetary context.

\subsubsection{BepiColombo}
The two top panels of Fig.~\ref{fig:IP_context_Bepi_PSP} (left) show the SEP event as observed by BepiColombo detected over a broad energy range by the MPO/SIXS, MPO/BERM, and Mio/SPM instruments with the time of the flare onset marked by the arrows at the top. Panel (1) shows the impulsive energetic electron event that reaches energies of at least 2~MeV.  The proton intensity-time profile (panel 2) shows a more gradual increase. The Mio/SPM observations show that the event was observed even at proton energies > 200 MeV. There is no plasma information available, but at the time  when particle intensities started to increase, the solar wind speed given by ENLIL is $\sim$400~km~s\textsuperscript{-1},  as shown by the pink line in panel (6). 

 Commonly, the in-situ identification of the passage of ICMEs is based on a set of signatures typically observed in magnetic field and plasma data as well as some other proxies, such as bi-directional suprathermal electron (BDE) profiles \citep[e.g.,][]{Zurbuchen2006, Kilpua2017}.  BepiColombo lacked plasma data, but the in-situ MPO-MAG magnetic field observations in panels (3)--(5) do not show any evidence of typical ICME signatures (e.g., enhanced field, low field variety, coherent field rotation) during the onset and rising phase of the SEP event. This agrees with the previously discussed ENLIL simulation results and confirms that there was no large-scale solar wind structure at BepiColombo that could have directly influenced the SEP time profiles. 

The increase in the  magnitude of the magnetic field observed by BepiColombo on 19 April marks the arrival of the ICME related to the SEP event. The IP shock arrives at 11:40~UT (vertical solid line), while ENLIL simulates the ICME arrival time $\sim$6 hours earlier. Unfortunately, SIXS has a data gap at that time so that a potential  low-energy particle intensity response to the shock passage (i.e., an energetic storm particle event) could not be studied in detail. However, BERM fluxes at $\sim$1.5-5.9~MeV do not  show a significant increase at the shock or a response to the passage of the ejecta, which follows the shock (gray shaded area). 

 The leading edge of the ejecta was observed at 13:57~UT on 19 April identified by a change in the magnetic field polarity along with the presence of coherent and organized magnetic field. Specifically, we observe a smooth and monotonic change of the magnetic field latitudinal and azimuthal angles shown in panels (4) and (5)  that lasted until 20 April 00:04~UT. No other structures are observed until the end of the period shown  in Fig. \ref{fig:IP_context_Bepi_PSP}.  

%
\subsubsection{Parker Solar Probe}

Panel (6) of Fig.~\ref{fig:IP_context_Bepi_PSP} (right) shows that at the time of the SEP event onset the solar wind speed at Parker Solar Probe is $\sim$320~km~s\textsuperscript{-1}. The SEP event has a very impulsive time profile both in the electrons shown in panel (1) and in the protons shown in panel (2). Compared with BepiColombo the event has a shorter duration, namely a faster decay. PSP/EPI-Hi/HET observes  observes intensity increases at electron energies above 2 MeV and protons above 50 MeV.

Based on the plasma and magnetic field data given by the SWEAP and FIELDS instruments, no IP structures  can be identified during the whole period shown in the right column of Fig. \ref{fig:IP_context_Bepi_PSP} location of Parker Solar Probe. This is in agreement with the ENLIL simulation results.
\subsubsection{Solar Orbiter}\label{sec:IP_solo}
Panels (1) and (2) of Fig.~\ref{fig:IP_context_SOLO_STA} (left) show the SEP event observed by Solar Orbiter. While the electron event is observed to reach energies up to $\sim$1~MeV, it is not as impulsive as the event observed by BepiColombo and Parker Solar Probe but shows more of a plateau-like profile. The intervening structures present at the time of the SEP onset, as suggested by ENLIL (Sect.~\ref{subsec:ENLIL_model}), might be associated with this behavior as they might hinder the SEP transport. This may also be the reason for the  low anisotropy observed at the onset of the event as described in Sect.~\ref{sec:solo_seps}. The energetic ion observations by Solar Orbiter/EPD/EPT  allow us to discern the initial phase of the event only at energies ${\gtrsim}400$~keV,  reaching energies up to ${\sim}60$~MeV as observed by EPD/HET.

The solar wind speed at the time of the electron event onset is $\sim$380~km~s\textsuperscript{-1}, as shown in panel (6), measured by the SWA instrument, which is well reproduced by ENLIL (pink line). In a later phase of the SEP event, Solar Orbiter observes several IP structures identified using MAG, SWA and the RPW instruments on board Solar Orbiter.  A first ejecta (first gray shaded area in the left column of Fig. \ref{fig:IP_context_SOLO_STA}) arrives at 05:24~UT on 18 April. While it does not affect the energetic  electron intensity time profiles or the high-energy ($\gtrsim$2 MeV) ion intensity time profiles, it seems to have acted as a particle barrier for low-energy ions. Only after its passage, at 14:18~UT on the same day, an increase in the ${\lesssim}400$~keV energy ions is observed. These particles are likely associated with  the solar eruption on 17 April. The ICME-driven shock associated with this eruption arrives at Solar Orbiter on April 17 at 20:20 UT on 19 April. The shock simulated by ENLIL arrives a about $\sim$30 minutes later than the measured shock as shown by the pink line in panels (3), (6-7) in the left column of Figure  \ref{fig:IP_context_SOLO_STA}.

The shock obliquity ($\theta_{Bn}$, namely the angle between the shock normal and the upstream magnetic field) is estimated at Solar Orbiter using the magnetic coplanarity method \citep[e.g.,][]{Paschmann2000}. A value of $\theta_{Bn} \sim (21 \pm 5)^\circ$ is computed, employing a systematic variation of the upstream and downstream averaging window lengths between 3 and 13 minutes, with the method described in \citet{Trotta2022}. The lack of plasma data around the shock crossing limits further analyses. However, using the novel method introduced by \citet{Gedalin2021}, an estimation for the Alfv\'enic Mach number using magnetic field only data yields $M_A \sim 1.8$, consistent with the fact that the shock passage has no significant influence on the energetic particle population at higher energies.

The low-energy ions keep rising until a peak is observed shortly before the shock passage, after which the intensities decrease. Right after the shock passage, another solar energetic electron event is observed on 20 April, which is not related to the event under study but originated from an M1.1 flare in AR~12816  (at S24E25 as seen from Earth), peaking at 23:42~UT on 19 April. The $\sim$1 MeV/nucleon ions associated with this new injection showed a large enrichment of  $^{3}$He, with a $^{3}$He/$^{4}$He ratio of, with a ratio of $\sim$5\% (not shown). The second ejecta, which arrives at 06:51 UT on 20 April, corresponds to the ICME associated with the 17 April SEP event. It is marked by a smooth magnetic field and monotonic and coherent rotations in the magnetic field angles. While the second energetic electron event shows a depression of fluxes during the ejecta passage, low-energy ions show an enhancement inside the ejecta during its first half and a decrease during the second half.   
%


%
\subsubsection{STEREO~A}
Observations of the SEP event at STEREO~A are shown in Fig.~\ref{fig:IP_context_SOLO_STA} (right). STEREO~A observes a clear electron event, however only  at near-relativistic ($\lesssim$400 keV) energies. Similarly to Solar Orbiter, the proton event is only well observed at higher energies, namely ${\gtrsim}2$~MeV. The maximum energy of the proton event seems to be lower than at the other spacecraft, barely reaching ${\sim}25$~MeV. At lower ion energies, SEPT observes an {|bf enhanced pre-event intensity background, (most likely due to the SIR as described below),} which might mask the SEP event.  

As shown by the magnetic field and plasma data in panels (3)--(7), the SEP event onset takes place during the passage of an SIR at STEREO~A, which is  indicated by the salmon-shaded vertical bar at the beginning of the time interval displayed in the right column of Fig. \ref{fig:IP_context_SOLO_STA}. The signatures indicate a glancing cross of the SIR structure, with only a very modest increase of the solar wind speed. The speed rises from ${\sim}400$ to ${\sim}450$~km~s\textsuperscript{-1}; sudden changes of the magnetic field polarity close to the stream interface (dashed vertical line), and drops in the magnetic field strength together with temperature increases (not shown) and proton density enhancement, which suggests that local reconnections are occurring. The ENLIL simulation also suggests an SIR arrival (not shown), but several hours earlier than observed, and infers a clearer intersection of the high-speed stream  with the spacecraft. At STEREO~A, no signatures of ICMEs are detected. This is in agreement with the ENLIL results. 

The lowest-energy ion channels of SEPT show a clear variation in their intensities happening right after the stream interface. At the same time, the thermal proton density drops.
%
The energetic electron increases observed after the data gap are associated to later SEP events that are not related to the event under study. 
\subsubsection{Earth}\label{sssec:insitu_L1}
Figure~\ref{fig:IP_context_Maven_L1} (left) shows the SEP event observed at near-Earth spacecraft. Similar to STEREO~A, Earth is embedded in the trailing portion of an SIR, after the stream interface passage, as measured by the MFI and SWE instruments on board Wind.  The arrival at Earth of the high-speed stream simulated by ENLIL arrives a few hours later than that actually measured. The rear boundary of the SIR is difficult to define, as there is not a clear step-like speed increase and the dynamic pressure does not show any clear peak. The reason behind this behavior might be that the solar wind arriving at Earth originates from multiple and complex coronal holes. There is a big southern coronal hole extending to the equator and some low-latitude large patchier holes (not shown). The solar wind speed at the onset of the particle event is ${\sim}600$~km~s\textsuperscript{-1}, as shown in panel (6). Panel (1) shows that only SOHO/EPHIN, which has a very low instrumental background, observes a clear but very gradual electron event at 0.25--0.7~MeV. The lower energies covered by Wind/3DP are showing an enhanced background that possibly masks the SEP event and may contain also ion contamination. This enhanced background, likely caused by the SIR, also dominates the low-energy ion observations by Wind/3DP. However, SOHO/EPHIN and ERNE show a proton event extending into the deca-MeV range, which is small, gradual, and clearly delayed with respect to the time of the flare.  
%
%
\subsubsection{Mars}
On 17 April 2021 Mars was located at a heliocentric distance of 1.6 au 22$^{\circ}$ west of the flaring active region at 225$^{\circ}$ Carrington longitude. The top two panels of the right column of Figure \ref{fig:IP_context_Maven_L1} show $\sim$60-210 keV electron and $\sim$70-7000 keV proton intensities as measured in different energy channels of MAVEN/SEP.
Panels (3)--(5) show only the ENLIL simulations of the magnetic field, as no measurements are available. The solar wind speed (6) and density (7) measurements by MEX/ASPERA-3/IMA are rather sparse, however, they show an overall good agreement with the ENLIL simulation for the solar wind speed. In this case, we also show the dashed pink lines corresponding to the background solar wind simulation, without including any CME. The separation of the solid and dashed lines  indicate the effects produced by the passage of the interplanetary structures, based only on ENLIL results. A first interplanetary shock is modeled to arrive at 09:00~UT on 18 April. Two pre-SEP event ICMEs arrive at 04:00~UT on 19 April and 10:00~UT on 21 April, as simulated by ENLIL, that might be the same ICMEs measured earlier by Solar Orbiter. Lastly, the interplanetary shock related to the SEP event under study impacts Mars at 10:00~UT on 22 April, based on both the simulation and on the increase in solar wind speed and density measured in situ. According to ENLIL, the shock is, however, not followed by an ejecta. The ICME flank might therefore have missed Mars.

It is difficult to associate the energetic particle increases observed by MAVEN with the 17 April SEP event. Nevertheless, an electron increase is observed in the higher-energy channels right after the flare (marked by an arrow). Although the onset times are very hard to determine and might suggest a too-early onset to account for the expected travel time of these electrons, a potential SEP contribution cannot be excluded. More likely, on the other hand, is that the CME-driven shock associated with the event has contributed to the electron and proton increases observed on 22 April because the peaks of the SEP increases agree well with the shock arrival time simulated by ENLIL. However, another possible source of this increase could be the same new SEP event as observed also by STEREO~A on 22 April  (at S24E25 as seen from Earth), which is magnetically well-connected with Mars during the period under study (see Fig.~\ref{fig:solar-mach_and_multi_sc_SEP}).

%
%
\subsection{SEP pitch-angle distributions and first arriving particles} \label{sec:sep_obs}
Figure~\ref{fig:solar-mach_and_multi_sc_SEP} (right) combines the SEP observations as measured by the five inner-heliospheric spacecraft and shows how strongly the event characteristics such as  intensity-time profiles, onset times, and peak intensities vary from observer to observer. Given the well-separated positions of these spacecraft shown in Fig.~\ref{fig:solar-mach_and_multi_sc_SEP} (left) and their varying separations with respect to the parent flare location, this is not unexpected. It has been found in earlier studies that the longitudinal distribution of peak intensities usually shows a decrease with increasing longitudinal separation angle from the associated flare longitude \citep[e.g.,][]{Lario2013, Dresing2014, Richardson2014}. These authors described the longitudinal peak-intensity distributions with Gaussian functions. However, due to the limitation of only three well-separated observers, these analyses suffered from large uncertainties. The new, larger spacecraft fleet will allow us to analyze these longitudinal distributions in a better way, however, instrument inter-calibrations, especially of the new mission's payload, are still pending. 
Looking at the $\sim$20--25~MeV proton peak intensities observed by each spacecraft (Fig.~\ref{fig:solar-mach_and_multi_sc_SEP}) we find, however, a deviation from the expected ordering of peak intensities with absolute longitudinal separation angle. Parker Solar Probe, which is slightly worse connected ($|\Delta\Phi|=72^{\circ}$) than Solar Orbiter ($|\Delta\Phi|=65^{\circ}$), observes not only a significantly higher-intensity event, but also a clearly more impulsive time profile. While the higher peak intensities at Parker Solar Probe could be explained by its smaller radial distance from the Sun as compared to Solar Orbiter (0.42~au vs. 0.84~au), the significantly different time profiles rather suggest a different connectivity to the SEP injection region.

In this section, we analyze the SEP observations in more detail to determine the timing of the first arriving particles, which allows us to relate the SEPs with their solar counterpart observations (see Sect.~\ref{sec:timing}). Furthermore, we analyze pitch-angle distributions (PADs) to characterize the degree of pitch-angle diffusion, namely the importance of transport effects.

\subsubsection{BepiColombo}\label{sec:bepi_seps}
\begin{figure*}
   \centering
      \includegraphics[width=0.49\textwidth]{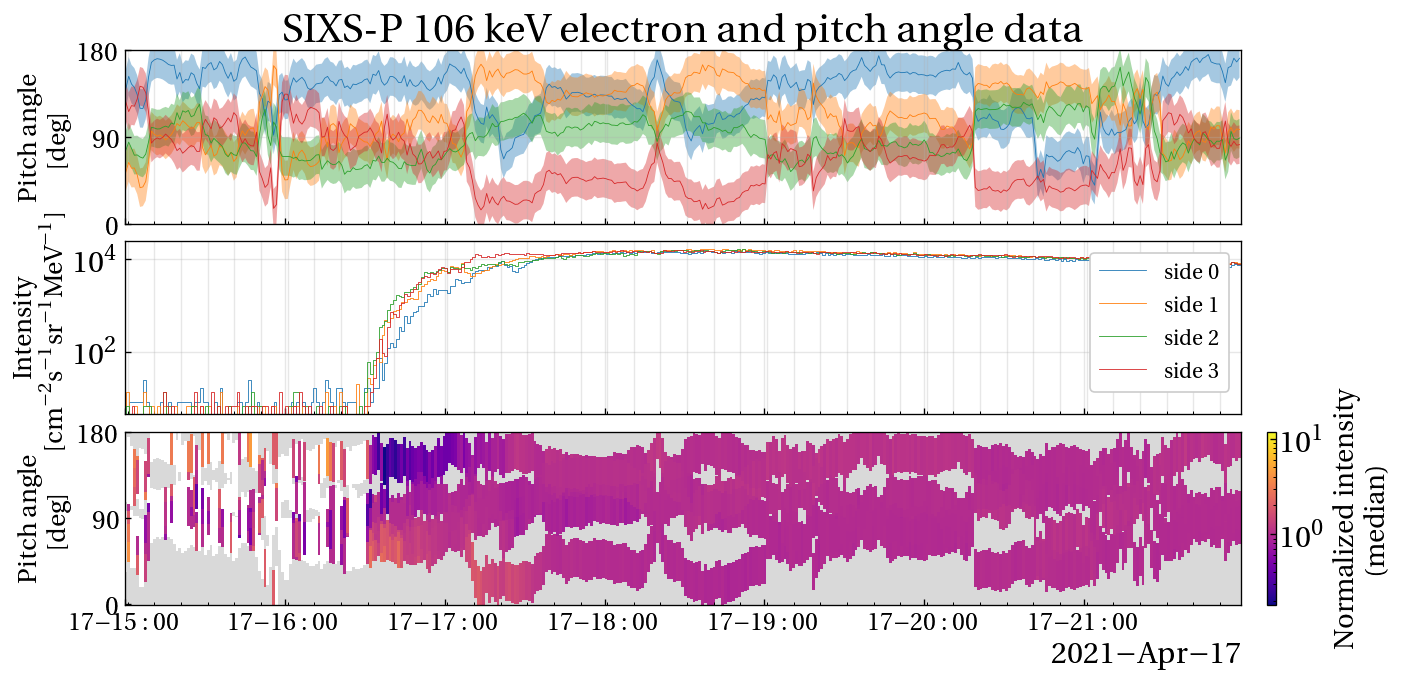}
      \includegraphics[width=0.49\textwidth]{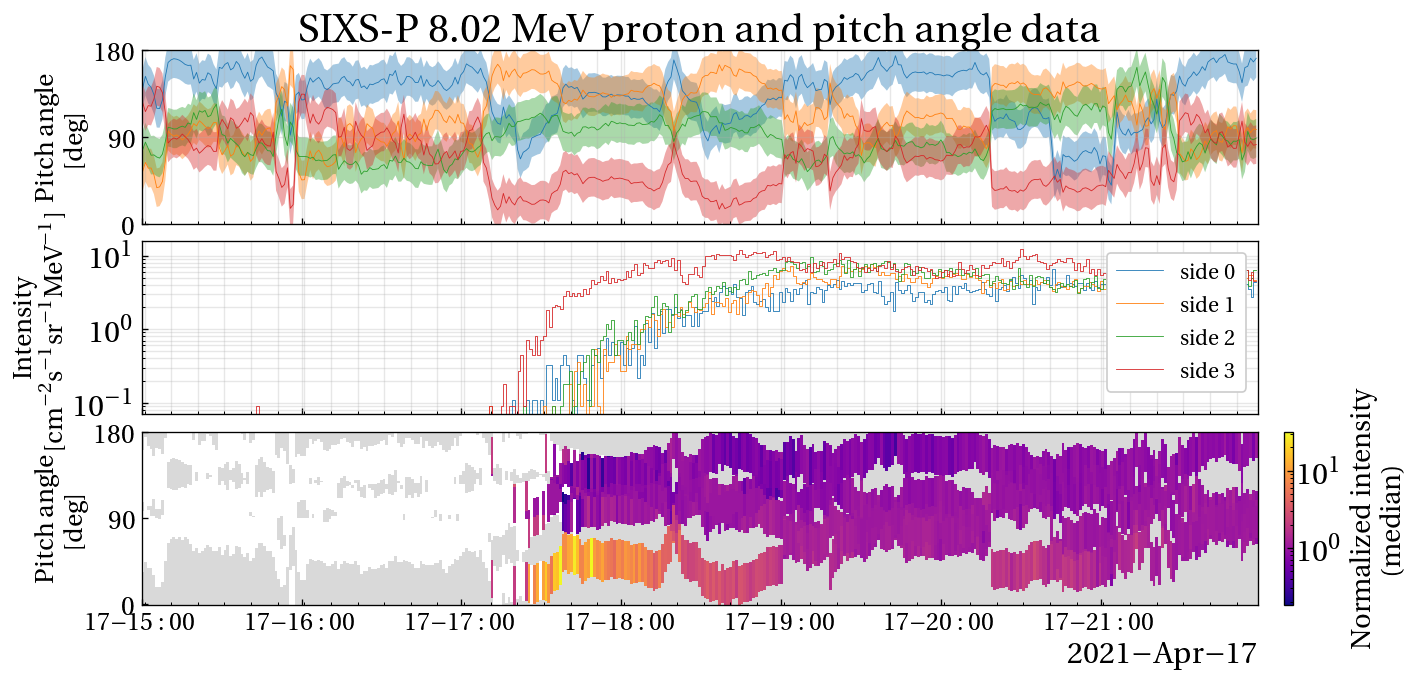}
      \caption{Pitch-angle distribution of 106~keV electrons (left) and 8.02~MeV protons (right) measured by BepiColombo/SIXS. Top: pitch-angle coverage of sides 0--3, middle: intensities measured in sides 0--3, bottom: pitch-angle distribution with color-coded intensities normalized to the median of each time step. Gray pitch-angle bins mark no pitch-angle coverage, while white bins are zero-count periods.
              }
         \label{fig:Bepi_PAD}
   \end{figure*}

BepiColombo detects the most intense event out of all observers.  This is expected based on both its closer radial distance from the Sun (0.63 au) and its fairly good connection to the associated flaring active region --$1^{\circ}$ ($79^{\circ}$) longitudinal (total)  separation angle between the flare site and the spacecraft magnetic footpoint (cf. Table~\ref{table:ADAPT-WSA connectivity}). BepiColombo also observes the earliest SEP onsets, e.g.\ 16:30 UT for 71~keV electrons, and the corresponding inferred injection times are the earliest out of all observers (see Sect.~\ref{sec:timing}).
Surprisingly, BepiColombo/SIXS detects a 5-minute earlier onset time for 71~keV electrons than for 960~keV electrons (see Table~\ref{table:timing}). Although these onset times almost agree within the error bars, the difference between the inferred injection times is significant. The much longer travel time of the lower energy electrons yields an 11-minutes earlier injection time compared to the $\sim$1~MeV electrons.
The first 25~MeV protons are detected at 17:00~UT$\pm4$~min, which corresponds to an inferred injection time situated between that of the low-energy electrons and the high-energy electrons. However, given the error bars, it would agree with both of the inferred electron injection times. 
 Figure~\ref{fig:Bepi_PAD} shows sectored energetic particle measurements by SIXS (middle panel) as well as the pitch-angles covered by the center of the four different viewing directions (top panel) and the intensity--PAD in the bottom panel. The left-hand figure shows ${\sim}100$~keV electrons and the right-hand figure shows 8~MeV protons.
Although the event is anisotropic, as can be seen by the different intensity levels observed by the different sides of the SIXS instrument, a velocity dispersion analysis did not yield meaningful results neither for electrons nor for protons. Therefore, we can only apply the time-shift analysis to infer the particle injection times at the Sun at specific energy bands. The proton anisotropy is stronger and the anisotropic phase is clearly longer for protons, lasting about two hours.
Unfortunately, the electron onset falls into a period of poor pitch-angle coverage of the sector where particles streaming from the Sun along the outward magnetic field would enter (pitch angle 0). The electron anisotropy could therefore be underestimated during the onset phase and this could lead also to a determination of too late electron onset times.

\subsubsection{Parker Solar Probe}\label{sec:psp_seps}
Based on the ${\sim}25$~MeV proton observations (see Fig.~\ref{fig:solar-mach_and_multi_sc_SEP}, right), Parker Solar Probe observes the second most intense event after BepiColombo. Because Parker Solar Probe's electron observations are not yet available in units of intensity, it is not possible to compare their intensity level with that of other spacecraft. 

\begin{figure*}
   \sidecaption
   \includegraphics[width=0.55\textwidth, trim=10 10 50 50, clip]{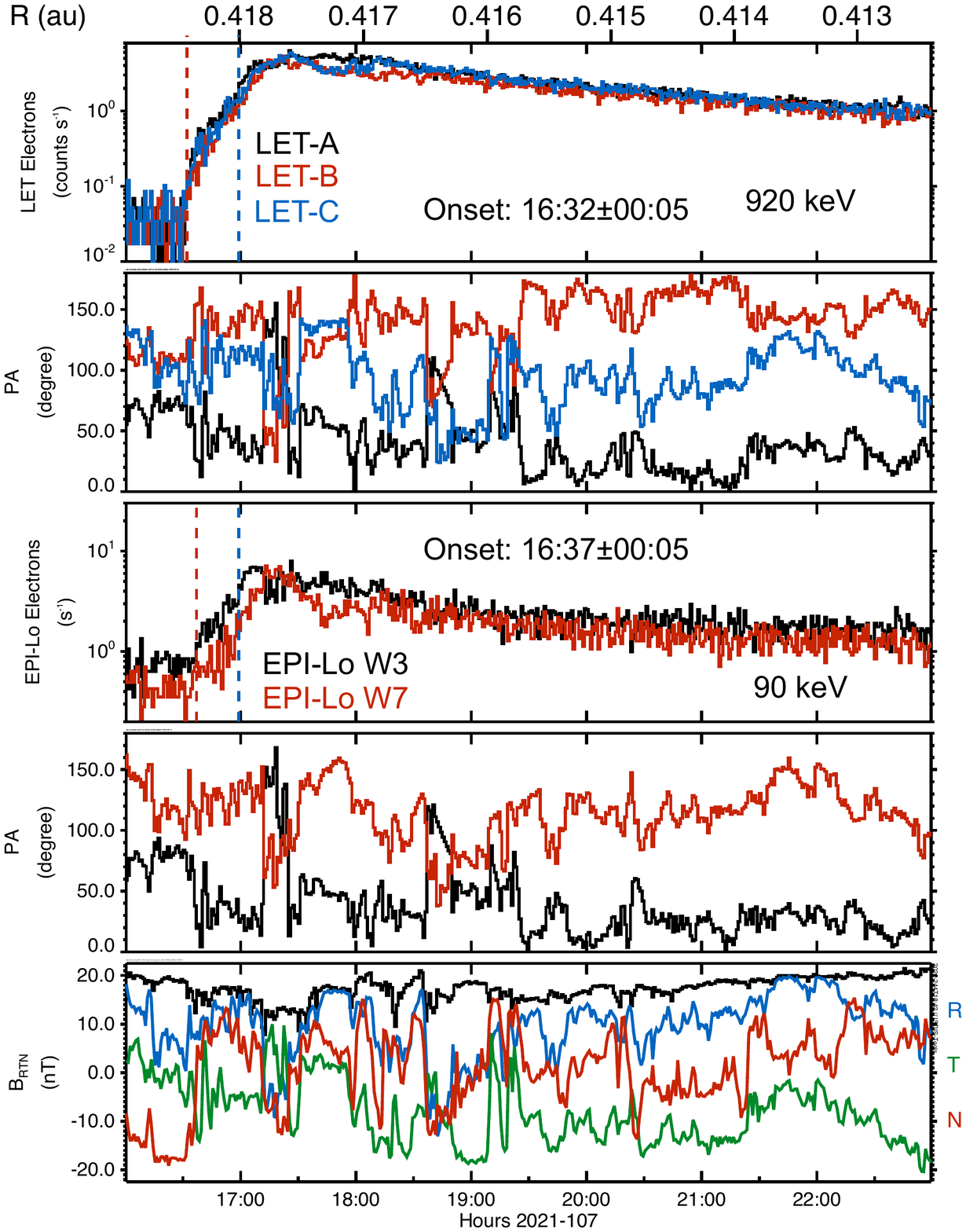}
   \includegraphics[width=0.45\textwidth, trim=0 0 70 50, clip]{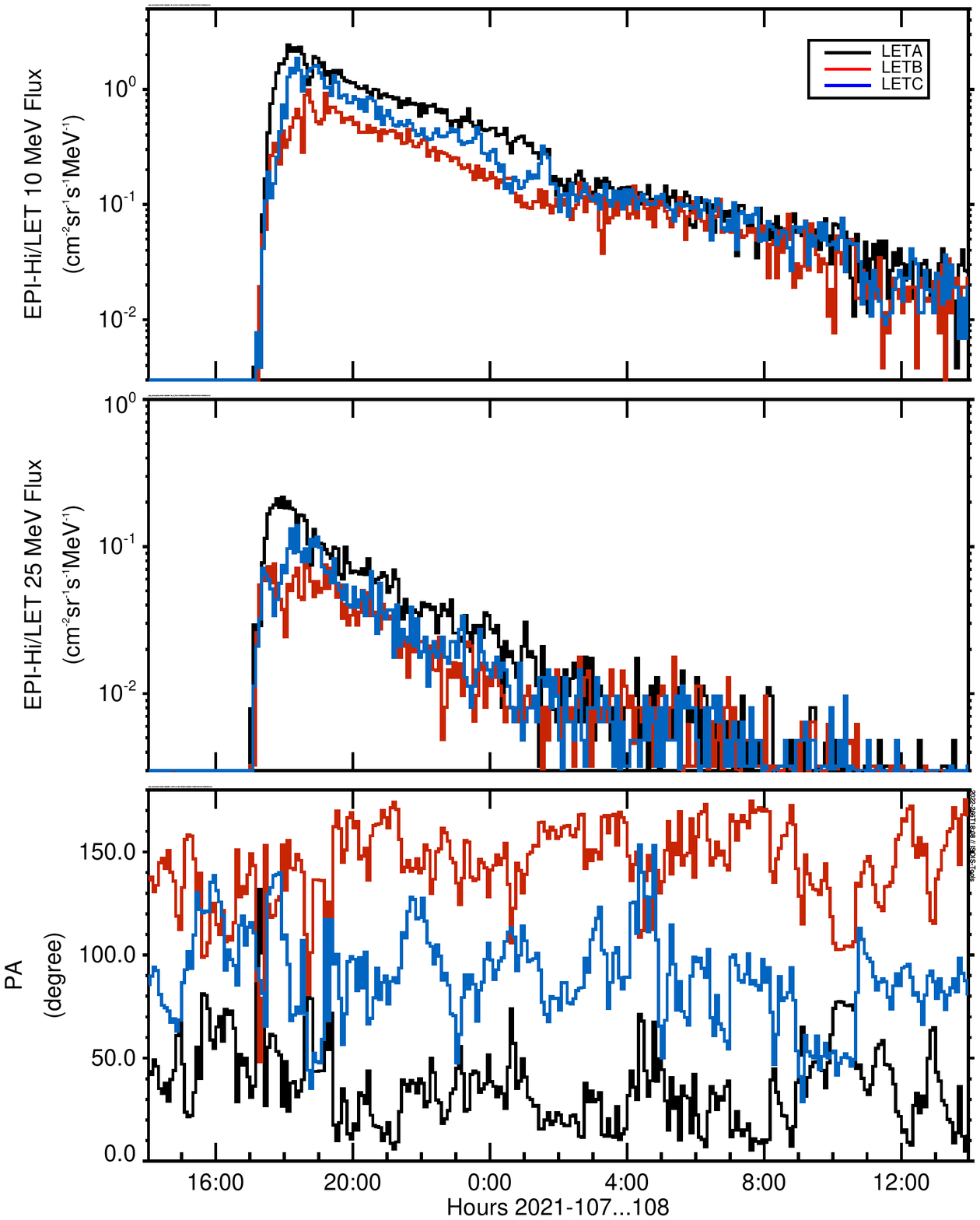}
      \caption{PSP/IS$\odot$IS observations of the onset of the energetic electron (left) and proton enhancement (right). Top left: time profile of $\sim$920~keV electrons observed by the three orthogonal EPI-Hi/LET telescope apertures. Left second panel: pitch angle of each of the EPI-Hi/LET apertures. Left middle: time profile of $\sim$90~keV electrons observed by EPI-Lo wedges 3 and 7 (sunward and anti-sunward facing, respectively). Left fourth panel: pitch angle of the boresight of EPI-Lo wedges 3 and 7. Left bottom: magnetic field magnitude and vector in RTN coordinates as measured by the PSP/FIELDS magnetometer. Top right: time profile of $\sim$10 MeV protons observed by the three orthogonal EPI-Hi/LET telescope apertures. Right middle: time profile of $\sim$25~MeV protons observed by the three orthogonal EPI-Hi/LET telescope apertures. Right bottom: pitch angle of each of the EPI-Hi/LET apertures.
              }
         \label{fig:psp_seps}
   \end{figure*}

Figure~\ref{fig:psp_seps} (left) shows the energetic electron observations at 920~keV (top panel) and 90~keV (third panel) in the different viewing directions as provided by EPI-Hi/LET and in two wedges of EPI-Lo, respectively. The determined onset times using the sunward-looking sectors and 5-min averaged data are marked by the red dashed lines. 

 While EPI-Hi does not provide the necessary time resolution to discern velocity dispersion in these relativistic electrons, the time resolution of EPI-Lo would be sufficient to discern velocity dispersion in the near-relativistic electrons. However, the limited statistics at these energies makes it challenging to conclude whether EPI-Lo observed electron velocity dispersion or not. Nevertheless, a small but significant anisotropy is present in the 90~keV electron observations, as denoted by the higher intensity of the sunward-looking wedge W3 (black) compared to the anti-sunward viewing wedge W7 (red).
 Still during the rising phase of the electron event, around 17:00~UT, we observe a second step, marked by the blue dashed line in the third panel of Fig.~\ref{fig:psp_seps} (left),  which is observed by both EPI-Lo and EPI-Hi. It does not correlate with any changes in the magnetic field and therefore seems not to be caused by a local effect.
 Later, around 17:30--18:00~UT, we observe a phase of stronger anisotropy consistent between EPI-Lo and EPI-Hi that appears to be tied to changes in the magnetic field.

Figure~\ref{fig:psp_seps} (right) shows proton observations at 10~MeV (top panel) and 25~MeV (middle panel) in different viewing directions as provided by EPI-Hi/LET. As for BepiColombo, the proton observations show a stronger anisotropy than the electrons, which also lasts longer (${>}6$~hours). In both the 10~MeV and 25~MeV time--intensity profiles, the sunward-facing aperture (LETA) shows the fastest onset and highest intensity.\\  
   
\begin{figure}[ht!]
    \includegraphics[width=0.5\textwidth, trim=0 0 0 0, clip]{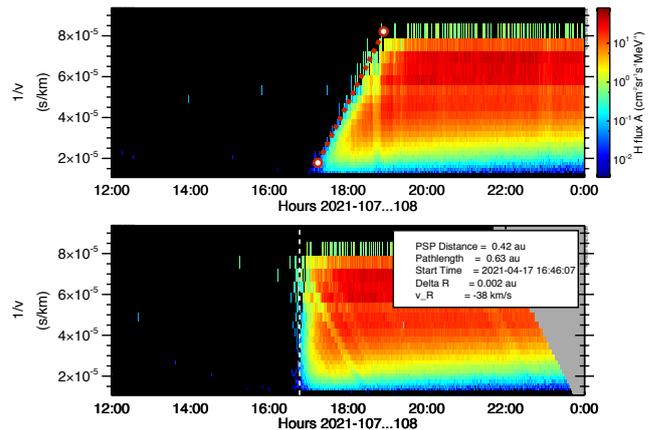}
      \caption{VDA of protons from PSP EPI-Hi/LETA from 1 to 30~MeV. Top panel: the red line is a ‘by-eye’ fit to the onset of the observed intensities as a function of 1/v and time. Bottom panel: the same data are plotted, but with the velocity dispersion removed. The legend provides the derived path length and injection times corresponding to the fit line.
              }
         \label{fig:PSP_VDA}
   \end{figure}

In contrast to the electron observations, the protons show a clear velocity dispersion. Figure~\ref{fig:PSP_VDA} shows a velocity dispersion analysis (VDA) that results in a path length of $L$ = 0.63~au traveled by the protons and an inferred proton injection time at 16:46~UT$\pm10$~min. Even considering the uncertainties, this injection time is significantly later than those determined for electrons through a time shift analysis (TSA), using the same path length, that result in 16:26~UT (16:30~UT) for 920~keV (90~keV).
   
%
%
\subsubsection{Solar Orbiter}\label{sec:solo_seps}
Solar Orbiter's magnetic footpoint at the Sun is similarly far separated in longitude from the flare location to that of Parker Solar Probe, but it is situated on the other side, namely west of the flare. Solar Orbiter observes significantly lower proton intensities than Parker Solar Probe (see Fig.~\ref{fig:solar-mach_and_multi_sc_SEP}, right). Furthermore, in contrast to BepiColombo and Parker Solar Probe, who observe an impulsive proton time profile, Solar Orbiter observes a gradual profile both in electrons and protons. While in case of CME-driven shock acceleration a more gradual time profile is expected for an observer situated to the east of the source region as compared to an observer situated to the west due to their different connections to the CME-driven shock front \cite[e.g.,][]{Cane1988}, the difference in their peak intensities is not expected to show such strong asymmetry \cite[e.g.,][]{Richardson2014}. However, Solar Orbiter's distance to the Sun, which is twice of Parker Solar Probe's distance, is expected to contribute to this intensity difference. 

\begin{figure*}
   \centering
   \includegraphics[width=0.49\textwidth]{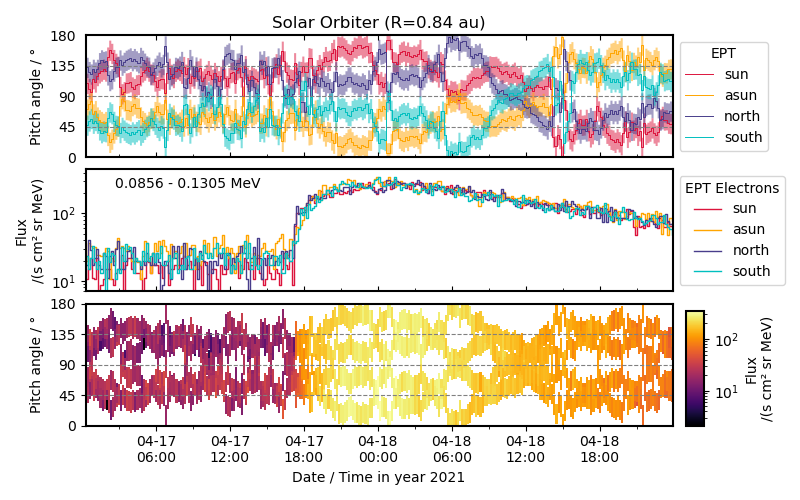}
   \includegraphics[width=0.49\textwidth]{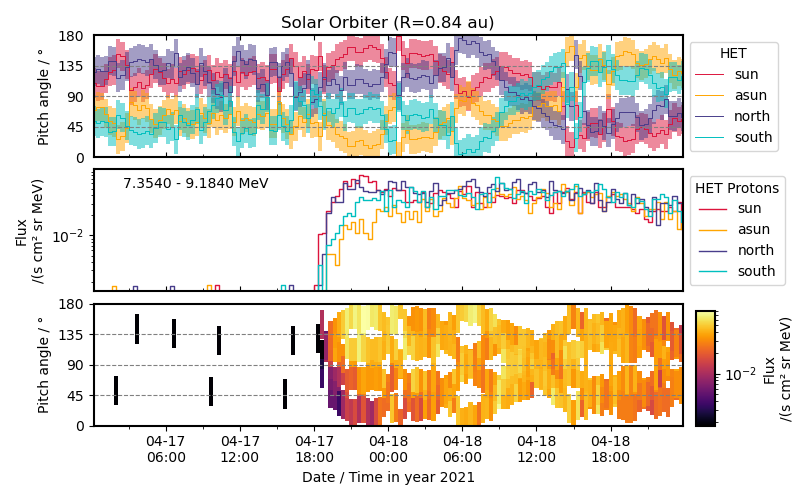}
      \caption{Pitch-angle distribution of 86-130~keV electrons (left) and 7.4-9.2~MeV protons (right) observed by Solar Orbiter/EPD-EPT and EPD-HET, respectively. Top: Pitch-angle coverage of the four different sensor apertures. Middle: Intensities observed by each field of view. Bottom: Pitch-angle distribution with color-coded intensities.
              }
         \label{fig:SolO_PAD}
   \end{figure*}

Energetic electron observations do not show any significant anisotropy, neither at lower energies as illustrated by the ${\sim}100$~keV electron PADs (Fig.~\ref{fig:SolO_PAD}, left) observed by Solar Orbiter/EPT, nor at MeV energies (not shown). The right-hand part of Fig.~\ref{fig:SolO_PAD} shows the PAD of ${\sim}8$~MeV protons as detected by Solar Orbiter/HET, which shows that the early phase of the MeV proton event is anisotropic for about seven hours, showing  higher fluxes in the sunward-looking telescope that corresponds to pitch angles near 180$^{\circ}$, consistent with the inward magnetic polarity (see also column 8 in Table~\ref{table:ADAPT-WSA connectivity}).

\begin{figure}
    \centering
    \includegraphics[width = 0.48\textwidth,]{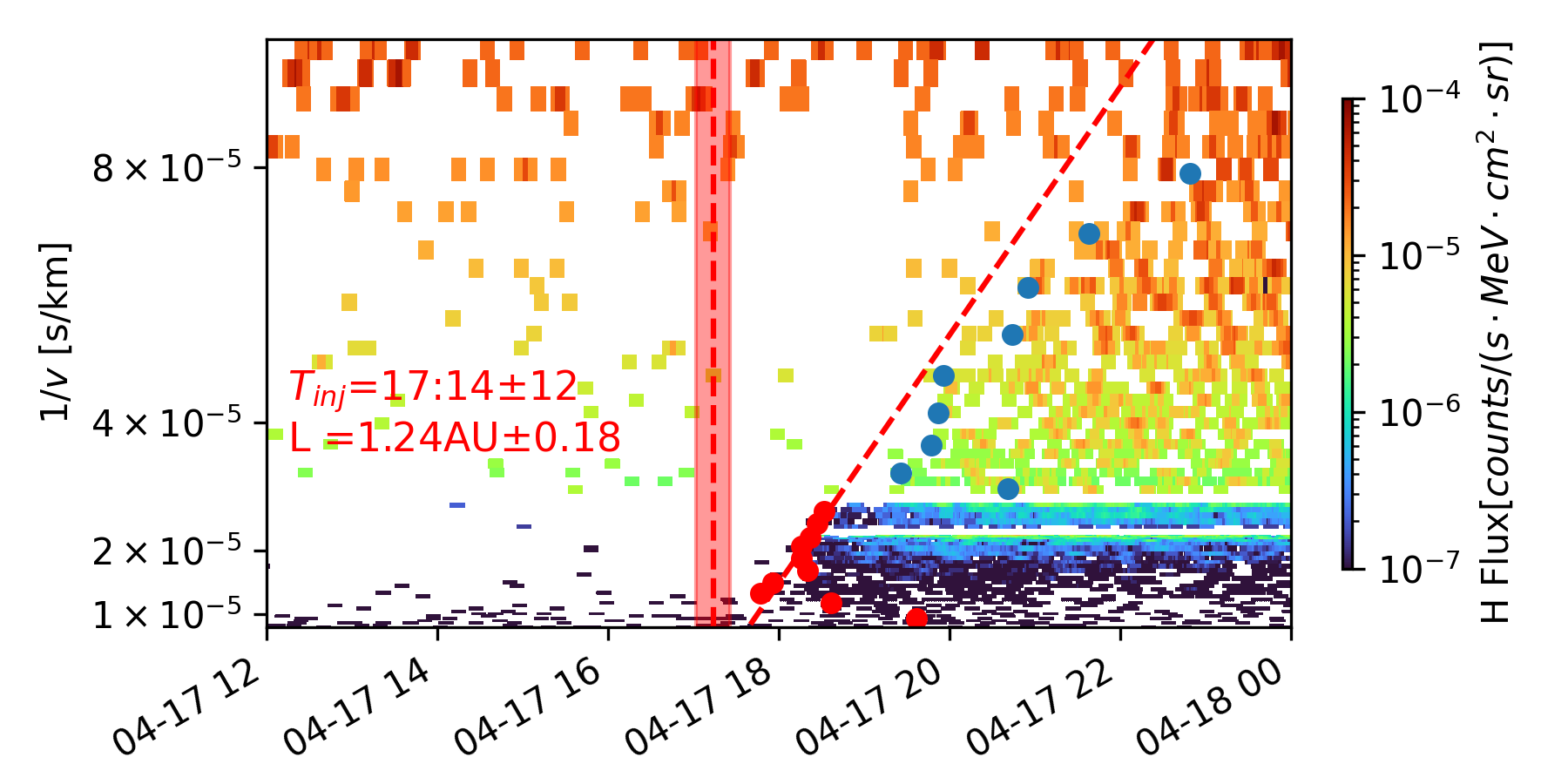}
    \caption{VDA of protons measured by Solar Orbiter/HET sun (red points) and EPT sun }(blue points, not included in the VDA fit). The vertical red line and shade represent the derived injection time and uncertainty.
    \label{fig:VDA_SOLO_proton}
\end{figure}

To perform a VDA, we determine the onset time based on the proton time profiles in the HET sunward telescope. Therefore, we use the energy channels between 7~MeV and 45~MeV and reconstruct the energy bins by combining every three proton channels. We then apply the Poisson-CUSUM method \citep{Huttunen2005} and derive the onset time in those new channels. The VDA for protons is based on these onset times and results in an inferred injection time at 17:14$\pm12$~min and a path length of $L$ = 1.24$\pm0.18$~au. We display the results in Fig.~\ref{fig:VDA_SOLO_proton} where we overplot the resulting VDA fit on the dynamic proton spectrogram.

For electrons, no VDA was possible because the onset times of many different energy channels were basically the same. A possible reason could be the rather poor pitch-angle coverage during the event onset not covering the direction along the magnetic field. We therefore determine inferred injection times for selected energy channels based on TSA assuming the same path length as derived from the proton VDA (see Table~\ref{table:timing}). 

The earliest arriving particles are MeV electrons with an onset time at 16:52$\pm15$~min (1.1--2.4~MeV) followed by near-relativistic (106~keV) electrons at 17:13$\pm2$~min. Both low and high-energy electron onset times lead to earlier solar injection times (16:41$\pm15$~min 16:55$\pm2$~min, respectively) than that obtained from proton VDA (see Sect.~\ref{sec:timing}).

\subsubsection{STEREO~A}\label{sec:sta_seps}
\begin{figure*}
   \centering
   \includegraphics[width=0.48\textwidth]{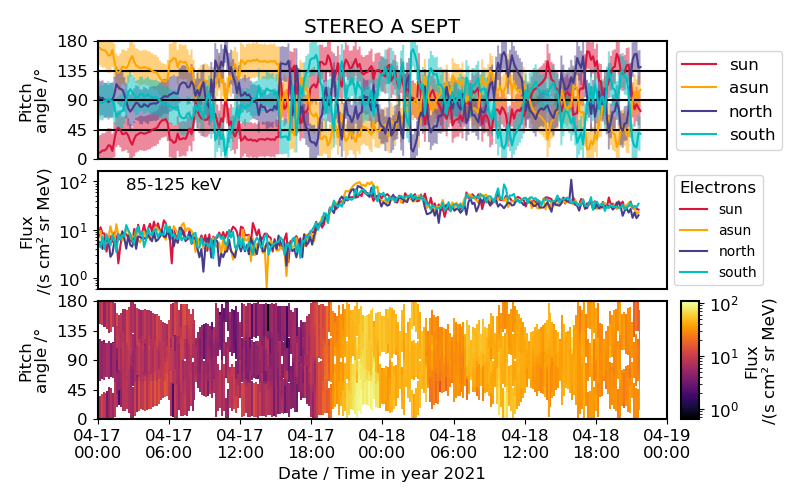}
   \includegraphics[width=0.51\textwidth]{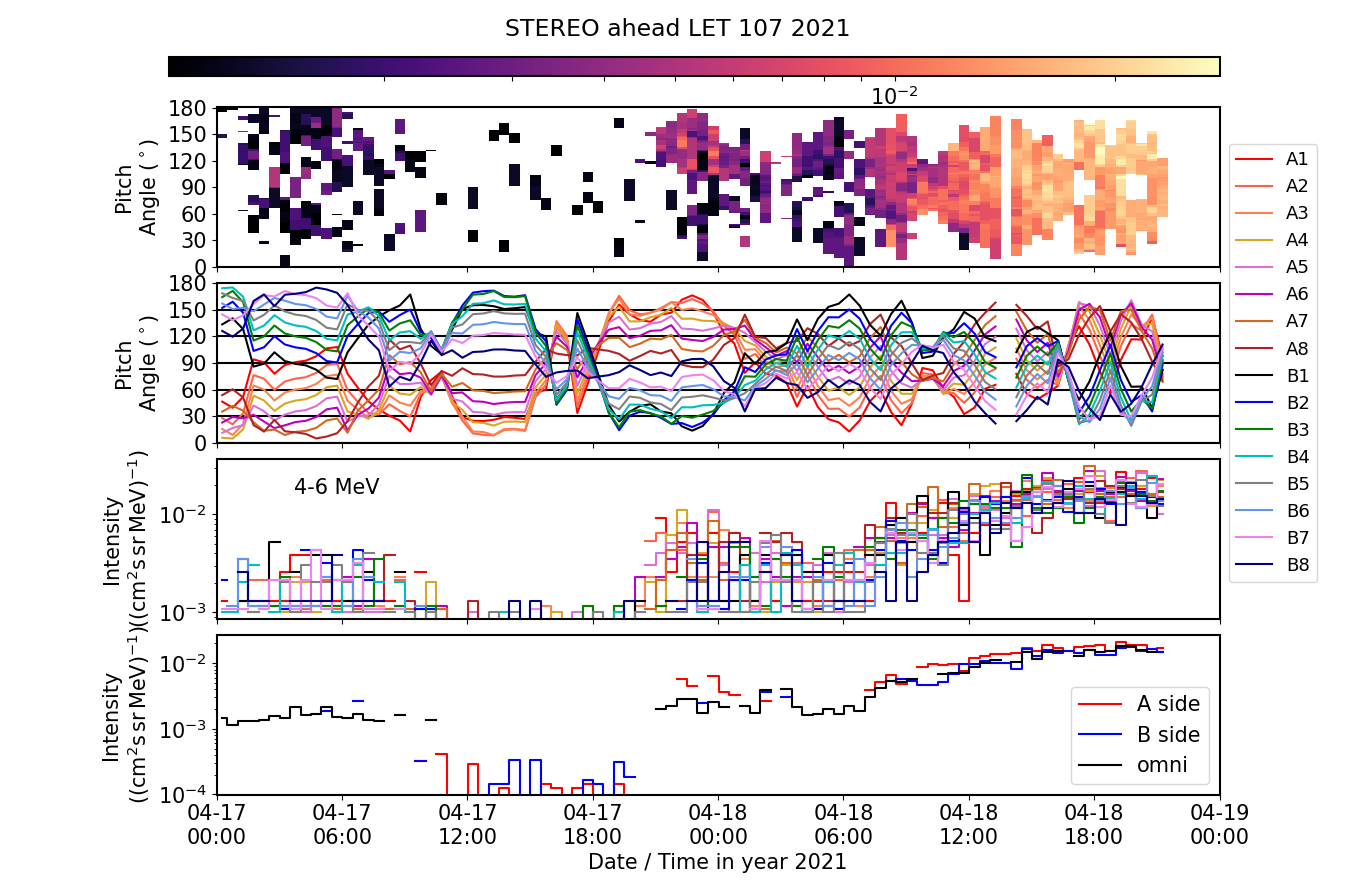}
      \caption{Pitch-angle distribution of 85-125~keV electrons (left) and 4-6~MeV protons (right) observed by STEREO~A/SEPT and LET, respectively. Left plot shows from top to bottom: Pitch-angle coverage of the different sensor apertures, intensities observed by each field of view, and pitch-angle distribution with color-coded intensities. Right plot shows from top to bottom: Pitch-angle distribution with color-coded intensities, pitch-angle pointing of the different LET sectors, intensities observed by each sector, and average intensities measured by the eight sectors on each side of the instrument.}
         \label{fig:STEREO_PAD}
   \end{figure*}

STEREO~A is a far-separated observer with $129^{\circ}$ ($109^{\circ}$) of longitudinal (total angular) separation between the flare location and the spacecraft's magnetic footpoint at the Sun  computed with ADAPT-WSA. It is therefore not surprising that the SEP event at STEREO~A is less intense than those explored so far, and that the  intensity-time profiles are more gradual and isotropic \citep[e.g.,][]{Dresing2014}.
Figure~\ref{fig:STEREO_PAD} (left) shows the electron PAD observed by STEREO~A/SEPT at ${\sim}100$~keV, which shows no anisotropy except during the time of the maximum, where the intensity in the anti-sunward sector is slightly higher. We note that, since the spacecraft was put upside-down after the superior solar conjunction in 2015, the Sun and anti-Sun sectors no longer point along the nominal Parker spiral but perpendicular to it. 

Figure~\ref{fig:STEREO_PAD} (right, third panel from top) shows the 4--6~MeV proton intensities observed in the 16 sectors of STEREO~A/LET. The top panel shows the color-coded intensity--PAD and the second panel shows the pitch-angles of the sector centers. The statistics in the single sectors are poor, which is why the bottom panel shows averaged intensities of the eight A and B sectors, respectively, and in black an average of all sectors. Interestingly, LET shows a double-peak time profile with the first  peak, starting shortly after 18:00~UT on 17 April, being much more anisotropic than the second  peak as almost no intensity is yet observed in the B-side sectors of LET. The depletion between the peaks at $\sim$6~UT on 18 April is not caused by poor pitch-angle coverage. Indeed, the pitch-angle coverage is better during this phase than during neighboring periods. As shown in Fig.~\ref{fig:IP_context_SOLO_STA} (right), there is no clear interplanetary structure that can be associated with this dip. We therefore argue that it is either caused by a change of the magnetic connection to the parent source region or a distinct new particle injection, which is also supported by the differently strong anisotropies during both peaks.

Due to the gradual nature of the event and rather poor statistics,  it was not possible to apply a VDA, and in order to determine onset times we had to average the data, leading to significant uncertainties. We obtain an onset at 18:25$\pm10$~min for 85--125~keV electrons and 19:30$\pm1$h for 13.6--23.8~MeV protons. Assuming a scatter-free propagation along a nominal Parker spiral  with a length of 1.16 au, this would translate to inferred injection times of 18:08$\pm10$~min for the electrons and 18:20$\pm1$~h for the protons.
The event is also less energetic at STEREO~A. Different to all other inner spacecraft, STEREO~A does not detect electrons in the MeV range and the event at 25~MeV protons is very weak (see Fig.~\ref{fig:solar-mach_and_multi_sc_SEP}, right). However, this could be also due to instrumental differences with STEREO~A/HET being less sensitive.

\subsubsection{Near-Earth Spacecraft}\label{sec:L1_seps}
As already discussed in Sect.~\ref{sssec:insitu_L1}, the SEP event at the Sun--Earth L1 point is only observed at high energies, both in electrons and protons. However, as the event is weak and very gradual, no VDA was possible and the determined onset times (see Table~\ref{table:timing}) suffer large uncertainties. Altogether, the observations of a gradual, delayed, and small event at Earth suggest that the event was only observed due to perpendicular particle diffusion \cite[e.g.,][]{Dresing2012} since there was probably no direct magnetic connection to a source region.


\section{Combined timing analysis and implications on the sources of the SEP event} \label{sec:timing}

   \begin{figure*}[ht!]
    \includegraphics[width=0.99\textwidth, trim=0 0 0 0, clip]{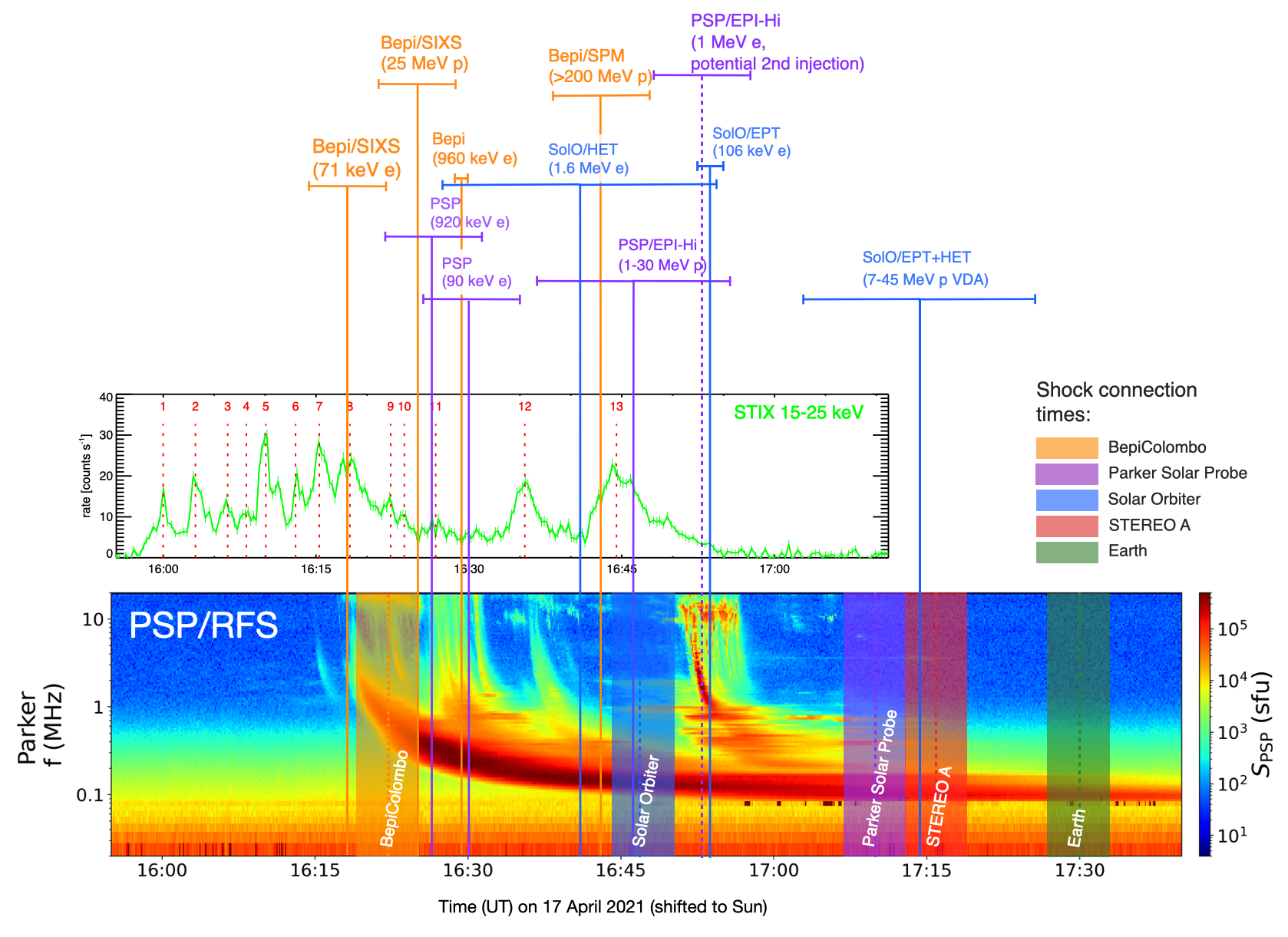}
      \caption{Inferred SEP injection times (vertical lines with temporal error bars on top) overplotted on the radio spectrogram as observed by PSP/RFS and the 15-25 keV X-ray observations by Solar Orbiter/STIX. All times have been shifted to the Sun by assuming the propagation time of the emission to the respective spacecraft. The shaded ranges mark the times when the spacecraft establish a magnetic connection with the CME-driven shock including an uncertainty of $\pm$3 min. 
              }
         \label{fig:radio_timing}
   \end{figure*}
\begin{table*}
\caption{Timing of solar phenomena and inferred SEP injection times t$_{\mbox{inj}}$. All times shifted to the Sun.}             
\label{table:timing}      
\centering          
\begin{tabular}{l l l l l}     
\hline   \\
 \multicolumn{2}{c}{Date / Time}  &   Observer / Instr. & Feature & Comment \\      
At the Sun & At observer & & & \\
 \hline \\
{\bf 17 April} & & & & \\

16:00	& 16:07    &	 Solar Orbiter/STIX & 1st nonthermal HXR peak & Major peak, very impulsive, \\
& & & & \quad <1 min duration \\			
16:16 &	16:24	& STA/SWAVES & type III burst \#1 onset & Also seen at PSP, Solar Orbiter,  \\
 & & & & \quad and Wind.\\

16:18$\pm4$min & 16:30$\pm4$min  &	BepiColombo/SIXS &	t$_{\mbox{inj}}$ of 71 keV el. & TSA, path length L=0.669 au \\		
16:18 &  16:26 UT &  Ground based/ & onset of decameter- & \\
& &   \quad e-CALLISO &  \quad type II burst & \\
16:22$\pm3$ min	& 16:30$\pm3$ min & BepiColombo & shock connection & time based on Earth \& STA obs.\\
16:25$\pm4$ min	& 17:00$\pm4$ min &	BepiColombo/SIXS &	t$_{\mbox{inj}}$ of 25 MeV p. & TSA, path length L=0.669 au \\
16:26 & 16:31    & PSP/RFS & type III burst  &	Series of five type III bursts;\\
& & &\quad group \#2 start & \quad all seen by PSP, highly polarized \\
& & & & \quad one also by SO and STA	\\
16:26 $\pm$5 min & 16:32$\pm$5 min	&	PSP/EPI-Hi & t$_{\mbox{inj}}$ of 920 keV el. & TSA, path length L=0.63 au	\\
16:28 &	16:36 &	SOHO/LASCO & CME 1st appearance &  E116S09; speed: v=880 km s$^{-1}$  \\
&&&&at $\sim$16~$R_{\odot}$\\
16:29$\pm1$min & 16:35$\pm1$min	&	BepiColombo/SIXS & t$_{\mbox{inj}}$ of 960 keV el. 	& TSA, path length L=0.669 au \\		
16:30$\pm5$min & 16:37$\pm5$min	&	PSP/EPI-Lo & t$_{\mbox{inj}}$ of $\sim$90 keV el. & TSA, path length L=0.63 au	\\
16:35 & 16:40 & PSP/RFS & type III burst & type II and III bursts,\\
 & & & \quad group \#3 start & \quad also seen at SolO, Wind, STA \\
16:35:30 &	16:42 &	Solar Orbiter/STIX &	nonthermal HXR peak \#12	& major late peak, \\
& & & & \quad $\sim$7 min duration \\

16:41$\pm15$min	& 16:52$\pm15$min &	Solar Orbiter/HET & t$_{\mbox{inj}}$ of 1.1-2.4 MeV el. &	TSA, path length L=1.24 au  \\		
16:43$\pm5$min &	16:53$\pm5$min &	BepiColombo/ &	t$_{\mbox{inj}}$ of $>$200 MeV p.  & Earliest onset seen in \\
 & & \quad Mio-SPM2& & \quad all SPM channels \\
& & & & \quad TSA, path length L=0.67 au\\

16:44:30 &	16:51:30 &	Solar Orbiter/STIX &	nonthermal HXR peak \#13 &	strongest peak at 25-50 keV; \\
& & & & \quad $\sim$10 min duration \\	
16:46$\pm10$min &  & PSP/EPI-Hi & t$_{\mbox{inj}}$ of 1-30 MeV protons &	VDA, resulting path\\
& & & & \quad length L=0.63$\pm0.05$ au\\
16:47$\pm3$ min	& 16:55$\pm3$ min & Solar Orbiter & shock connection & time based on Earth \& STA obs.\\
16:49 & 16:54 & PSP/RFS & type III burst \#4 start & type II and III bursts,\\
 & & & & \quad highly polarized\\
16:54$\pm5$min & 16:58$\pm5$min	&	PSP/EPI-Hi & time of potential 2nd inj. & TSA, path length L=0.43 au	\\
& & & \quad of 920 keV of el. & \\
16:55$\pm2$min & 17:13$\pm2$min  &	Solar Orbiter/ & t$_{\mbox{inj}}$ of 106 keV el. & TSA, path length L=1.24 au \\	
 & & \quad EPT-North & & \\[4pt]
17:11$\pm3$ min	& 17:19$\pm3$ min & PSP & shock connection & time based on Earth \& STA obs.\\
17:14$\pm12$min &   & 	Solar Orbiter/ & t$_{\mbox{inj}}$ of 7-45 MeV p. &	VDA, resulting path\\		
& &  \quad EPT+HET &&  \quad length L=1.24$\pm$0.18 au \\
17:16$\pm3$ min	& 17:24$\pm3$ min & STEREO~A & shock connection & time based on Earth \& STA obs.\\
17:30$\pm3$ min	& 17:38$\pm3$ min & Earth & shock connection & time based on Earth \& STA obs.\\
18:08$\pm10$min	& 18:25$\pm10$min	&	STA/SEPT-North & t$_{\mbox{inj}}$ of 85-125 keV el. &	TSA, path length L=1.16 au \\
18:20$\pm1$h & 19:30$\pm1$h	&	STA/HET & t$_{\mbox{inj}}$ of 13.6-23.8 MeV p. &	TSA, path length L=1.16 au \\
22:03$\pm2.5$h & 22:15$\pm2.5$h	&	SOHO/EPHIN & t$_{\mbox{inj}}$ of 0.25-0.7 MeV el. &	TSA, path length L=1.23 au \\
& & & & \\
{\bf 18 April} & & & & \\
4:07$\pm2$h & 5:00$\pm2$h	&	SOHO/ERNE & t$_{\mbox{inj}}$ of 13-25 MeV p. &	TSA, path length L=1.23 au \\
\hline
\end{tabular}
\end{table*}



Table~\ref{table:timing} presents a timeline of the main features of the 17 April 2021 SEP event with all observations times at the different spacecraft (column 2) being shifted to the Sun (column 1). This means, remote-sensing observations are corrected for the varying light travel times based on the different spacecraft distances, and energetic particle onset times are used to infer the corresponding injection times at the Sun. Studying the SEP event as observed by the multiple spacecraft implies the use of a multitude of instruments that provide different energy ranges and channel widths, different instrumental backgrounds, varying signal-to-noise ratios based on their locations with respect to the SEP source region, as well as different local interplanetary conditions that can influence the SEP observations (see Sect.~\ref{sec:in_situ_obs}). All these factors make a comparison of the SEP observations at the different spacecraft challenging. For example, for many spacecraft locations or species, a VDA was not possible (see Sect.~\ref{sec:sep_obs}). In these cases, we apply a simple TSA to infer the SEP injection times and use an energy channel showing a clear onset time. This implies that we sometimes have to use different energy ranges to infer SEP injection times. 

Table~\ref{table:timing} (see also Fig.~\ref{fig:radio_timing}) shows that the first feature of the event was the start of the HRX flare, which happened already at 16:00~UT. The flare then continued for almost an hour showing multiple HXR peaks with the last and strongest one at 16:44:30~UT (see Sect.~\ref{sec:flare}). The first radio type~III burst onset (TIII(1)) was only observed 16 minutes after the start of the flare, followed by the first type~II (TII(1)) burst observation at 16:18~UT (see also Sect.~\ref{sec:radio_obs}). 
Around this time (16:18$\pm$4min), the first SEPs were inferred to be injected towards BepiColombo as derived from the 71~keV electron onset.  The protons of about 25 MeV were injected were injected later towards BepiColombo at 16:25$\pm$4min, temporally situated between TIII(1) and TIII(2), at the end of TII(1). BepiColombo's SPM instrument on board Mio was even able to detect $>200$ MeV protons, which arrived, however, significantly delayed with an inferred injection time at 16:43$\pm$5min. 
Surprisingly, we determine a significant later injection time for $\sim$1~MeV electrons at 16:29$\pm$1min, as compared to the 71~keV electrons, which happened during TIII(2). Due to the impulsive, high-intensity event observed by BepiColombo/SIXS the onset times are well-defined carrying only small uncertainties, which are assumed to be the same for the inferred injection times. This strongly suggests that not only the electrons and protons observed at BepiColombo are related to different injection episodes, but also that the near-relativistic and relativistic particles suggest different injection times, a feature which was also observed during the 9 October 2021 SEP event \citep{Jebaraj23}.

Figure~\ref{fig:radio_timing} illustrates the inferred SEP injection times (vertical lines) in comparison with the HXR flare observations taken by Solar Orbiter/STIX and the radio observations by PSP/RFS. In contrast to Table~\ref{table:timing}, Fig.~\ref{fig:radio_timing} only displays those injection times that were inferred to happen during the early phase of the event, namely during the radio active phase. Therefore, only times corresponding to BepiColombo, Parker Solar Probe, and Solar Orbiter, the three best-connected spacecraft, are included. 

In the case of Parker Solar Probe, we find later injection times compared to BepiColombo and a significantly earlier injection of electrons compared to protons. Both relativistic and near-relativistic electrons are inferred to be injected during TIII(2) at  16:26$\pm 5$min ($\sim$1~MeV) and 16:30$\pm$5min ($\sim$90~keV). Because TIII(2) was found to be strongly directed towards Parker Solar Probe, this association is not surprising. We do not find evidence of SEPs related with TIII(1) to reach Parker Solar Probe's location. However, the inferred injection time of a step-like feature in the rising phase of Parker Solar Probe's electron event (see Sect.~\ref{sec:psp_seps}) shows a temporal correlation with TIII(4), the second type III burst, which shows a strong directivity towards Parker Solar Probe (cf.\ Sect.~\ref{sec:radio_obs}). 
The clear velocity dispersion observed by Parker Solar Probe for deka-MeV protons yields an injection time at 16:46$\pm$10min, which is temporally situated between TIII(3) and TIII(4). This suggests, similar to BepiColombo observations, that electrons and protons were injected during different episodes and could possibly be related to different acceleration mechanisms and locations.

As discussed in Sect.~\ref{sec:IP_solo} the interplanetary conditions between Solar Orbiter and the Sun were disturbed by minor transient events, likely affecting the SEP transport and leading to the comparatively lower peak intensities and less well-defined onsets. This could also lead to delayed SEP onsets and consequently yield too-late-inferred injection times, especially when using TSA,  which we do for the electrons observed by Solar Orbiter. Nevertheless, using the same path length as derived from proton VDA we find the same pattern of electrons inferred to be injected earlier at 16:41$\pm15$min ($\sim$1.6~MeV) and 16:55$\pm2$min ($\sim$100~keV) than protons for which we were able to perform a VDA yielding an injection time at 17:14$\pm12$min  for protons between 7 and 45 MeV. Only the 106~keV electron injection time could potentially be  related to a radio feature, that is, TIII(4). The inferred proton injection time is about 20 minutes later than the last type III burst (TIII(4)). 

The onset times of SEPs at STEREO and Earth are so delayed and uncertain that we cannot infer a direct connection with any of the early activity phenomena of the event, as shown in Fig.~\ref{fig:radio_timing}. Furthermore, the injection times of the three best-connected spacecraft (BepiColombo, Parker Solar Probe, and Solar Orbiter) spread already over the whole radio active time period of about 40 min. This implies that all four type~III radio bursts could mark distinct SEP injections that have contributed to the global multi-spacecraft SEP event. The different directions of these radio bursts (see Sect.~\ref{sec:radio_obs}) furthermore opens the possibility that the multiple injection episodes were differently important for the different observer locations.        

The vertical shaded regions in Fig.~\ref{fig:radio_timing} denote the times (including uncertainties) at which a magnetic connection with the CME-driven shock was established with each of the five inner-heliosphere spacecraft according to the analysis reported in Sect.~\ref{sec:shock_obs} and summarized in Table~\ref{table:shock_connectivity}. Although we find the shock to potentially connect already early and at low heights with all the five spacecraft locations, several inferred injection times happened already before, suggesting that the shock was not the main accelerator of these first arriving particles.
For BepiColombo, the 71~keV electron injection time and that of the 25~MeV protons (taking into account the uncertainty ranges) agree with the shock connection time. Relativistic electrons are found to be injected later, making a shock-related source still possible. For Parker Solar Probe, which has a comparatively late shock connection time at 17:11$\pm3$, all SEP injection times are inferred to happen significantly earlier. In contrast, for Solar Orbiter a sole shock source could be justified as the shock connection time happens during the first inferred injection time (taking into account the large error bar), which is the one of the MeV electrons, and well before the inferred injections of the ${\sim}100$~keV electrons and that of the protons. For STEREO~A and Earth, the shock connection times happened earlier than any inferred injection times, but because of the strongly delayed and uncertain onset times, it is not possible to pin down a clear role of the shock against the potentially involved transport effects.

\section{Interplanetary transport modeling}
\label{Sec:modelling}
  \begin{figure*}
   \centering
    \includegraphics[width=0.43\textwidth]{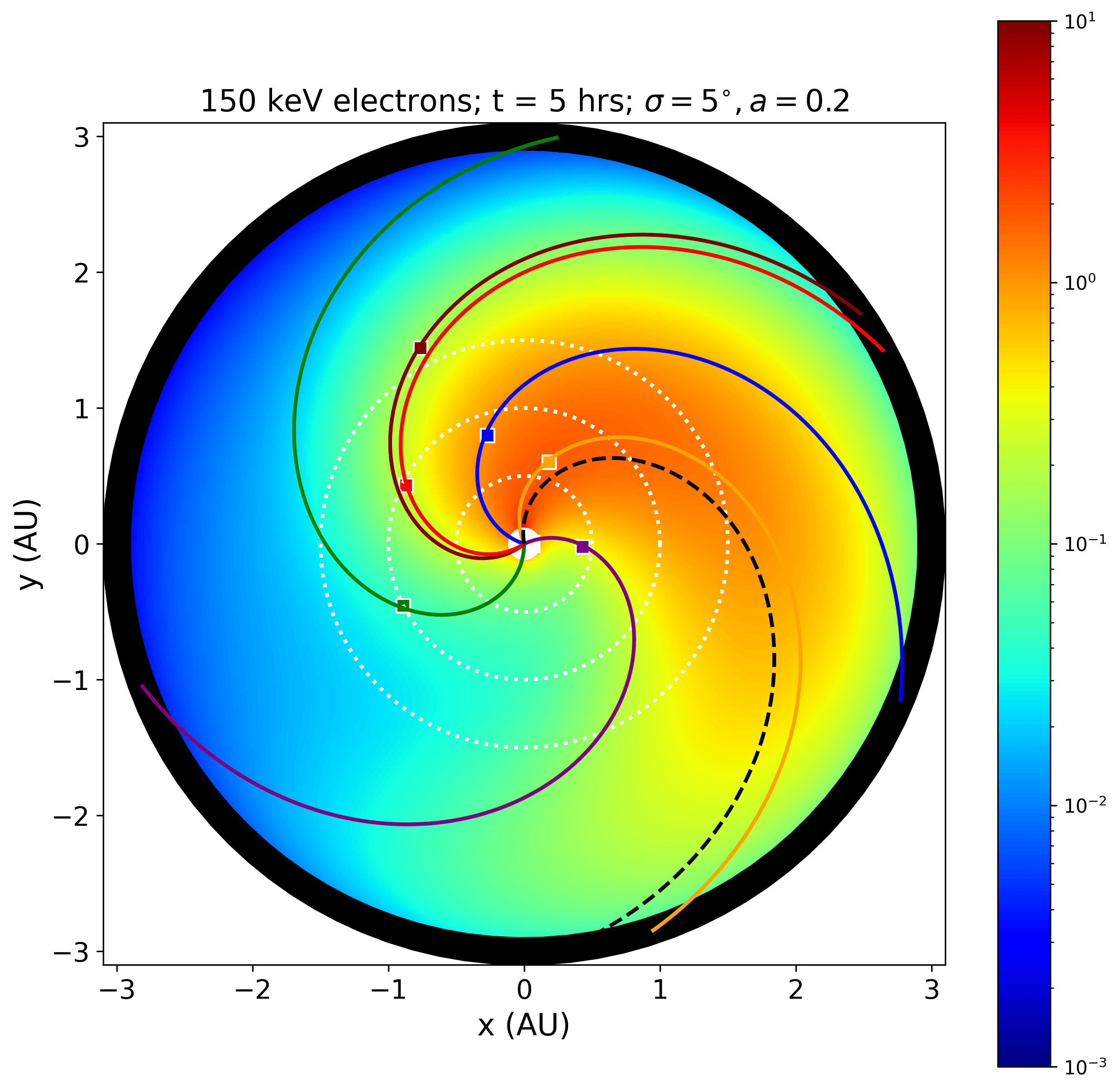}
    \includegraphics[width=0.43\textwidth]{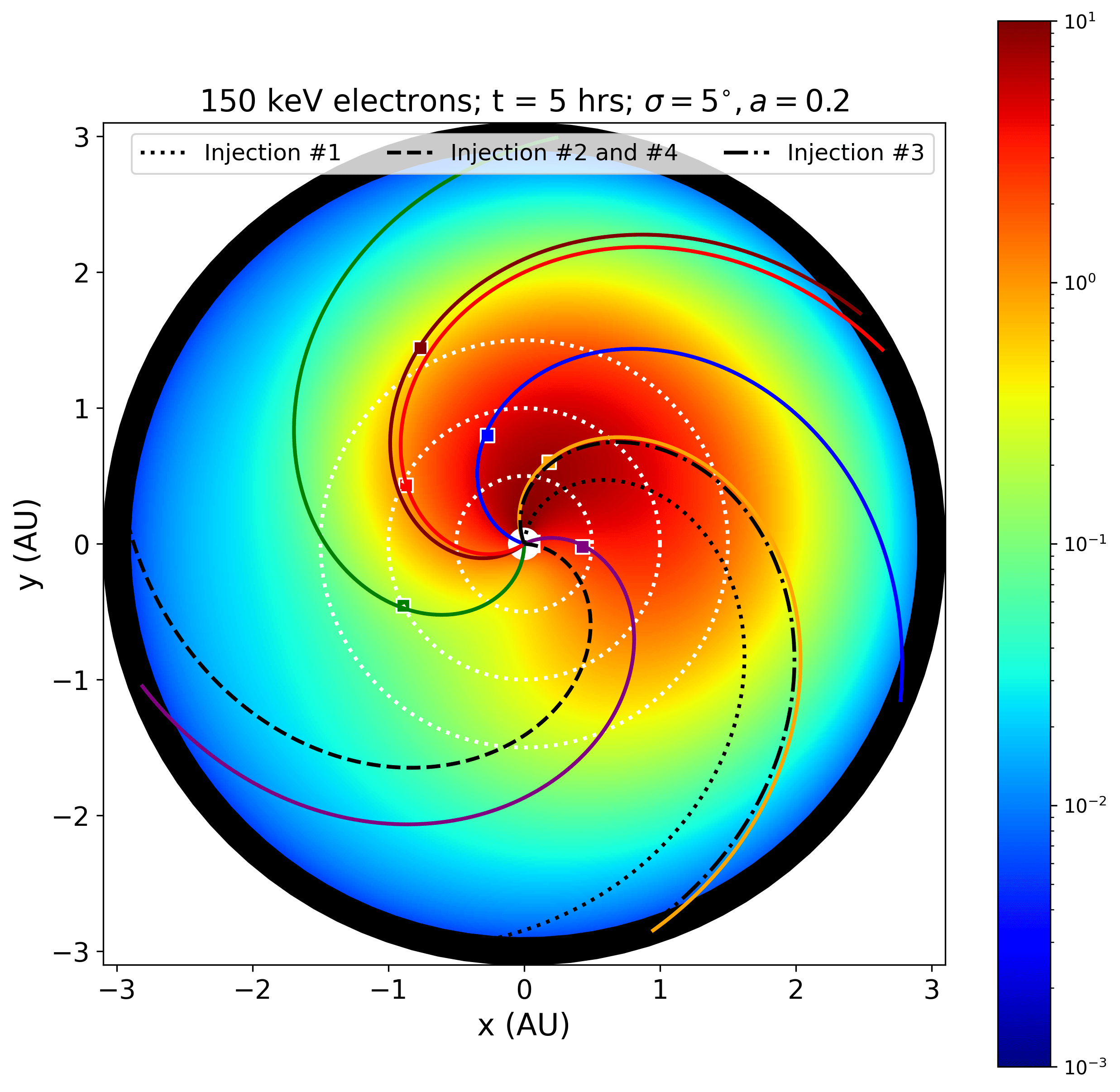}

   \includegraphics[width=0.43\textwidth]{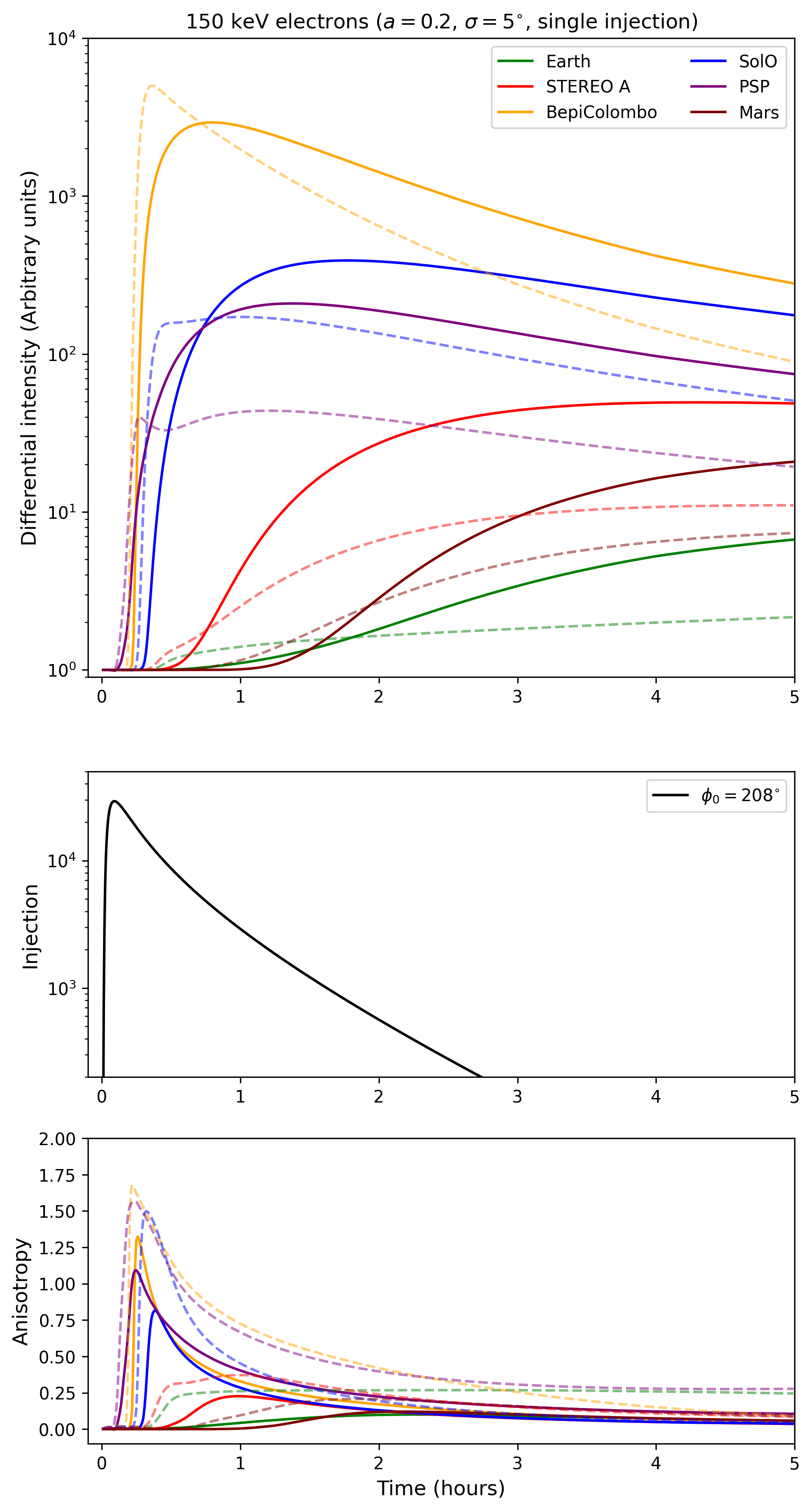}
   \includegraphics[width=0.43\textwidth]{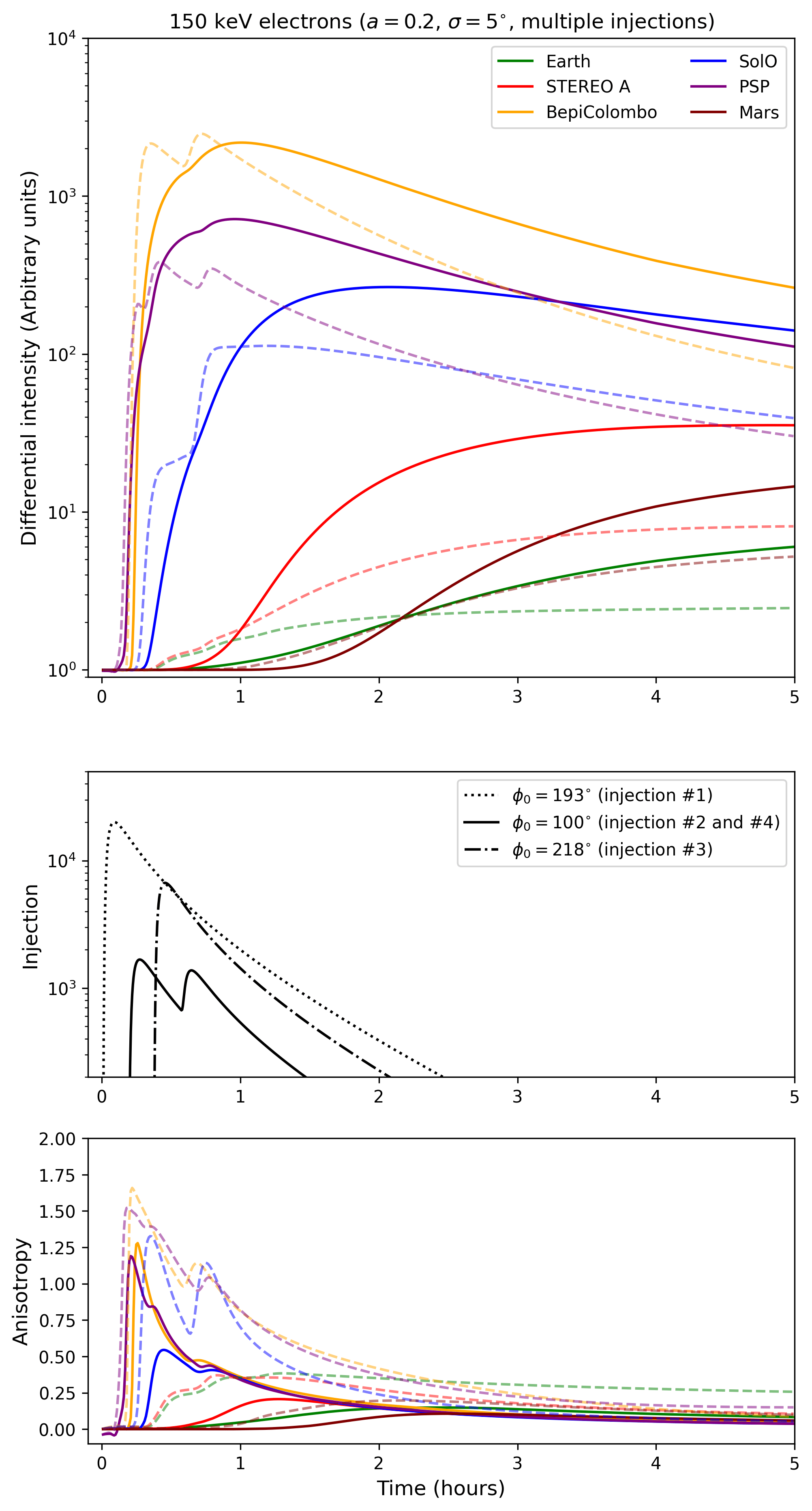}
      \caption{Transport modeling results for 150~keV electrons. The left panels represent the standard case of a single SEP injection into interplanetary space, while the right panels are for multiple injections. The top panels are normalized contour plots of the SEP intensity at five hours after the initial injection, while the bottom panels show the resulting particle intensities, as a function of time, at a number of spacecraft positions, the different SEP injections, and the resulting particle anisotropies. More details are given in the main text.
              }
         \label{fig:model_time_profiles}
   \end{figure*}

In this section, we present simulations of the interplanetary transport of SEPs using the spatially 2D model of \citet{Strauss2015}. Simulations are performed for 150~keV electrons to qualitatively illustrate the transport concepts discussed in this work  like the role of perpendicular diffusion vs. a direct magnetic connection to the source region. As input to the model we implement the pitch-angle and perpendicular diffusion coefficients used by \citet{Strauss2017,Strauss2020} that are based on  fundamental turbulent quantities and optimized to reproduce the catalog of widespread  events from \citet{Dresing2014}. The top left panel of Fig.~\ref{fig:model_time_profiles} shows a contour plot of the omni-directional intensity  of 150 keV electrons, calculated from this model when assuming a single SEP source, at five hours after particle injection. Here, the position of the different spacecraft are shown, along with their magnetic connectivity to the inner model boundary, assumed to be the Alfv\'en surface approximately located at $r \sim 10$  $R_{\odot}$ (0.05~au). The dashed magnetic field line connects to the maximum of the injected SEP distribution, which was assumed to be a Gaussian spatial distribution with a broadness of $5^{\circ}$. Below this,  we show the temporal profile of the omni-directional intensity at the different spacecraft, the assumed profile of the SEP source, and lastly the corresponding particle anisotropies. As in previous work, we assume a Reid--Axford profile for the SEP source with an acceleration time of 0.1 hours and an escape time of one hour. For the temporal profile of the differential intensity two sets of solutions are shown: The first, in dashed lines, are model solutions using the default set-up of \citet{Strauss2017}, while the solid curves are solutions where the parallel mean-free-path is reduced, in an admittedly ad-hoc fashion, by a factor of 5. This is done to account for the possibly more disturbed nature of the inner heliosphere during this event as discussed in Sect.~\ref{sec:in_situ_obs}, in contrast to the basic model that assumes quiet solar minimum conditions. Of course, a smaller parallel mean-free-path, namely more pitch-angle scattering, leads to a slower rise to maximum, a slower decay phase, and a smaller level of anisotropy. The exact levels of the transport parameters appropriate to reproduce this specific event will be the topic of a future, more detailed, modeling endeavor. Interestingly, and the main conclusion from this basic set-up, is the fact that the model, while reproducing the general trends observed by most of the spacecraft, consistently under-estimates the SEP intensity at the position of the Parker Solar Probe spacecraft for a range of transport parameters.

As a possible remedy for this discrepancy, the right panels of Fig.~\ref{fig:model_time_profiles} show the modeling scenario of multiple SEP sources, releasing particles into the inner heliosphere at different positions. The dotted and dashed lines in the top panel show the position of these four injection sources where these are chosen to approximately correspond to the inferred position of the four observed radio bursts from Sect.~\ref{sec:radio_obs}. The magnitude of the four injections are, however, not well constrained and chosen here rather arbitrarily to roughly correspond to the measurements. The normalization of these injections are chosen such that the total fluence of the electrons introduced into the heliosphere is the same for the left and right panels. For the default model set-up (i.e., the dashed curves), the different injections are visible in the calculated temporal profiles of the magnetically well-connected spacecraft, while any such prominent peaks disappear for the case of more pitch-angle scattering. More importantly, the level of the simulated profile at Parker Solar Probe is also now more consistent with the observations.

\section{Discussion} \label{sec:discussion}

The SEP event associated with the flare--CME on 17 April 2021 was observed by five well-separated spacecraft in the inner heliosphere, with additional constrains provided by observations at Mars. The multi-vantage point observations portray a complex picture of the event, which involves significantly different characteristics of both the energetic electron and the proton/ion event and an asymmetry in the longitudinal distribution of their intensities. We find evidence that the reason for the wide SEP spread involves a number of different mechanisms with varying importance for different vantage points, which we will discuss in the following.

While the associated CME was relatively slow and narrow (speed: $\sim$880~km~s$^{-1}$ and width: $\sim$46$^{\circ}$, see Sect.~\ref{sec:CME}) as compared to other widespread SEP events with high-energy particles \citep[e.g.,][]{Lario2017, Kouloumvakos2019}, the solar flare  emission in HXR was exceptionally long-lasting (one hour) and complex (cf.\ Sect.~\ref{sec:flare}). The radio event was also equally long-lasting (Sect.~\ref{sec:radio_obs}) and showed four distinct type III burst groups indicating particle injection episodes over a period of about 40 min. Several type II bursts were also observed, which suggests particle acceleration at the different flank regions of the associated shock. Although the event was observed by a fleet of five well-separated spacecraft, no full optical coverage of the solar surface was available, leaving some sectors of potential flare locations unobserved. However, our comprehensive analysis of the available X-ray, EUV, white-light, and radio observations suggests that the event was caused by activity related to a single source active region at the Sun (cf.\ Sect.~\ref{sec:remote_obs}). 

The earliest SEP onsets and inferred injection times are found for BepiColombo, which was the best-connected spacecraft to the flaring active region. Significantly further, but similarly far separated from the active region in heliolongitude were Parker Solar Probe (east of the flare) and Solar Orbiter (west of the flare). However, both spacecraft observed dramatically different SEP characteristics suggesting a longitudinal asymmetry: Parker Solar Probe observed a more intense and impulsive event while Solar Orbiter observed a more gradual, less intense and delayed event. While the intensity difference could also be explained by the different radial distances of the spacecraft, the strongly different intensity-time profiles rather suggest a different magnetic connection to the source region, which could be different portions of the CME-driven shock front \cite[e.g.,][]{Cane1988}  and/or combination of differently directed SEP injections. 

Our detailed radio analysis reveals that two of the four  observed TIII radio burst episodes were directed significantly more towards Parker Solar Probe as compared to the other radio bursts. These two injection episodes are likely the main source of the electron event at Parker Solar Probe, which is also supported by the inferred injection times (cf.\ Sect.~\ref{sec:timing}) and the results of the transport modeling (Sect.~\ref{Sec:modelling}). These injections may also have contributed to the Parker Solar Probe proton event as we infer an earlier injection of protons than the shock connection time (cf. Sect.~\ref{sec:timing}). This would also explain the comparatively high intensities observed at Parker Solar Probe when compared to Solar Orbiter. However, the long-lasting anisotropy of the proton event  at Parker Solar Probe of about nine hours (see Fig.~\ref{fig:psp_seps}), which is not observed in the case of the electrons, suggests a long-lasting proton injection, most likely related to the CME-driven shock. 
A similar picture is presented by Solar Orbiter, at which a clear anisotropy, lasting about seven hours, was observed for protons but not for electrons  (Fig. \ref{fig:SolO_PAD}), suggesting that the shock played an important role in creating the proton event at Solar Orbiter.
Also, BepiColombo observes a long-lasting anisotropy for protons (about seven hours), while electrons only show significant anisotropies during the rise phase of the event  (Fig. \ref{fig:Bepi_PAD}), which is in agreement with a short, likely flare-related, injection. While an initial flare contribution to the proton event cannot be excluded for BepiColombo and Parker Solar Probe, a later shock contribution is, therefore, likely.

An even less intense and more delayed SEP event was observed by STEREO~A and near-Earth spacecraft, which were far separated in heliolongitude ($126^{\circ}$ for STEREO~A and $144^{\circ}$ for Earth, see Table~\ref{table:sep_angles}). At both locations, no significant anisotropies were observed for electrons and only at STEREO~A a short anisotropic period during the early phase of the event was visible in LET proton measurements  (Fig. \ref{fig:STEREO_PAD}). Although we determine potential shock connection times for both positions already before any inferred injection time of SEPs reaching STEREO~A or Earth, the observed significantly lower intensities and missing anisotropies suggest that no direct connection with the shock nor flare-related injection was established but rather that perpendicular diffusion was involved in distributing the SEPs. However, the short anisotropic period observed by STEREO/LET for 4-6 MeV protons could be the trace of a shorter-lasting connection to the shock. 
The presence of interplanetary structures such as pre-event ICMEs and the SIRs at the spacecraft locations could further have modified the magnetic field topology and enhanced scattering conditions, which likely contributed to the unclear and delayed SEP onset times. 
However, a wider SEP injection region either provided by the extended shock front of 180$^{\circ}$ or the four type III burst-related injection episodes, which covered a longitudinal angle of about $110^{\circ}$, may have been a key ingredient in producing the widespread SEP event reaching also STEREO~A and Earth. Especially TIII(2) and TIII(4), marking injections close to the longitudinal location angle of Parker Solar Probe (see Sect.~\ref{sec:radio_obs}), would provide another, significantly closer injection region for the Earth's location, facilitated over the western limb, as compared to the location of the flare. This could explain the comparatively early electron onset time detected by SOHO/EPHIN around 22:00 UT on 17 April. 

At Mars, which was magnetically well-connected with STEREO~A during the onset of the event (see Fig.~\ref{fig:solar-mach_and_multi_sc_SEP}, left), we did not observe a clear SEP increase associated with the early phase of the event. However, the associated CME-driven shock could have reached Mars on 22 April and an energetic particle increase was observed, which could have been related to the shock.

\section{Summary and conclusions} \label{sec:summary}

The 17 April 2021 SEP event is the second widespread event of solar cycle 25 and the first one that was ever observed by five well-separated space missions in the inner heliosphere (within 1~au) constrained also by observations at Mars. 
It is an energetic event showing electrons up to the MeV range and 25~MeV protons reaching all inner heliospheric spacecraft positions, which span a longitudinal range of $210^{\circ}$. BepiColombo observations by Mio/SPM even show the presence of >200~MeV protons.
The closest observer to the Sun was Parker Solar Probe ($r=0.42$~au) followed by BepiColombo ($r=0.63$~au) and Solar Orbiter ($r=0.84$~au). 
As outlined in Sect.~\ref{sec:discussion}, the interplanetary SEP event was likely formed by a combination of different processes with varying importance at different spacecraft positions. For instance, the observations suggest a different origin of the electron and proton SEP event. This is most clear in the case of the three best-connected observers, BepiColombo, Parker Solar Probe, and Solar Orbiter:  at Parker Solar Probe and Solar Orbiter we find significantly earlier inferred injection times for electrons (at all energies) than for $\sim$25 MeV protons. At BepiColombo only the near-relativistic electron injection is found to be earlier than that of the protons. Also the much longer-lasting anisotropies observed in the proton event compared to electrons suggest an extended injection for protons only.  Furthermore, different spacecraft were likely fed by different injections related to the various radio features with different injection directions as suggested by the radio directivity analyses (see Sect.~\ref{sec:radio_obs}). The timing analysis (see Sect.~\ref{table:timing}) shows that BepiColombo detected electrons injected already during the first episode (TIII(1)), while Parker Solar Probe likely detected only electrons from the later episodes, mainly from TIII(2) and TIII(4), which were directed towards Parker Solar Probe. 

A possible alternative source of the two type III groups, namely TIII(2) and TIII(4) observed by Parker Solar Probe, could be the shock wave. In Sect.~\ref{sec:radio_obs}, we show that TIII(2) and TIII(4) were strongly polarized. This high degree of polarization indicates that the source is a region with strong magnetic fields. A highly compressive shock wave may provide such conditions where the electron beams are accelerated via a shock drift acceleration mechanism \citep[SDA; ][]{Ball01,Mann2018}. The energy gained by the electrons in these cases may also dependent on other factors such as the upstream electron distribution. If we were to assume that the thermal electrons \citep[$\sim$1\% of the speed of light, ][]{Halekas20} are being accelerated, then the maximum energy gain in a short period through SDA can be a factor of 14 which leads to a leads to 14\% speed of light or $\sim$10 keV. However, a small portion of the tail electrons may be accelerated to higher energies. Our multi-spacecraft analysis further emphasizes that the location where the TIII(2) and TIII(4) originated was in the direction of Parker Solar Probe. 
There is a strong possibility that some of the type III bursts within TIII(4) were accelerated by the shock wave since they are observed to be emanating from TII(HB) (cf. Sect.~\ref{sec:radio_obs}). However, the lack of meter-decameter wave measurements limits us from corroborating the generation of TIII(2) by the shock wave. In order to grasp the mechanisms of electron acceleration in the corona, full meter-decameter measurement would be necessary \citep[cf.][]{Jebaraj23}.  

Due to the not yet available Parker Solar Probe electron measurements in units of intensity, we cannot compare the electron intensity levels of Parker Solar Probe with other spacecraft. However, because the time profiles at Parker Solar Probe are much more impulsive and peaked for both electrons and protons compared to Solar Orbiter, situated at a comparable absolute longitudinal separation angle, it is plausible that the overall ordering of intensities observed at the different spacecraft is similar for electrons and protons. Based on this assumption we performed initial interplanetary transport modeling of this event for electrons (see Sect.~\ref{Sec:modelling}) supporting the idea that SEPs were released from several different source longitudes. These show that without a SEP source near the magnetic footpoint of Parker Solar Probe, the measured intensity at that spacecraft cannot be reproduced by the model, independently of the adopted transport coefficients. While these simulations show promise, a more detailed modeling study is required, taking the disturbed nature of the interplanetary medium into consideration while sufficiently optimizing the transport coefficients used in the model. The detailed 3D modeling of the particle transport will be based on the treatment developed by \cite{Droege2010}, which employs focused particle transport along the large-scale heliospheric magnetic field as well as diffusion perpendicular to the field. We will also take into account the disturbances in the large-scale magnetic field caused by preceding CMEs as comes out from the observations and predictions from the ENLIL model used in the current version of the paper. Such an extensive study is currently ongoing.

An important feature, which is not observed for the electron event, is the presence of long-lasting periods of proton anisotropies as observed by BepiColombo, Parker Solar Probe, and Solar Orbiter. This requires an extended, likely shock-associated proton injection. However, a flare contribution to the early phase of the proton event at these spacecraft cannot be excluded and is especially likely for Parker Solar Probe, for which the shock-connection time is determined to be established only after the inferred proton injection time. In the case when the scattering of protons, in particular through 90$^\circ$ pitch angle, is reduced a long-lasting anisotropy can arise as well. 

In the case of the two farthest separated observers, STEREO~A at a longitudinal separation angle of $126^{\circ}$, and Earth at $144^{\circ}$, SEP intensities were significantly lower, showing a more gradual profile and significantly delayed onsets, which suggests that these observers did not establish a direct magnetic connection with any of the potential SEP source regions. Missing anisotropies together with the aforementioned characteristics suggest that perpendicular diffusion was involved in distributing the SEPs to these far separated longitudes. The presence of interplanetary structures such as the pre-event ICMEs and the SIRs may have contributed to modifying the magnetic field topology and enhancing scattering conditions leaving also room for a potential direct magnetic connection that was masked by a strongly disturbed parallel transport. However, even in the case of perpendicular transport being involved, we consider it likely that the widespread SEP observations were supported by an extended injection region. This could either have been provided by the extended shock front ($\sim180^{\circ}$) or by the different injection directions marked by the four radio type III burst episodes covering in total a longitudinal range of about $110^{\circ}$. A likely evidence for an extended shock front is the presence of multiple type II radio bursts, namely TII(1), TII(2), and TII(HB) (see Sect.~\ref{sec:radio_obs}), which are emitted at different locations on the expanding shock front. Our analysis of the radio intensity and directivity suggests that the sources of TII(1) and TII(2) were directed towards Solar Orbiter and STEREO~A, while that of TII(HB) was clearly directed towards Parker Solar Probe. 

The study of the 17 April 2021 widespread SEP event allowed us to perform a comprehensive multi-spacecraft analysis combining remote-sensing and in-situ observations of six well-separated observer positions and taking full advantage of the various complementary data sets. The advanced spacecraft fleet enabled us to characterize signatures of a very complex SEP event, which would not have been possible with fewer observers.
We were able to identify significant differences between the electron and proton SEP event, as observed by the different spacecraft, with a more likely flare association of the electron event and a more likely shock source for the proton event. However, a mixing of both cannot be excluded.
Thanks to the position of Parker Solar Probe, we were able to observe otherwise hidden SEP features that highlight the role of significantly different injection directions of the four different injection episodes, which we consider a new scenario that has to be taken into account as a potential contributor to widespread events. 

Future case studies of additional widespread events with the currently available spacecraft fleet will hopefully allow us to further characterize the necessary ingredients of widespread events and the different scenarios that are able to produce these rather rare events.


\begin{acknowledgements}
      Solar Orbiter is a space mission of international collaboration between ESA and NASA, operated by ESA. The STIX instrument is an international collaboration between Switzerland, Poland, France, Czech Republic, Germany, Austria, Ireland, and Italy. We acknowledge funding by the European Union’s Horizon 2020 research and innovation program under grant agreements No.\ 101004159 (SERPENTINE), and No.\ 870405 (EUHFORIA 2.0).
      BepiColombo is a joint ESA -- JAXA science mission with instruments and contributions directly funded by ESA Member States and JAXA.
      Parker Solar Probe was designed, built, and is now operated by the Johns Hopkins Applied Physics Laboratory as part of NASA’s Living with a Star (LWS) program (contract NNN06AA01C). Support from the LWS management and technical team has played a critical role in the success of the Parker Solar Probe mission.
      Work in the University of Turku was performed under the umbrella of Finnish Centre of Excellence in Research of Sustainable Space (Academy of Finland Grant No.\ 336809). N.D.\ is grateful for support by the Turku Collegium for Science, Medicine and Technology of the University of Turku, Finland. N.D.\ and I.C.J.\ are grateful for support by the Academy of Finland (SHOCKSEE, grant No.\ 346902).
      L.R.G.\ thanks Toni Galvin for her assistance in the use of STEREO/PLASTIC data and Leila Mays, Dusan Odstrcil, Nick Arge, and Shaela Jones-Mecholsky regarding the use of WSA-ENLIL model. 
      The UAH team acknowledges the financial support by the Spanish Ministerio de Ciencia, Innovación y Universidades FEDER/MCIU/AEI Projects ESP2017-88436-R and PID2019-104863RB-I00/AEI/10.13039/501100011033. 
      I.C.J.\ acknowledges funding by the BRAIN-be project SWiM (Solar Wind Modelling with EUHFORIA for the new heliospheric missions).
      A.K.\ acknowledges support from NASA's NNN06AA01C (SO-SIS Phase-E) contract.
      V.K.\ acknowledges the support by NASA under grants No.\ 18-2HSWO218\_2-0010 and 19-HSR-19\_2-0143.
      E.P.\ acknowledges support from NASA's PSP-GI (grant No.\ 80NSSC22K0349), O2R (grant No.\ 80NSSC20K0285), and LWS-SC (grant No.\ 80NSSC22K0893) programmes.
      E.A.\ acknowledges support from the Academy of Finland (Postdoctoral Researcher Grant 322455).
      W.D.\  and Y.K.\ acknowledge ISSI for the possibility to discuss the questions related to particle propagation in interplanetary space during the meeting of the team No.\ 459 (led by G.~Li and L.~Wang).
      B.S.-C.\ acknowledges support through UK-STFC Ernest Rutherford Fellowship ST/V004115/1 and STFC grant ST/V000209/1.
      The work of F.S.\ was supported by DLR grant No.\ 50 OT 1904.
      N.W.\ acknowledges support from NASA program NNH17ZDA001N-LWS and from the Research Foundation - Flanders (FWO-Vlaanderen, fellowship No.\ 1184319N).
      C.O.L.\ acknowledges support from NASA's LWS Program (Grant No. 80NSSC21K1325) and the MAVEN project funded through the NASA Mars Exploration Program. 
      C.O.L.\ and C.M.S.C.\ acknowledge support from the IMPACT Investigation by the NASA Heliophysics Division through the STEREO Project Office at NASA GSFC (Grant No. 80NSSC18K1446).
      M.L.\ acknowledges support from the Italian Space Agency and the National Institute of Astrophysics through the ASI-INAF n. 2020-35-HH.0 agreement for the development of the ASPIS prototype of scientific data centre for Space Weather.
      ENLIL simulation results have been provided by the CCMC at NASA Goddard Space Flight Center (GSFC) through their public Runs on Request system (\url{http://ccmc.gsfc.nasa.gov}; run ID Laura\_Rodriguez-Garcia\_041322\_SH\_1). 
      The WSA model was developed by N.~Arge, currently at GSFC, and the ENLIL Model was developed by D.~Odstrcil, currently at George Mason University.
\end{acknowledgements}

\bibliographystyle{aa} 
\bibliography{references}

\begin{thebibliography}{155}
\expandafter\ifx\csname natexlab\endcsname\relax\def\natexlab#1{#1}\fi

\bibitem[{{Acu{\~n}a} {et~al.}(2008){Acu{\~n}a}, {Curtis}, {Scheifele},
  {Russell}, {Schroeder}, {Szabo}, \& {Luhmann}}]{Acuna2008}
{Acu{\~n}a}, M.~H., {Curtis}, D., {Scheifele}, J.~L., {et~al.} 2008, \ssr, 136,
  203

\bibitem[{Agostinelli {et~al.}(2003)Agostinelli, Allison, Amako, Apostolakis,
  Araujo, Arce, Asai, Axen, Banerjee, Barrand, Behner, Bellagamba, Boudreau,
  Broglia, Brunengo, Burkhardt, Chauvie, Chuma, Chytracek, Cooperman, Cosmo,
  Degtyarenko, Dell'Acqua, Depaola, Dietrich, Enami, Feliciello, Ferguson,
  Fesefeldt, Folger, Foppiano, Forti, Garelli, Giani, Giannitrapani, Gibin,
  {Gómez Cadenas}, González, {Gracia Abril}, Greeniaus, Greiner, Grichine,
  Grossheim, Guatelli, Gumplinger, Hamatsu, Hashimoto, Hasui, Heikkinen,
  Howard, Ivanchenko, Johnson, Jones, Kallenbach, Kanaya, Kawabata, Kawabata,
  Kawaguti, Kelner, Kent, Kimura, Kodama, Kokoulin, Kossov, Kurashige, Lamanna,
  Lampén, Lara, Lefebure, Lei, Liendl, Lockman, Longo, Magni, Maire,
  Medernach, Minamimoto, {Mora de Freitas}, Morita, Murakami, Nagamatu,
  Nartallo, Nieminen, Nishimura, Ohtsubo, Okamura, O'Neale, Oohata, Paech,
  Perl, Pfeiffer, Pia, Ranjard, Rybin, Sadilov, {Di Salvo}, Santin, Sasaki,
  Savvas, Sawada, Scherer, Sei, Sirotenko, Smith, Starkov, Stoecker, Sulkimo,
  Takahata, Tanaka, Tcherniaev, {Safai Tehrani}, Tropeano, Truscott, Uno,
  Urban, Urban, Verderi, Walkden, Wander, Weber, Wellisch, Wenaus, Williams,
  Wright, Yamada, Yoshida, \& Zschiesche}]{Agostinelli2003}
Agostinelli, S., Allison, J., Amako, K., {et~al.} 2003, Nuclear Instruments and
  Methods in Physics Research Section A: Accelerators, Spectrometers, Detectors
  and Associated Equipment, 506, 250

\bibitem[{{Altschuler} \& {Newkirk}(1969)}]{Altschuler1969}
{Altschuler}, M.~D. \& {Newkirk}, G. 1969, \solphys, 9, 131

\bibitem[{{Arge} {et~al.}(2013){Arge}, {Henney}, {Hernandez}, {Toussaint},
  {Koller}, \& {Godinez}}]{Arge2013}
{Arge}, C.~N., {Henney}, C.~J., {Hernandez}, I.~G., {et~al.} 2013, in Solar
  Wind 13, ed. G.~P. {Zank}, J.~{Borovsky}, R.~{Bruno}, J.~{Cirtain},
  S.~{Cranmer}, H.~{Elliott}, J.~{Giacalone}, W.~{Gonzalez}, G.~{Li},
  E.~{Marsch}, E.~{Moebius}, N.~{Pogorelov}, J.~{Spann}, \&
  O.~{Verkhoglyadova}, Vol. CS-1539, Am. Inst. Phys., Melville, 11--14

\bibitem[{{Arge} {et~al.}(2010){Arge}, {Henney}, {Koller}, {Compeau}, {Young},
  {MacKenzie}, {Fay}, \& {Harvey}}]{ADAPT_Arge2009}
{Arge}, C.~N., {Henney}, C.~J., {Koller}, J., {et~al.} 2010, in Solar Wind 12,
  ed. M.~{Maksimovic}, K.~{Issautier}, N.~{Meyer-Vernet}, M.~M.~{Moncuquet}, \&
  F.~{Pantellini}, Vol. CS-1216, Am. Inst. Phys., Melville, 343--346

\bibitem[{{Arge} {et~al.}(2011){Arge}, {Henney}, {Koller}, {Toussaint},
  {Harvey}, \& {Young}}]{Arge2010}
{Arge}, C.~N., {Henney}, C.~J., {Koller}, J., {et~al.} 2011, in J. Phys., ed.
  N.~V. {Pogorelov}, E.~{Audit}, \& G.~P. {Zank}, Vol. CS-444, Astron. Soc.
  Pacific, San Francisco, 99

\bibitem[{{Arge} {et~al.}(2004){Arge}, {Luhmann}, {Odstr{\v{c}}il},
  {Schrijver}, \& {Li}}]{Arge2004}
{Arge}, C.~N., {Luhmann}, J.~G., {Odstr{\v{c}}il}, D., {Schrijver}, C.~J., \&
  {Li}, Y. 2004, J. Atmos. Solar-Terr. Phys., 66, 1295

\bibitem[{{Arge} {et~al.}(2003){Arge}, {Odstr{\v{c}}il}, {Pizzo}, \&
  {Mayer}}]{Arge2003a}
{Arge}, C.~N., {Odstr{\v{c}}il}, D., {Pizzo}, V.~J., \& {Mayer}, L.~R. 2003, in
  Solar Wind Ten, ed. M.~{Velli}, R.~{Bruno}, F.~{Malara}, \& B.~{Bucci}, Vol.
  CS-679, Am. Inst. Phys., Melville, 190--193

\bibitem[{{Arge} \& {Pizzo}(2000)}]{ArgePizzo2000}
{Arge}, C.~N. \& {Pizzo}, V.~J. 2000, \jgr, 105, 10465

\bibitem[{{Bale} {et~al.}(2016){Bale}, {Goetz}, {Harvey}, {Turin}, {Bonnell},
  {Dudok de Wit}, {Ergun}, {MacDowall}, {Pulupa}, {Andre}, {Bolton},
  {Bougeret}, {Bowen}, {Burgess}, {Cattell}, {Chandran}, {Chaston}, {Chen},
  {Choi}, {Connerney}, {Cranmer}, {Diaz-Aguado}, {Donakowski}, {Drake},
  {Farrell}, {Fergeau}, {Fermin}, {Fischer}, {Fox}, {Glaser}, {Goldstein},
  {Gordon}, {Hanson}, {Harris}, {Hayes}, {Hinze}, {Hollweg}, {Horbury},
  {Howard}, {Hoxie}, {Jannet}, {Karlsson}, {Kasper}, {Kellogg}, {Kien},
  {Klimchuk}, {Krasnoselskikh}, {Krucker}, {Lynch}, {Maksimovic}, {Malaspina},
  {Marker}, {Martin}, {Martinez-Oliveros}, {McCauley}, {McComas}, {McDonald},
  {Meyer-Vernet}, {Moncuquet}, {Monson}, {Mozer}, {Murphy}, {Odom},
  {Oliverson}, {Olson}, {Parker}, {Pankow}, {Phan}, {Quataert}, {Quinn},
  {Ruplin}, {Salem}, {Seitz}, {Sheppard}, {Siy}, {Stevens}, {Summers}, {Szabo},
  {Timofeeva}, {Vaivads}, {Velli}, {Yehle}, {Werthimer}, \&
  {Wygant}}]{Bale2016}
{Bale}, S.~D., {Goetz}, K., {Harvey}, P.~R., {et~al.} 2016, \ssr, 204, 49

\bibitem[{{Ball} \& {Melrose}(2001)}]{Ball01}
{Ball}, L. \& {Melrose}, D.~B. 2001, \pasa, 18, 361

\bibitem[{{Balmaceda} {et~al.}(2018){Balmaceda}, {Vourlidas}, {Stenborg}, \&
  {Dal Lago}}]{balmaceda2018}
{Balmaceda}, L.~A., {Vourlidas}, A., {Stenborg}, G., \& {Dal Lago}, A. 2018,
  \apj, 863, 57

\bibitem[{{Barabash} {et~al.}(2006){Barabash}, {Lundin}, {Andersson},
  {Brinkfeldt}, {Grigoriev}, {Gunell}, {Holmstr{\"o}m}, {Yamauchi}, {Asamura},
  {Bochsler}, {Wurz}, {Cerulli-Irelli}, {Mura}, {Milillo}, {Maggi}, {Orsini},
  {Coates}, {Linder}, {Kataria}, {Curtis}, {Hsieh}, {Sandel}, {Frahm},
  {Sharber}, {Winningham}, {Grande}, {Kallio}, {Koskinen}, {Riihel{\"a}},
  {Schmidt}, {S{\"a}les}, {Kozyra}, {Krupp}, {Woch}, {Livi}, {Luhmann},
  {McKenna-Lawlor}, {Roelof}, {Williams}, {Sauvaud}, {Fedorov}, \&
  {Thocaven}}]{Barabash2006}
{Barabash}, S., {Lundin}, R., {Andersson}, H., {et~al.} 2006, \ssr, 126, 113

\bibitem[{{Benkhoff} {et~al.}(2021){Benkhoff}, {Murakami}, {Baumjohann},
  {Besse}, {Bunce}, {Casale}, {Cremosese}, {Glassmeier}, {Hayakawa}, {Heyner},
  {Hiesinger}, {Huovelin}, {Hussmann}, {Iafolla}, {Iess}, {Kasaba},
  {Kobayashi}, {Milillo}, {Mitrofanov}, {Montagnon}, {Novara}, {Orsini},
  {Quemerais}, {Reininghaus}, {Saito}, {Santoli}, {Stramaccioni}, {Sutherland},
  {Thomas}, {Yoshikawa}, \& {Zender}}]{Benkhoff2021}
{Benkhoff}, J., {Murakami}, G., {Baumjohann}, W., {et~al.} 2021, \ssr, 217, 90

\bibitem[{{Benz} \& {Thejappa}(1988)}]{Benz88}
{Benz}, A.~O. \& {Thejappa}, G. 1988, \aap, 202, 267

\bibitem[{{Bougeret} {et~al.}(2008){Bougeret}, {Goetz}, {Kaiser}, {Bale},
  {Kellogg}, {Maksimovic}, {Monge}, {Monson}, {Astier}, {Davy}, {Dekkali},
  {Hinze}, {Manning}, {Aguilar-Rodriguez}, {Bonnin}, {Briand}, {Cairns},
  {Cattell}, {Cecconi}, {Eastwood}, {Ergun}, {Fainberg}, {Hoang}, {Huttunen},
  {Krucker}, {Lecacheux}, {MacDowall}, {Macher}, {Mangeney}, {Meetre},
  {Moussas}, {Nguyen}, {Oswald}, {Pulupa}, {Reiner}, {Robinson}, {Rucker},
  {Salem}, {Santolik}, {Silvis}, {Ullrich}, {Zarka}, \&
  {Zouganelis}}]{2008SSRv..136..487B}
{Bougeret}, J.~L., {Goetz}, K., {Kaiser}, M.~L., {et~al.} 2008, \ssr, 136, 487

\bibitem[{{Bougeret} {et~al.}(1995){Bougeret}, {Kaiser}, {Kellogg}, {Manning},
  {Goetz}, {Monson}, {Monge}, {Friel}, {Meetre}, {Perche}, {Sitruk}, \&
  {Hoang}}]{1995SSRv...71..231B}
{Bougeret}, J.~L., {Kaiser}, M.~L., {Kellogg}, P.~J., {et~al.} 1995, \ssr, 71,
  231

\bibitem[{{Brown}(1963)}]{Brown1963}
{Brown}, R.~G. 1963, {Smoothing forecasting and prediction in discrete time
  series} (Prentice Hall, New Jersey), 468pp

\bibitem[{Brueckner {et~al.}(1995)Brueckner, Howard, Koomen, Korendyke,
  Michels, Moses, Socker, Dere, Lamy, Llebaria, Bout, Schwenn, Simnett,
  Bedford, \& Eyles}]{Brueckner1995}
Brueckner, G.~E., Howard, R.~A., Koomen, M.~J., {et~al.} 1995, \solphys, 162,
  357

\bibitem[{{Cane} {et~al.}(1988){Cane}, {Reames}, \& {von
  Rosenvinge}}]{Cane1988}
{Cane}, H.~V., {Reames}, D.~V., \& {von Rosenvinge}, T.~T. 1988, \jgr, 93, 9555

\bibitem[{Chicarro(2004)}]{chicarro2004}
Chicarro, A. 2004, Mars Express: the Scientific Payload, 1240, v

\bibitem[{{Dalla} {et~al.}(2020){Dalla}, {de Nolfo, G. A.}, {Bruno, A.},
  {Giacalone, J.}, {Laitinen, T.}, {Thomas, S.}, {Battarbee, M.}, \& {Marsh, M.
  S.}}]{Dalla2020}
{Dalla}, S., {de Nolfo, G. A.}, {Bruno, A.}, {et~al.} 2020, A\&A, 639, A105

\bibitem[{{Domingo} {et~al.}(1995){Domingo}, {Fleck}, \&
  {Poland}}]{Domingo1995}
{Domingo}, V., {Fleck}, B., \& {Poland}, A.~I. 1995, \solphys, 162, 1

\bibitem[{{Downs} {et~al.}(2021){Downs}, {Warmuth}, {Long}, {Bloomfield},
  {Kwon}, {Veronig}, {Vourlidas}, \& {Vr{\v{s}}nak}}]{Downs2021}
{Downs}, C., {Warmuth}, A., {Long}, D.~M., {et~al.} 2021, \apj, 911, 118

\bibitem[{{Dresing} {et~al.}(2014){Dresing}, {G{\'o}mez-Herrero}, {Heber},
  {Klassen}, {Malandraki}, {Dr{\"o}ge}, \& {Kartavykh}}]{Dresing2014}
{Dresing}, N., {G{\'o}mez-Herrero}, R., {Heber}, B., {et~al.} 2014, \aap, 567,
  A27

\bibitem[{{Dresing} {et~al.}(2012){Dresing}, {G{\'o}mez-Herrero}, {Klassen},
  {Heber}, {Kartavykh}, \& {Dr{\"o}ge}}]{Dresing2012}
{Dresing}, N., {G{\'o}mez-Herrero}, R., {Klassen}, A., {et~al.} 2012, \solphys,
  281, 281

\bibitem[{{Dr{\"o}ge} {et~al.}(2016){Dr{\"o}ge}, {Kartavykh}, {Dresing}, \&
  {Klassen}}]{Droege2016}
{Dr{\"o}ge}, W., {Kartavykh}, Y.~Y., {Dresing}, N., \& {Klassen}, A. 2016,
  \apj, 826, 134

\bibitem[{Dr\"{o}ge {et~al.}(2010)Dr\"{o}ge, Kartavykh, Klecker, \&
  Kovaltsov}]{Droege2010}
Dr\"{o}ge, W., Kartavykh, Y.~Y., Klecker, B., \& Kovaltsov, G.~A. 2010, \apj,
  709, 912

\bibitem[{{Dumbovi{\'c}} {et~al.}(2019){Dumbovi{\'c}}, {Guo}, {Temmer}, {Mays},
  {Veronig}, {Heinemann}, {Dissauer}, {Hofmeister}, {Halekas}, {M{\"o}stl},
  {Amerstorfer}, {Hinterreiter}, {Banjac}, {Herbst}, {Wang}, {Holzknecht},
  {Leitner}, \& {Wimmer─Schweingruber}}]{Dumbovic2019}
{Dumbovi{\'c}}, M., {Guo}, J., {Temmer}, M., {et~al.} 2019, \apj, 880, 18

\bibitem[{{Dumbovi{\'c}} {et~al.}(2021){Dumbovi{\'c}}, {Veronig},
  {Podladchikova}, {Thalmann}, {Chikunova}, {Dissauer}, {Magdaleni{\'c}},
  {Temmer}, {Guo}, \& {Samara}}]{Dumbovic2021}
{Dumbovi{\'c}}, M., {Veronig}, A.~M., {Podladchikova}, T., {et~al.} 2021, \aap,
  652, A159

\bibitem[{{Eyles} {et~al.}(2009){Eyles}, {Harrison}, {Davis}, {Waltham},
  {Shaughnessy}, {Mapson-Menard}, {Bewsher}, {Crothers}, {Davies}, {Simnett},
  {Howard}, {Moses}, {Newmark}, {Socker}, {Halain}, {Defise}, {Mazy}, \&
  {Rochus}}]{Eyles2009}
{Eyles}, C.~J., {Harrison}, R.~A., {Davis}, C.~J., {et~al.} 2009, \solphys,
  254, 387

\bibitem[{{Fletcher} {et~al.}(2011){Fletcher}, {Dennis}, {Hudson}, {Krucker},
  {Phillips}, {Veronig}, {Battaglia}, {Bone}, {Caspi}, {Chen}, {Gallagher},
  {Grigis}, {Ji}, {Liu}, {Milligan}, \& {Temmer}}]{Fletcher2011}
{Fletcher}, L., {Dennis}, B.~R., {Hudson}, H.~S., {et~al.} 2011, Space Sci.
  Rev., 159, 19

\bibitem[{Forbush(1946)}]{Forbush1946}
Forbush, S.~E. 1946, Phys. Rev., 70, 771

\bibitem[{{Fox} {et~al.}(2016){Fox}, {Velli}, {Bale}, {Decker}, {Driesman},
  {Howard}, {Kasper}, {Kinnison}, {Kusterer}, {Lario}, {Lockwood}, {McComas},
  {Raouafi}, \& {Szabo}}]{Fox2016}
{Fox}, N.~J., {Velli}, M.~C., {Bale}, S.~D., {et~al.} 2016, \ssr, 204, 7

\bibitem[{{Frassati} {et~al.}(2022){Frassati}, {Laurenza}, {Bemporad}, {West},
  {Mancuso}, {Susino}, {Alberti}, \& {Romano}}]{Frassati2022}
{Frassati}, F., {Laurenza}, M., {Bemporad}, A., {et~al.} 2022, \apj, 926, 227

\bibitem[{Galvin {et~al.}(2008)Galvin, Kistler, Popecki, Farrugia, Simunac,
  Ellis, M\"{o}bius, Lee, Boehm, Carroll, Crawshaw, Conti, Demaine, Ellis,
  Gaidos, Googins, Granoff, Gustafson, Heirtzler, King, Knauss, Levasseur,
  Longworth, Singer, Turco, Vachon, Vosbury, Widholm, Blush, Karrer, Bochsler,
  Daoudi, Etter, Fischer, Jost, Opitz, Sigrist, Wurz, Klecker, Ertl,
  Seidenschwang, Wimmer-Schweingruber, Koeten, Thompson, \&
  Steinfeld}]{Galvin2008}
Galvin, A.~B., Kistler, L.~M., Popecki, M.~A., {et~al.} 2008, \ssr, 136, 437

\bibitem[{Gedalin {et~al.}(2021)Gedalin, Russell, \& Dimmock}]{Gedalin2021}
Gedalin, M., Russell, C.~T., \& Dimmock, A.~P. 2021, Journal of Geophysical
  Research: Space Physics, 126, e2021JA029519, e2021JA029519 2021JA029519

\bibitem[{{Gieseler} {et~al.}(2022){Gieseler}, {Dresing}, {Palmroos}, {von
  Forstner}, {Price}, {Vainio}, {Kouloumvakos}, {Rodr{\'\i}guez-Garc{\'\i}a},
  {Trotta}, {G{\'e}not}, {Masson}, {Roth}, \& {Veronig}}]{Gieseler2022}
{Gieseler}, J., {Dresing}, N., {Palmroos}, C., {et~al.} 2022, arXiv e-prints,
  arXiv:2210.00819

\bibitem[{{Ginzburg} \& {Zhelezniakov}(1958)}]{Ginzburg58}
{Ginzburg}, V.~L. \& {Zhelezniakov}, V.~V. 1958, \sovast, 2, 653

\bibitem[{{G{\'o}mez-Herrero} {et~al.}(2015){G{\'o}mez-Herrero}, {Dresing},
  {Klassen}, {Heber}, {Lario}, {Agueda}, {Malandraki}, {Blanco},
  {Rodr{\'{\i}}guez-Pacheco}, \& {Banjac}}]{Gomez-Herrero2015}
{G{\'o}mez-Herrero}, R., {Dresing}, N., {Klassen}, A., {et~al.} 2015, \apj,
  799, 55

\bibitem[{{Grigis} \& {Benz}(2004)}]{Grigis2004}
{Grigis}, P.~C. \& {Benz}, A.~O. 2004, \aap, 426, 1093

\bibitem[{{Halekas} {et~al.}(2020){Halekas}, {Whittlesey}, {Larson},
  {McGinnis}, {Maksimovic}, {Berthomier}, {Kasper}, {Case}, {Korreck},
  {Stevens}, {Klein}, {Bale}, {MacDowall}, {Pulupa}, {Malaspina}, {Goetz}, \&
  {Harvey}}]{Halekas20}
{Halekas}, J.~S., {Whittlesey}, P., {Larson}, D.~E., {et~al.} 2020, \apjs, 246,
  22

\bibitem[{Harvey {et~al.}(1996)Harvey, Hill, Hubbard, Kennedy, Leibacher,
  Pintar, Gilman, Noyes, Title, Toomre, Ulrich, Bhatnagar, Kennewell,
  Marquette, Patrón, Saá, \& Yasukawa}]{Harvey1996}
Harvey, J.~W., Hill, F., Hubbard, R.~P., {et~al.} 1996, Science, 272, 1284

\bibitem[{Heyner {et~al.}(2021)Heyner, Auster, Forna{\c{c}}on, Carr, Richter,
  Mieth, Kolhey, Exner, Motschmann, Baumjohann, Matsuoka, Magnes, Berghofer,
  Fischer, Plaschke, Nakamura, Narita, Delva, Volwerk, Balogh, Dougherty,
  Horbury, Langlais, Mandea, Masters, Oliveira, S{\'{a}}nchez-Cano, Slavin,
  Vennerstr{\o}m, Vogt, Wicht, \& Glassmeier}]{Heyner2021}
Heyner, D., Auster, H.-U., Forna{\c{c}}on, K.-H., {et~al.} 2021, Space Science
  Reviews, 217, 52

\bibitem[{{Hickmann} {et~al.}(2015){Hickmann}, {Godinez}, {Henney}, \&
  {Arge}}]{Hickmann2015}
{Hickmann}, K.~S., {Godinez}, H.~C., {Henney}, C.~J., \& {Arge}, C.~N. 2015,
  \solphys, 290, 1105

\bibitem[{{Hill} {et~al.}(2017){Hill}, {Mitchell}, {Andrews}, {Cooper},
  {Gurnee}, {Hayes}, {Layman}, {McNutt}, {Nelson}, {Parker}, {Schlemm},
  {Stokes}, {Begley}, {Boyle}, {Burgum}, {Do}, {Dupont}, {Gold}, {Haggerty},
  {Hoffer}, {Hutcheson}, {Jaskulek}, {Krimigis}, {Liang}, {London}, {Noble},
  {Roelof}, {Seifert}, {Strohbehn}, {Vandegriff}, \& {Westlake}}]{Hill2017}
{Hill}, M.~E., {Mitchell}, D.~G., {Andrews}, G.~B., {et~al.} 2017, J. Geophys.
  Res. Space Phys., 122, 1513

\bibitem[{{Horbury} {et~al.}(2020){Horbury}, {O'Brien}, {Carrasco Blazquez},
  {Bendyk}, {Brown}, {Hudson}, {Evans}, {Oddy}, {Carr}, {Beek}, {Cupido},
  {Bhattacharya}, {Dominguez}, {Matthews}, {Myklebust}, {Whiteside}, {Bale},
  {Baumjohann}, {Burgess}, {Carbone}, {Cargill}, {Eastwood}, {Erd{\"o}s},
  {Fletcher}, {Forsyth}, {Giacalone}, {Glassmeier}, {Goldstein}, {Hoeksema},
  {Lockwood}, {Magnes}, {Maksimovic}, {Marsch}, {Matthaeus}, {Murphy},
  {Nakariakov}, {Owen}, {Owens}, {Rodriguez-Pacheco}, {Richter}, {Riley},
  {Russell}, {Schwartz}, {Vainio}, {Velli}, {Vennerstrom}, {Walsh},
  {Wimmer-Schweingruber}, {Zank}, {M{\"u}ller}, {Zouganelis}, \&
  {Walsh}}]{Horbury2020}
{Horbury}, T.~S., {O'Brien}, H., {Carrasco Blazquez}, I., {et~al.} 2020, \aap,
  642, A9

\bibitem[{Howard {et~al.}(2008)Howard, Moses, Vourlidas, Newmark, Socker,
  Plunkett, Korendyke, Cook, Hurley, Davila, Thompson, {St Cyr}, Mentzell,
  Mehalick, Lemen, Wuelser, Duncan, Tarbell, Wolfson, Moore, Harrison, Waltham,
  Lang, Davis, Eyles, Mapson-Menard, Simnett, Halain, Defise, Mazy, Rochus,
  Mercier, Ravet, Delmotte, Auchere, Delaboudiniere, Bothmer, Deutsch, Wang,
  Rich, Cooper, Stephens, Maahs, Baugh, McMullin, \& Carter}]{Howard2008}
Howard, R.~A., Moses, J.~D., Vourlidas, A., {et~al.} 2008, \ssr, 136, 67

\bibitem[{Huovelin {et~al.}(2020)Huovelin, Vainio, Kilpua, Lehtolainen,
  Korpela, Esko, Muinonen, Bunce, Martindale, Grande, Andersson, Nenonen,
  Lehti, Schmidt, Genzer, Vihavainen, Saari, Peltonen, Valtonen, Talvioja,
  Portin, Narendranath, Jarvinen, Okada, Milillo, Laurenza, Heino, \&
  Oleynik}]{Huovelin2020}
Huovelin, J., Vainio, R., Kilpua, E., {et~al.} 2020, Space Science Reviews, 216

\bibitem[{{Huttunen-Heikinmaa} {et~al.}(2005){Huttunen-Heikinmaa}, {Valtonen},
  \& {Laitinen}}]{Huttunen2005}
{Huttunen-Heikinmaa}, K., {Valtonen}, E., \& {Laitinen}, T. 2005, \aap, 442,
  673

\bibitem[{Jakosky {et~al.}(2015)Jakosky, Grebowsky, Luhmann, \&
  Brain}]{Jakosky2015}
Jakosky, B.~M., Grebowsky, J.~M., Luhmann, J.~G., \& Brain, D.~A. 2015,
  Geophysical Research Letters, 42, 8791

\bibitem[{{Jebaraj} {et~al.}(2023{\natexlab{a}}){Jebaraj}, {Koulooumvakos},
  {Dresing}, {Warmuth}, {Wijsen}, {Palmroos}, {Gieseler}, {Vainio}, {Krupar},
  {Magdalenic}, {Wiegelmann}, {Schuller}, {Battaglia}, \& {Fedeli}}]{Jebaraj23}
{Jebaraj}, I.~C., {Koulooumvakos}, A., {Dresing}, N., {et~al.}
  2023{\natexlab{a}}, arXiv e-prints, arXiv:2301.03650

\bibitem[{{Jebaraj} {et~al.}(2021){Jebaraj}, {Kouloumvakos}, {Magdalenic},
  {Rouillard}, {Mann}, {Krupar}, \& {Poedts}}]{Jebaraj21}
{Jebaraj}, I.~C., {Kouloumvakos}, A., {Magdalenic}, J., {et~al.} 2021, \aap,
  654, A64

\bibitem[{{Jebaraj} {et~al.}(2023{\natexlab{b}}){Jebaraj}, {Magdalenic},
  {Krasnoselskikh}, {Krupar}, \& {Poedts}}]{Jebaraj22}
{Jebaraj}, I.~C., {Magdalenic}, J., {Krasnoselskikh}, V., {Krupar}, V., \&
  {Poedts}, S. 2023{\natexlab{b}}, \aap, 670, A20

\bibitem[{{Jebaraj} {et~al.}(2020){Jebaraj}, {Magdaleni{\'c}}, {Podladchikova},
  {Scolini}, {Pomoell}, {Veronig}, {Dissauer}, {Krupar}, {Kilpua}, \&
  {Poedts}}]{Jebaraj20}
{Jebaraj}, I.~C., {Magdaleni{\'c}}, J., {Podladchikova}, T., {et~al.} 2020,
  \aap, 639, A56

\bibitem[{{Kaiser} {et~al.}(2008){Kaiser}, {Kucera}, {Davila}, {St. Cyr},
  {Guhathakurta}, \& {Christian}}]{Kaiser2008}
{Kaiser}, M.~L., {Kucera}, T.~A., {Davila}, J.~M., {et~al.} 2008, \ssr, 136, 5

\bibitem[{{Kasper} {et~al.}(2016){Kasper}, {Abiad}, {Austin}, {Balat-Pichelin},
  {Bale}, {Belcher}, {Berg}, {Bergner}, {Berthomier}, {Bookbinder}, {Brodu},
  {Caldwell}, {Case}, {Chandran}, {Cheimets}, {Cirtain}, {Cranmer}, {Curtis},
  {Daigneau}, {Dalton}, {Dasgupta}, {DeTomaso}, {Diaz-Aguado}, {Djordjevic},
  {Donaskowski}, {Effinger}, {Florinski}, {Fox}, {Freeman}, {Gallagher},
  {Gary}, {Gauron}, {Gates}, {Goldstein}, {Golub}, {Gordon}, {Gurnee}, {Guth},
  {Halekas}, {Hatch}, {Heerikuisen}, {Ho}, {Hu}, {Johnson}, {Jordan},
  {Korreck}, {Larson}, {Lazarus}, {Li}, {Livi}, {Ludlam}, {Maksimovic},
  {McFadden}, {Marchant}, {Maruca}, {McComas}, {Messina}, {Mercer}, {Park},
  {Peddie}, {Pogorelov}, {Reinhart}, {Richardson}, {Robinson}, {Rosen},
  {Skoug}, {Slagle}, {Steinberg}, {Stevens}, {Szabo}, {Taylor}, {Tiu}, {Turin},
  {Velli}, {Webb}, {Whittlesey}, {Wright}, {Wu}, \& {Zank}}]{Kasper2016}
{Kasper}, J.~C., {Abiad}, R., {Austin}, G., {et~al.} 2016, \ssr, 204, 131

\bibitem[{{Kay} {et~al.}(2020){Kay}, {Mays}, \& {Verbeke}}]{Kay2020}
{Kay}, C., {Mays}, M.~L., \& {Verbeke}, C. 2020, Space Weather, 18, e02382

\bibitem[{{Khotyaintsev} {et~al.}(2021){Khotyaintsev}, {Graham}, {Vaivads},
  {Steinvall}, {Edberg}, {Eriksson}, {Johansson}, {Sorriso-Valvo},
  {Maksimovic}, {Bale}, {Chust}, {Krasnoselskikh}, {Kretzschmar},
  {Lorf{\`e}vre}, {Plettemeier}, {Sou{\v{c}}ek}, {Steller},
  {{\v{S}}tver{\'a}k}, {Tr{\'a}vn{\'\i}{\v{c}}ek}, {Vecchio}, {Horbury},
  {O'Brien}, {Evans}, \& {Angelini}}]{Khotyaintsev2021}
{Khotyaintsev}, Y.~V., {Graham}, D.~B., {Vaivads}, A., {et~al.} 2021, \aap,
  656, A19

\bibitem[{{Kienreich} {et~al.}(2009){Kienreich}, {Temmer}, \&
  {Veronig}}]{Kienreich2009}
{Kienreich}, I.~W., {Temmer}, M., \& {Veronig}, A.~M. 2009, \apjl, 703, L118

\bibitem[{{Kilpua} {et~al.}(2017){Kilpua}, {Koskinen}, \&
  {Pulkkinen}}]{Kilpua2017}
{Kilpua}, E., {Koskinen}, H. E.~J., \& {Pulkkinen}, T.~I. 2017, Living Reviews
  in Solar Physics, 14, 5

\bibitem[{{Klein} \& {Dalla}(2017)}]{Klein2017}
{Klein}, K.-L. \& {Dalla}, S. 2017, \ssr, 212, 1107

\bibitem[{{Kollhoff} {et~al.}(2021){Kollhoff}, {Kouloumvakos}, {Lario},
  {Dresing}, {G{\'o}mez-Herrero}, {Rodr{\'\i}guez-Garc{\'\i}a}, {Malandraki},
  {Richardson}, {Posner}, {Klein}, {Pacheco}, {Klassen}, {Heber}, {Cohen},
  {Laitinen}, {Cernuda}, {Dalla}, {Espinosa Lara}, {Vainio}, {K{\"o}berle},
  {K{\"u}hl}, {Xu}, {Berger}, {Eldrum}, {Br{\"u}dern}, {Laurenza}, {Kilpua},
  {Aran}, {Rouillard}, {Bu{\v{c}}{\'\i}k}, {Wijsen}, {Pomoell},
  {Wimmer-Schweingruber}, {Martin}, {B{\"o}ttcher}, {Freiherr von Forstner},
  {Terasa}, {Boden}, {Kulkarni}, {Ravanbakhsh}, {Yedla}, {Janitzek},
  {Rodr{\'\i}guez-Pacheco}, {Prieto Mateo}, {S{\'a}nchez Prieto}, {Parra
  Espada}, {Rodr{\'\i}guez Polo}, {Mart{\'\i}nez Hell{\'\i}n}, {Carcaboso},
  {Mason}, {Ho}, {Allen}, {Bruce Andrews}, {Schlemm}, {Seifert}, {Tyagi},
  {Lees}, {Hayes}, {Bale}, {Krupar}, {Horbury}, {Angelini}, {Evans}, {O'Brien},
  {Maksimovic}, {Khotyaintsev}, {Vecchio}, {Steinvall}, \&
  {Asvestari}}]{Kollhoff2021}
{Kollhoff}, A., {Kouloumvakos}, A., {Lario}, D., {et~al.} 2021, \aap, 656, A20

\bibitem[{{Kouloumvakos} {et~al.}(2022){Kouloumvakos}, {Kwon},
  {Rodr{\'\i}guez-Garc{\'\i}a}, {Lario}, {Dresing}, {Kilpua}, {Vainio},
  {T{\"o}r{\"o}k}, {Plotnikov}, {Rouillard}, {Downs}, {Linker}, {Malandraki},
  {Pinto}, {Riley}, \& {Allen}}]{Kouloumvakos2022}
{Kouloumvakos}, A., {Kwon}, R.~Y., {Rodr{\'\i}guez-Garc{\'\i}a}, L., {et~al.}
  2022, \aap, 660, A84

\bibitem[{{Kouloumvakos} {et~al.}(2021){Kouloumvakos}, {Rouillard}, {Warmuth},
  {Magdalenic}, {Jebaraj}, {Mann}, {Vainio}, \& {Monstein}}]{Kouloumvakos21}
{Kouloumvakos}, A., {Rouillard}, A., {Warmuth}, A., {et~al.} 2021, \apj, 913,
  99

\bibitem[{{Kouloumvakos} {et~al.}(2019){Kouloumvakos}, {Rouillard}, {Wu},
  {Vainio}, {Vourlidas}, {Plotnikov}, {Afanasiev}, \&
  {{\"O}nel}}]{Kouloumvakos2019}
{Kouloumvakos}, A., {Rouillard}, A.~P., {Wu}, Y., {et~al.} 2019, \apj, 876, 80

\bibitem[{{Krasnoselskikh} {et~al.}(1985){Krasnoselskikh}, {Kruchina},
  {Volokitin}, \& {Thejappa}}]{Krasnoselskikh85}
{Krasnoselskikh}, V.~V., {Kruchina}, E.~N., {Volokitin}, A.~S., \& {Thejappa},
  G. 1985, \aap, 149, 323

\bibitem[{{Krucker} {et~al.}(2020){Krucker}, {Hurford}, {Grimm}, {K{\"o}gl},
  {Gr{\"o}belbauer}, {Etesi}, {Casadei}, {Csillaghy}, {Benz}, {Arnold},
  {Molendini}, {Orleanski}, {Schori}, {Xiao}, {Kuhar}, {Hochmuth}, {Felix},
  {Schramka}, {Marcin}, {Kobler}, {Iseli}, {Dreier}, {Wiehl}, {Kleint},
  {Battaglia}, {Lastufka}, {Sathiapal}, {Lapadula}, {Bednarzik}, {Birrer},
  {Stutz}, {Wild}, {Marone}, {Skup}, {Cichocki}, {Ber}, {Rutkowski}, {Bujwan},
  {Juchnikowski}, {Winkler}, {Darmetko}, {Michalska}, {Seweryn}, {Bia{\l}ek},
  {Osica}, {Sylwester}, {Kowalinski}, {{\'S}cis{\l}owski}, {Siarkowski},
  {St{\k{e}}{\'s}licki}, {Mrozek}, {Podg{\'o}rski}, {Meuris}, {Limousin},
  {Gevin}, {Le Mer}, {Brun}, {Strugarek}, {Vilmer}, {Musset}, {Maksimovi{\'c}},
  {F{\'a}rn{\'\i}k}, {Koz{\'a}{\v{c}}ek}, {Ka{\v{s}}parov{\'a}}, {Mann},
  {{\"O}nel}, {Warmuth}, {Rendtel}, {Anderson}, {Bauer}, {Dionies}, {Paschke},
  {Pl{\"u}schke}, {Woche}, {Schuller}, {Veronig}, {Dickson}, {Gallagher},
  {Maloney}, {Bloomfield}, {Piana}, {Massone}, {Benvenuto}, {Massa},
  {Schwartz}, {Dennis}, {van Beek}, {Rodr{\'\i}guez-Pacheco}, \&
  {Lin}}]{Krucker2020}
{Krucker}, S., {Hurford}, G.~J., {Grimm}, O., {et~al.} 2020, \aap, 642, A15

\bibitem[{{Krupar} {et~al.}(2014){Krupar}, {Maksimovic}, {Santolik}, {Cecconi},
  \& {Kruparova}}]{2014SoPh..289.4633K}
{Krupar}, V., {Maksimovic}, M., {Santolik}, O., {Cecconi}, B., \& {Kruparova},
  O. 2014, \solphys, 289, 4633

\bibitem[{{Kwon} \& {Vourlidas}(2017)}]{kwon2017}
{Kwon}, R.-Y. \& {Vourlidas}, A. 2017, \apj, 836, 246

\bibitem[{Laitinen {et~al.}(2013)Laitinen, Dalla, \& Marsh}]{Laitinen2013}
Laitinen, T., Dalla, S., \& Marsh, M.~S. 2013, \apjl, 773, L29

\bibitem[{Lario {et~al.}(2013)Lario, Aran, G{\'o}mez-Herrero, Dresing, Heber,
  Ho, Decker, \& Roelof}]{Lario2013}
Lario, D., Aran, A., G{\'o}mez-Herrero, R., {et~al.} 2013, \apj, 767, 41

\bibitem[{{Lario} {et~al.}(2017){Lario}, {Kwon}, {Richardson}, {Raouafi},
  {Thompson}, {von Rosenvinge}, {Mays}, {M{\"a}kel{\"a}}, {Xie}, {Bain},
  {Zhang}, {Zhao}, {Cane}, {Papaioannou}, {Thakur}, \& {Riley}}]{Lario2017}
{Lario}, D., {Kwon}, R.-Y., {Richardson}, I.~G., {et~al.} 2017, \apj, 838, 51

\bibitem[{{Lario} {et~al.}(2016){Lario}, {Kwon}, {Vourlidas}, {Raouafi},
  {Haggerty}, {Ho}, {Anderson}, {Papaioannou}, {G{\'o}mez-Herrero}, {Dresing},
  \& {Riley}}]{Lario2016}
{Lario}, D., {Kwon}, R.-Y., {Vourlidas}, A., {et~al.} 2016, \apj, 819, 72

\bibitem[{{Lario} {et~al.}(2022{\natexlab{a}}){Lario}, {Wijsen}, {Kwon},
  {S{\'a}nchez-Cano}, {Richardson}, {Pacheco}, {Palmerio}, {Stevens}, {Szabo},
  {Heyner}, {Dresing}, {G{\'o}mez-Herrero}, {Carcaboso}, {Aran}, {Afanasiev},
  {Vainio}, {Riihonen}, {Poedts}, {Br{\"u}den}, {Xu}, \&
  {Kollhoff}}]{Lario2022}
{Lario}, D., {Wijsen}, N., {Kwon}, R.~Y., {et~al.} 2022{\natexlab{a}}, \apj,
  934, 55

\bibitem[{{Lario} {et~al.}(2022{\natexlab{b}}){Lario}, {Wijsen}, {Kwon},
  {S{\'a}nchez-Cano}, {Richardson}, {Pacheco}, {Palmerio}, {Stevens}, {Szabo},
  {Heyner}, {Dresing}, {G{\'o}mez-Herrero}, {Carcaboso}, {Aran}, {Afanasiev},
  {Vainio}, {Riihonen}, {Poedts}, {Br{\"u}den}, {Xu}, \&
  {Kollhoff}}]{2022Lario}
{Lario}, D., {Wijsen}, N., {Kwon}, R.~Y., {et~al.} 2022{\natexlab{b}}, \apj,
  934, 55

\bibitem[{{Larson} {et~al.}(2015){Larson}, {Lillis}, {Lee}, {Dunn}, {Hatch},
  {Robinson}, {Glaser}, {Chen}, {Curtis}, {Tiu}, {Lin}, {Luhmann}, \&
  {Jakosky}}]{Larson2015}
{Larson}, D.~E., {Lillis}, R.~J., {Lee}, C.~O., {et~al.} 2015, \ssr, 195, 153

\bibitem[{{Ledenev} \& {Messerotti}(1999)}]{Ledenev99}
{Ledenev}, V.~G. \& {Messerotti}, M. 1999, \solphys, 185, 193

\bibitem[{{Lee} {et~al.}(2013){Lee}, {Arge}, {Odstr{\v{c}}il}, {Millward},
  {Pizzo}, {Quinn}, \& {Henney}}]{Lee2013}
{Lee}, C.~O., {Arge}, C.~N., {Odstr{\v{c}}il}, D., {et~al.} 2013, \solphys,
  285, 349

\bibitem[{{Lemen} {et~al.}(2012){Lemen}, {Title}, {Akin}, {Boerner}, {Chou},
  {Drake}, {Duncan}, {Edwards}, {Friedlaender}, {Heyman}, {Hurlburt}, {Katz},
  {Kushner}, {Levay}, {Lindgren}, {Mathur}, {McFeaters}, {Mitchell}, {Rehse},
  {Schrijver}, {Springer}, {Stern}, {Tarbell}, {Wuelser}, {Wolfson}, {Yanari},
  {Bookbinder}, {Cheimets}, {Caldwell}, {Deluca}, {Gates}, {Golub}, {Park},
  {Podgorski}, {Bush}, {Scherrer}, {Gummin}, {Smith}, {Auker}, {Jerram},
  {Pool}, {Soufli}, {Windt}, {Beardsley}, {Clapp}, {Lang}, \&
  {Waltham}}]{Lemen2012}
{Lemen}, J.~R., {Title}, A.~M., {Akin}, D.~J., {et~al.} 2012, \solphys, 275, 17

\bibitem[{{Lepping} {et~al.}(1995){Lepping}, {Ac{\~{u}}na}, {Burlaga},
  {Farrell}, {Slavin}, {Schatten}, {Mariani}, {Ness}, {Neubauer}, {Whang},
  {Byrnes}, {Kennon}, {Panetta}, {Scheifele}, \& {Worley}}]{Lepping1995}
{Lepping}, R.~P., {Ac{\~{u}}na}, M.~H., {Burlaga}, L.~F., {et~al.} 1995, \ssr,
  71, 207

\bibitem[{{Lin} {et~al.}(1995){Lin}, {Anderson}, {Ashford}, {Carlson},
  {Curtis}, {Ergun}, {Larson}, {McFadden}, {McCarthy}, {Parks}, {R{\`e}me},
  {Bosqued}, {Coutelier}, {Cotin}, {D'Uston}, {Wenzel}, {Sanderson}, {Henrion},
  {Ronnet}, \& {Paschmann}}]{Lin1995}
{Lin}, R.~P., {Anderson}, K.~A., {Ashford}, S., {et~al.} 1995, \ssr, 71, 125

\bibitem[{{Luhmann} {et~al.}(2008){Luhmann}, {Curtis}, {Schroeder}, {McCauley},
  {Lin}, {Larson}, {Bale}, {Sauvaud}, {Aoustin}, {Mewaldt}, {Cummings},
  {Stone}, {Davis}, {Cook}, {Kecman}, {Wiedenbeck}, {von Rosenvinge}, {Acuna},
  {Reichenthal}, {Shuman}, {Wortman}, {Reames}, {Mueller-Mellin}, {Kunow},
  {Mason}, {Walpole}, {Korth}, {Sanderson}, {Russell}, \&
  {Gosling}}]{Luhmann2008}
{Luhmann}, J.~G., {Curtis}, D.~W., {Schroeder}, P., {et~al.} 2008, \ssr, 136,
  117

\bibitem[{{Maksimovic} {et~al.}(2020){Maksimovic}, {Bale}, {Chust},
  {Khotyaintsev}, {Krasnoselskikh}, {Kretzschmar}, {Plettemeier}, {Rucker},
  {Sou{\v{c}}ek}, {Steller}, {{\v{S}}tver{\'a}k}, {Tr{\'a}vn{\'\i}{\v{c}}ek},
  {Vaivads}, {Chaintreuil}, {Dekkali}, {Alexandrova}, {Astier}, {Barbary},
  {B{\'e}rard}, {Bonnin}, {Boughedada}, {Cecconi}, {Chapron}, {Chariet},
  {Collin}, {de Conchy}, {Dias}, {Gu{\'e}guen}, {Lamy}, {Leray}, {Lion},
  {Malac-Allain}, {Matteini}, {Nguyen}, {Pantellini}, {Parisot}, {Plasson},
  {Thijs}, {Vecchio}, {Fratter}, {Bellouard}, {Lorf{\`e}vre}, {Danto},
  {Julien}, {Guilhem}, {Fiachetti}, {Sanisidro}, {Laffaye}, {Gonzalez},
  {Pontet}, {Qu{\'e}ruel}, {Jannet}, {Fergeau}, {Brochot}, {Cassam-Chenai},
  {Dudok de Wit}, {Timofeeva}, {Vincent}, {Agrapart}, {Delory}, {Turin},
  {Jeandet}, {Leroy}, {Pellion}, {Bouzid}, {Katra}, {Piberne}, {Recart},
  {Santol{\'\i}k}, {Kolma{\v{s}}ov{\'a}}, {Krupa{\v{r}}},
  {Krupa{\v{r}}ov{\'a}}, {P{\'\i}{\v{s}}a}, {Uhl{\'\i}{\v{r}}}, {L{\'a}n},
  {Ba{\v{s}}e}, {Ahl{\`e}n}, {Andr{\'e}}, {Bylander}, {Cripps}, {Cully},
  {Eriksson}, {Jansson}, {Johansson}, {Karlsson}, {Puccio},
  {B{\v{r}}{\'\i}nek}, {{\"O}ttacher}, {Panchenko}, {Berthomier}, {Goetz},
  {Hellinger}, {Horbury}, {Issautier}, {Kontar}, {Krucker}, {Le Contel},
  {Louarn}, {Martinovi{\'c}}, {Owen}, {Retino}, {Rodr{\'\i}guez-Pacheco},
  {Sahraoui}, {Wimmer-Schweingruber}, {Zaslavsky}, \&
  {Zouganelis}}]{Maksimovic20b}
{Maksimovic}, M., {Bale}, S.~D., {Chust}, T., {et~al.} 2020, \aap, 642, A12

\bibitem[{{Maksimovic} {et~al.}(2021){Maksimovic}, {Sou{\v{c}}ek}, {Chust},
  {Khotyaintsev}, {Kretzschmar}, {Bonnin}, {Vecchio}, {Alexandrova}, {Bale},
  {B{\'e}rard}, {Brochot}, {Edberg}, {Eriksson}, {Hadid}, {Johansson},
  {Karlsson}, {Katra}, {Krasnoselskikh}, {Krupa{\v{r}}}, {Lion},
  {Lorf{\`e}vre}, {Matteini}, {Nguyen}, {P{\'\i}{\v{s}}a}, {Piberne},
  {Plettemeier}, {Rucker}, {Santol{\'\i}k}, {Steinvall}, {Steller},
  {{\v{S}}tver{\'a}k}, {Tr{\'a}vn{\'\i}{\v{c}}ek}, {Vaivads}, {Zaslavsky},
  {Chaintreuil}, {Dekkali}, {Astier}, {Barbary}, {Boughedada}, {Cecconi},
  {Chapron}, {Collin}, {Dias}, {Gu{\'e}guen}, {Lamy}, {Leray}, {Malac-Allain},
  {Pantellini}, {Parisot}, {Plasson}, {Thijs}, {Fratter}, {Bellouard}, {Danto},
  {Julien}, {Guilhem}, {Fiachetti}, {Sanisidro}, {Laffaye}, {Gonzalez},
  {Pontet}, {Qu{\'e}ruel}, {Jannet}, {Fergeau}, {Dudok de Wit}, {Vincent},
  {Agrapart}, {Pragout}, {Bergerard-Timofeeva}, {Delory}, {Turin}, {Jeandet},
  {Leroy}, {Pellion}, {Bouzid}, {Recart}, {Kolma{\v{s}}ov{\'a}},
  {Krupa{\v{r}}ov{\'a}}, {Uhl{\'\i}{\v{r}}}, {L{\'a}n}, {Ba{\v{s}}e},
  {Andr{\'e}}, {Bylander}, {Cripps}, {Cully}, {Jansson}, {Puccio},
  {B{\v{r}}{\'\i}nek}, {Ottacher}, {Angelini}, {Berthomier}, {Evans}, {Goetz},
  {Hellinger}, {Horbury}, {Issautier}, {Kontar}, {Le Contel}, {Louarn},
  {Martinovi{\'c}}, {M{\"u}ller}, {O'Brien}, {Owen}, {Retino},
  {Rodr{\'\i}guez-Pacheco}, {Sahraoui}, {Sanchez}, {Walsh},
  {Wimmer-Schweingruber}, \& {Zouganelis}}]{Maksimovic2021}
{Maksimovic}, M., {Sou{\v{c}}ek}, J., {Chust}, T., {et~al.} 2021, \aap, 656,
  A41

\bibitem[{{Mann} \& {Classen}(1995)}]{Mann95c}
{Mann}, G. \& {Classen}, H.~T. 1995, \aap, 304, 576

\bibitem[{{Mann} \& {Klassen}(2005)}]{Mann2005}
{Mann}, G. \& {Klassen}, A. 2005, \aap, 441, 319

\bibitem[{{Mann} {et~al.}(2018){Mann}, {Melnik}, {Rucker}, {Konovalenko}, \&
  {Brazhenko}}]{Mann2018}
{Mann}, G., {Melnik}, V.~N., {Rucker}, H.~O., {Konovalenko}, A.~A., \&
  {Brazhenko}, A.~I. 2018, \aap, 609, A41

\bibitem[{{Massa} {et~al.}(2019){Massa}, {Piana}, {Massone}, \&
  {Benvenuto}}]{Massa2019}
{Massa}, P., {Piana}, M., {Massone}, A.~M., \& {Benvenuto}, F. 2019, \aap, 624,
  A130

\bibitem[{{Masson} {et~al.}(2012){Masson}, {D{\'e}moulin}, {Dasso}, \&
  {Klein}}]{2012Masson}
{Masson}, S., {D{\'e}moulin}, P., {Dasso}, S., \& {Klein}, K.~L. 2012, \aap,
  538, A32

\bibitem[{{Mays} {et~al.}(2015){Mays}, {Taktakishvili}, {Pulkkinen},
  {MacNeice}, {Rast{\"a}tter}, {Odstrcil}, {Jian}, {Richardson}, {LaSota},
  {Zheng}, \& {Kuznetsova}}]{Mays2015}
{Mays}, M.~L., {Taktakishvili}, A., {Pulkkinen}, A., {et~al.} 2015, \solphys,
  290, 1775

\bibitem[{{McComas} {et~al.}(2016){McComas}, {Alexander}, {Angold}, {Bale},
  {Beebe}, {Birdwell}, {Boyle}, {Burgum}, {Burnham}, {Christian}, {Cook},
  {Cooper}, {Cummings}, {Davis}, {Desai}, {Dickinson}, {Dirks}, {Do}, {Fox},
  {Giacalone}, {Gold}, {Gurnee}, {Hayes}, {Hill}, {Kasper}, {Kecman}, {Klemic},
  {Krimigis}, {Labrador}, {Layman}, {Leske}, {Livi}, {Matthaeus}, {McNutt},
  {Mewaldt}, {Mitchell}, {Nelson}, {Parker}, {Rankin}, {Roelof}, {Schwadron},
  {Seifert}, {Shuman}, {Stokes}, {Stone}, {Vandegriff}, {Velli}, {von
  Rosenvinge}, {Weidner}, {Wiedenbeck}, \& {Wilson}}]{McComas2016}
{McComas}, D.~J., {Alexander}, N., {Angold}, N., {et~al.} 2016, \ssr, 204, 187

\bibitem[{{McGregor} {et~al.}(2008){McGregor}, {Hughes}, {Arge}, \&
  {Owens}}]{McGregor2008}
{McGregor}, S.~L., {Hughes}, W.~J., {Arge}, C.~N., \& {Owens}, M.~J. 2008, J.
  Geophys. Res. Space Phys., 113, A08112

\bibitem[{{Melrose}(1985)}]{Melrose85}
{Melrose}, D.~B. 1985, in Solar Radiophysics: Studies of Emission from the Sun
  at Metre Wavelengths, ed. D.~J. {McLean} \& N.~R. {Labrum} (Cambridge
  University Press), 177--210

\bibitem[{{Melrose} {et~al.}(1978){Melrose}, {Dulk}, \& {Smerd}}]{Melrose78}
{Melrose}, D.~B., {Dulk}, G.~A., \& {Smerd}, S.~F. 1978, \aap, 66, 315

\bibitem[{{Mewaldt} {et~al.}(2008){Mewaldt}, {Cohen}, {Cook}, {Cummings},
  {Davis}, {Geier}, {Kecman}, {Klemic}, {Labrador}, {Leske}, {Miyasaka},
  {Nguyen}, {Ogliore}, {Stone}, {Radocinski}, {Wiedenbeck}, {Hawk}, {Shuman},
  {von Rosenvinge}, \& {Wortman}}]{Mewaldt2008}
{Mewaldt}, R.~A., {Cohen}, C.~M.~S., {Cook}, W.~R., {et~al.} 2008, \ssr, 136,
  285

\bibitem[{{Meyer-Vernet} {et~al.}(2017){Meyer-Vernet}, {Issautier}, \&
  {Moncuquet}}]{Meyer-Vernet2017}
{Meyer-Vernet}, N., {Issautier}, K., \& {Moncuquet}, M. 2017, J. Geophys. Res.
  Space Phys., 122, 7925

\bibitem[{{M{\"u}ller} {et~al.}(2020){M{\"u}ller}, {St. Cyr}, {Zouganelis},
  {Gilbert}, {Marsden}, {Nieves-Chinchilla}, {Antonucci}, {Auch{\`e}re},
  {Berghmans}, {Horbury}, {Howard}, {Krucker}, {Maksimovic}, {Owen}, {Rochus},
  {Rodriguez-Pacheco}, {Romoli}, {Solanki}, {Bruno}, {Carlsson}, {Fludra},
  {Harra}, {Hassler}, {Livi}, {Louarn}, {Peter}, {Sch{\"u}hle}, {Teriaca}, {del
  Toro Iniesta}, {Wimmer-Schweingruber}, {Marsch}, {Velli}, {De Groof},
  {Walsh}, \& {Williams}}]{Muller2020}
{M{\"u}ller}, D., {St. Cyr}, O.~C., {Zouganelis}, I., {et~al.} 2020, \aap, 642,
  A1

\bibitem[{{M{\"u}ller-Mellin} {et~al.}(2008){M{\"u}ller-Mellin},
  {B{\"o}ttcher}, {Falenski}, {Rode}, {Duvet}, {Sanderson}, {Butler},
  {Johlander}, \& {Smit}}]{Muller-Mellin2008}
{M{\"u}ller-Mellin}, R., {B{\"o}ttcher}, S., {Falenski}, J., {et~al.} 2008,
  \ssr, 136, 363

\bibitem[{M\"{u}ller-Mellin {et~al.}(1995)M\"{u}ller-Mellin, Kunow, Fleissner,
  Pehlke, Rode, R\"{o}schmann, Scharmberg, Sierks, Rusznyak, Mckenna-Lawlor,
  Elendt, Sequeiros, Meziat, Sanchez, Medina, Peral, Witte, Marsden, \&
  Henrion}]{Muller-Mellin1995}
M\"{u}ller-Mellin, R., Kunow, H., Fleissner, V., {et~al.} 1995, \solphys, 162,
  483

\bibitem[{{Murakami} {et~al.}(2020){Murakami}, {Hayakawa}, {Ogawa}, {Matsuda},
  {Seki}, {Kasaba}, {Saito}, {Yoshikawa}, {Kobayashi}, {Baumjohann},
  {Matsuoka}, {Kojima}, {Yagitani}, {Moncuquet}, {Wahlund}, {Delcourt},
  {Hirahara}, {Barabash}, {Korablev}, \& {Fujimoto}}]{Murakami2020}
{Murakami}, G., {Hayakawa}, H., {Ogawa}, H., {et~al.} 2020, \ssr, 216, 113

\bibitem[{{Odstrcil} {et~al.}(2004){Odstrcil}, {Riley}, \&
  {Zhao}}]{Odstrcil2004}
{Odstrcil}, D., {Riley}, P., \& {Zhao}, X.~P. 2004, J. Geophys. Res. Space
  Phys., 109, A02116

\bibitem[{{Ogilvie} {et~al.}(1995){Ogilvie}, {Chornay}, {Fritzenreiter},
  {Hunsaker}, {Keller}, {Lobell}, {Miller}, {Scudder}, {Sittler}, {Torbert},
  {Bodet}, {Needell}, {Lazarus}, {Steinberg}, {Tappan}, {Mavretic}, \&
  {Gergin}}]{Ogilvie1995}
{Ogilvie}, K.~W., {Chornay}, D.~J., {Fritzenreiter}, R.~J., {et~al.} 1995,
  \ssr, 71, 55

\bibitem[{{Ogilvie} \& {Desch}(1997)}]{Ogilvie1997}
{Ogilvie}, K.~W. \& {Desch}, M.~D. 1997, Advances in Space Research, 20, 559

\bibitem[{{Ontiveros} \& {Vourlidas}(2009)}]{ontiveros2009}
{Ontiveros}, V. \& {Vourlidas}, A. 2009, \apj, 693, 267

\bibitem[{{Owen} {et~al.}(2020){Owen}, {Bruno}, {Livi}, {Louarn}, {Al Janabi},
  {Allegrini}, {Amoros}, {Baruah}, {Barthe}, {Berthomier}, {Bordon},
  {Brockley-Blatt}, {Brysbaert}, {Capuano}, {Collier}, {DeMarco}, {Fedorov},
  {Ford}, {Fortunato}, {Fratter}, {Galvin}, {Hancock}, {Heirtzler}, {Kataria},
  {Kistler}, {Lepri}, {Lewis}, {Loeffler}, {Marty}, {Mathon}, {Mayall}, {Mele},
  {Ogasawara}, {Orlandi}, {Pacros}, {Penou}, {Persyn}, {Petiot}, {Phillips},
  {P{\v{r}}ech}, {Raines}, {Reden}, {Rouillard}, {Rousseau}, {Rubiella},
  {Seran}, {Spencer}, {Thomas}, {Trevino}, {Verscharen}, {Wurz}, {Alapide},
  {Amoruso}, {Andr{\'e}}, {Anekallu}, {Arciuli}, {Arnett}, {Ascolese},
  {Bancroft}, {Bland}, {Brysch}, {Calvanese}, {Castronuovo},
  {{\v{C}}erm{\'a}k}, {Chornay}, {Clemens}, {Coker}, {Collinson}, {D'Amicis},
  {Dandouras}, {Darnley}, {Davies}, {Davison}, {De Los Santos}, {Devoto},
  {Dirks}, {Edlund}, {Fazakerley}, {Ferris}, {Frost}, {Fruit}, {Garat},
  {G{\'e}not}, {Gibson}, {Gilbert}, {de Giosa}, {Gradone}, {Hailey}, {Horbury},
  {Hunt}, {Jacquey}, {Johnson}, {Lavraud}, {Lawrenson}, {Leblanc}, {Lockhart},
  {Maksimovic}, {Malpus}, {Marcucci}, {Mazelle}, {Monti}, {Myers}, {Nguyen},
  {Rodriguez-Pacheco}, {Phillips}, {Popecki}, {Rees}, {Rogacki}, {Ruane},
  {Rust}, {Salatti}, {Sauvaud}, {Stakhiv}, {Stange}, {Stubbs}, {Taylor},
  {Techer}, {Terrier}, {Thibodeaux}, {Urdiales}, {Varsani}, {Walsh}, {Watson},
  {Wheeler}, {Willis}, {Wimmer-Schweingruber}, {Winter}, {Yardley}, \&
  {Zouganelis}}]{Owen2020}
{Owen}, C.~J., {Bruno}, R., {Livi}, S., {et~al.} 2020, \aap, 642, A16

\bibitem[{{Palmerio} {et~al.}(2021){Palmerio}, {Kilpua}, {Witasse}, {Barnes},
  {S{\'a}nchez-Cano}, {Weiss}, {Nieves-Chinchilla}, {M{\"o}stl}, {Jian},
  {Mierla}, {Zhukov}, {Guo}, {Rodriguez}, {Lowrance}, {Isavnin}, {Turc},
  {Futaana}, \& {Holmstr{\"o}m}}]{2021Palmerio}
{Palmerio}, E., {Kilpua}, E. K.~J., {Witasse}, O., {et~al.} 2021, Space
  Weather, 19, e2020SW002654

\bibitem[{{Palmerio} {et~al.}(2022){Palmerio}, {Lee}, {Mays}, {Luhmann},
  {Lario}, {S{\'a}nchez-Cano}, {Richardson}, {Vainio}, {Stevens}, {Cohen},
  {Steinvall}, {M{\"o}stl}, {Weiss}, {Nieves-Chinchilla}, {Li}, {Larson},
  {Heyner}, {Bale}, {Galvin}, {Holmstr{\"o}m}, {Khotyaintsev}, {Maksimovic}, \&
  {Mitrofanov}}]{Palmerio2022}
{Palmerio}, E., {Lee}, C.~O., {Mays}, M.~L., {et~al.} 2022, Space Weather, 20,
  e2021SW002993

\bibitem[{{Paschmann} \& {Schwartz}(2000)}]{Paschmann2000}
{Paschmann}, G. \& {Schwartz}, S.~J. 2000, in ESA Special Publication, Vol.
  449, Cluster-II Workshop Multiscale / Multipoint Plasma Measurements, ed.
  R.~A. {Harris}, 99

\bibitem[{{Patsourakos} \& {Vourlidas}(2009)}]{Patsourakos2009}
{Patsourakos}, S. \& {Vourlidas}, A. 2009, \apjl, 700, L182

\bibitem[{{Pesnell} {et~al.}(2012){Pesnell}, {Thompson}, \&
  {Chamberlin}}]{Pesnell2012}
{Pesnell}, W.~D., {Thompson}, B.~J., \& {Chamberlin}, P.~C. 2012, \solphys,
  275, 3

\bibitem[{{Piantschitsch} {et~al.}(2018){Piantschitsch}, {Vr{\v{s}}nak},
  {Hanslmeier}, {Lemmerer}, {Veronig}, {Hernandez-Perez}, \&
  {{\v{C}}alogovi{\'c}}}]{Piantschitsch2018}
{Piantschitsch}, I., {Vr{\v{s}}nak}, B., {Hanslmeier}, A., {et~al.} 2018, \apj,
  857, 130

\bibitem[{{Pinto} {et~al.}(2022){Pinto}, {Sanchez-Cano}, {Moissl}, {Benkhoff},
  {Cardoso}, {Gon{\c{c}}alves}, {Assis}, {Vainio}, {Oleynik}, {Lehtolainen},
  {Grande}, \& {Marques}}]{Pinto2021}
{Pinto}, M., {Sanchez-Cano}, B., {Moissl}, R., {et~al.} 2022, \ssr, 218, 54

\bibitem[{{Podladchikova} \& {Berghmans}(2005)}]{Podladchikova2005}
{Podladchikova}, O. \& {Berghmans}, D. 2005, \solphys, 228, 265

\bibitem[{{Podladchikova} {et~al.}(2019){Podladchikova}, {Veronig}, {Dissauer},
  {Temmer}, \& {Podladchikova}}]{Podladchikova2019}
{Podladchikova}, T., {Veronig}, A.~M., {Dissauer}, K., {Temmer}, M., \&
  {Podladchikova}, O. 2019, \apj, 877, 68

\bibitem[{{Pulupa} {et~al.}(2020){Pulupa}, {Bale}, {Badman}, {Bonnell}, {Case},
  {de Wit}, {Goetz}, {Harvey}, {Hegedus}, {Kasper}, {Korreck},
  {Krasnoselskikh}, {Larson}, {Lecacheux}, {Livi}, {MacDowall}, {Maksimovic},
  {Malaspina}, {Mart{\'\i}nez Oliveros}, {Meyer-Vernet}, {Moncuquet},
  {Stevens}, \& {Whittlesey}}]{Pulupa20}
{Pulupa}, M., {Bale}, S.~D., {Badman}, S.~T., {et~al.} 2020, \apjs, 246, 49

\bibitem[{{Pulupa} {et~al.}(2017){Pulupa}, {Bale}, {Bonnell}, {Bowen},
  {Carruth}, {Goetz}, {Gordon}, {Harvey}, {Maksimovic},
  {Mart{\'\i}nez-Oliveros}, {Moncuquet}, {Saint-Hilaire}, {Seitz}, \&
  {Sundkvist}}]{Pulupa2017}
{Pulupa}, M., {Bale}, S.~D., {Bonnell}, J.~W., {et~al.} 2017, J. Geophys. Res.
  Space Phys., 122, 2836

\bibitem[{{Reames}(2021)}]{Reames2021}
{Reames}, D.~V. 2021, {Solar Energetic Particles. A Modern Primer on
  Understanding Sources, Acceleration and Propagation}, Vol. 978 (Springer
  Cham)

\bibitem[{{Reiner} {et~al.}(2007){Reiner}, {Fainberg}, {Kaiser}, \&
  {Bougeret}}]{Reiner07}
{Reiner}, M.~J., {Fainberg}, J., {Kaiser}, M.~L., \& {Bougeret}, J.~L. 2007,
  \solphys, 241, 351

\bibitem[{{Richardson} {et~al.}(2014){Richardson}, {von Rosenvinge}, {Cane},
  {Christian}, {Cohen}, {Labrador}, {Leske}, {Mewaldt}, {Wiedenbeck}, \&
  {Stone}}]{Richardson2014}
{Richardson}, I.~G., {von Rosenvinge}, T.~T., {Cane}, H.~V., {et~al.} 2014,
  \solphys, 289, 3059

\bibitem[{{Rodr{\'\i}guez-Garc{\'\i}a}
  {et~al.}(2021){Rodr{\'\i}guez-Garc{\'\i}a}, {G{\'o}mez-Herrero},
  {Zouganelis}, {Balmaceda}, {Nieves-Chinchilla}, {Dresing}, {Dumbovi{\'c}},
  {Nitta}, {Carcaboso}, {dos Santos}, {Jian}, {Mays}, {Williams}, \&
  {Rodr{\'\i}guez-Pacheco}}]{2021Rodriguez-Garcia}
{Rodr{\'\i}guez-Garc{\'\i}a}, L., {G{\'o}mez-Herrero}, R., {Zouganelis}, I.,
  {et~al.} 2021, \aap, 653, A137

\bibitem[{{Rodr\'{\i}guez-Garc\'{\i}a}
  {et~al.}(2022){Rodr\'{\i}guez-Garc\'{\i}a}, {Nieves-Chinchilla, T.},
  {G\'omez-Herrero, R.}, {Zouganelis, I.}, {Vourlidas, A.}, {Balmaceda, L. A.},
  {Dumbovi\'{}c, M.}, {Jian, L. K.}, {Mays, L.}, {Carcaboso, F.}, {dos Santos,
  L. F. G.}, \& {Rodr\'{\i}guez-Pacheco, J.}}]{Rodriguez-Garcia2022CME}
{Rodr\'{\i}guez-Garc\'{\i}a}, L., {Nieves-Chinchilla, T.}, {G\'omez-Herrero,
  R.}, {et~al.} 2022, A\&A, 662, A45

\bibitem[{{Rodr{\'\i}guez-Pacheco} {et~al.}(2020){Rodr{\'\i}guez-Pacheco},
  {Wimmer-Schweingruber}, {Mason}, {Ho}, {S{\'a}nchez-Prieto}, {Prieto},
  {Mart{\'\i}n}, {Seifert}, {Andrews}, {Kulkarni}, {Panitzsch}, {Boden},
  {B{\"o}ttcher}, {Cernuda}, {Elftmann}, {Espinosa Lara}, {G{\'o}mez-Herrero},
  {Terasa}, {Almena}, {Begley}, {B{\"o}hm}, {Blanco}, {Boogaerts}, {Carrasco},
  {Castillo}, {da Silva Fari{\~n}a}, {de Manuel Gonz{\'a}lez}, {Drews},
  {Dupont}, {Eldrum}, {Gordillo}, {Guti{\'e}rrez}, {Haggerty}, {Hayes},
  {Heber}, {Hill}, {J{\"u}ngling}, {Kerem}, {Knierim}, {K{\"o}hler}, {Kolbe},
  {Kulemzin}, {Lario}, {Lees}, {Liang}, {Mart{\'\i}nez Hell{\'\i}n}, {Meziat},
  {Montalvo}, {Nelson}, {Parra}, {Paspirgilis}, {Ravanbakhsh}, {Richards},
  {Rodr{\'\i}guez-Polo}, {Russu}, {S{\'a}nchez}, {Schlemm}, {Schuster},
  {Seimetz}, {Steinhagen}, {Tammen}, {Tyagi}, {Varela}, {Yedla}, {Yu},
  {Agueda}, {Aran}, {Horbury}, {Klecker}, {Klein}, {Kontar}, {Krucker},
  {Maksimovic}, {Malandraki}, {Owen}, {Pacheco}, {Sanahuja}, {Vainio},
  {Connell}, {Dalla}, {Dr{\"o}ge}, {Gevin}, {Gopalswamy}, {Kartavykh},
  {Kudela}, {Limousin}, {Makela}, {Mann}, {{\"O}nel}, {Posner}, {Ryan},
  {Soucek}, {Hofmeister}, {Vilmer}, {Walsh}, {Wang}, {Wiedenbeck}, {Wirth}, \&
  {Zong}}]{Rodriguez-Pacheco2020}
{Rodr{\'\i}guez-Pacheco}, J., {Wimmer-Schweingruber}, R.~F., {Mason}, G.~M.,
  {et~al.} 2020, \aap, 642, A7

\bibitem[{Rouillard {et~al.}(2012)Rouillard, Sheeley, Tylka, Vourlidas, Ng,
  Rakowski, Cohen, Mewaldt, Mason, Reames, Savani, StCyr, \&
  Szabo}]{Rouillard2012}
Rouillard, A.~P., Sheeley, N.~R., Tylka, A., {et~al.} 2012, \apj, 752, 44

\bibitem[{{Schatten} {et~al.}(1969){Schatten}, {Wilcox}, \&
  {Ness}}]{Schatten1969}
{Schatten}, K.~H., {Wilcox}, J.~M., \& {Ness}, N.~F. 1969, \solphys, 6, 442

\bibitem[{{Schmidt} \& {Cairns}(2014)}]{Schmidt14}
{Schmidt}, J.~M. \& {Cairns}, I.~H. 2014, J. Geophys. Res. Space Phys., 119, 69

\bibitem[{{Strauss} {et~al.}(2020){Strauss}, {Dresing}, {Kollhoff}, \&
  {Br{\"u}dern}}]{Strauss2020}
{Strauss}, R.~D., {Dresing}, N., {Kollhoff}, A., \& {Br{\"u}dern}, M. 2020,
  \apj, 897, 24

\bibitem[{{Strauss} \& {Fichtner}(2015)}]{Strauss2015}
{Strauss}, R.~D. \& {Fichtner}, H. 2015, \apj, 801, 29

\bibitem[{{Strauss} {et~al.}(2017){Strauss}, {Dresing}, \&
  {Engelbrecht}}]{Strauss2017}
{Strauss}, R.~D.~T., {Dresing}, N., \& {Engelbrecht}, N.~E. 2017, \apj, 837, 43

\bibitem[{{Thejappa} {et~al.}(2012){Thejappa}, {MacDowall}, \&
  {Bergamo}}]{Thejappa12}
{Thejappa}, G., {MacDowall}, R.~J., \& {Bergamo}, M. 2012, \apj, 745, 187

\bibitem[{{Thernisien} {et~al.}(2009){Thernisien}, {Vourlidas}, \&
  {Howard}}]{Thernisien2009}
{Thernisien}, A., {Vourlidas}, A., \& {Howard}, R.~A. 2009, \solphys, 256, 111

\bibitem[{{Thernisien} {et~al.}(2006){Thernisien}, {Howard}, \&
  {Vourlidas}}]{Thernisien2006GCS}
{Thernisien}, A.~F.~R., {Howard}, R.~A., \& {Vourlidas}, A. 2006, \apj, 652,
  763

\bibitem[{{Tkachenko} {et~al.}(2021){Tkachenko}, {Krasnoselskikh}, \&
  {Voshchepynets}}]{Tkachenko21}
{Tkachenko}, A., {Krasnoselskikh}, V., \& {Voshchepynets}, A. 2021, \apj, 908,
  126

\bibitem[{{Torsti} {et~al.}(1995){Torsti}, {Valtonen}, {Lumme}, {Peltonen},
  {Eronen}, {Louhola}, {Riihonen}, {Schultz}, {Teittinen}, {Ahola}, {Holmlund},
  {Kelh{\"a}}, {Lepp{\"a}l{\"a}}, {Ruuska}, \& {Str{\"o}mmer}}]{Torsti1995}
{Torsti}, J., {Valtonen}, E., {Lumme}, M., {et~al.} 1995, \solphys, 162, 505

\bibitem[{Trotta {et~al.}(2022)Trotta, Vuorinen, Hietala, Horbury, Dresing,
  Gieseler, Kouloumvakos, Price, Valentini, Kilpua, \& Vainio}]{Trotta2022}
Trotta, D., Vuorinen, L., Hietala, H., {et~al.} 2022, Front. Astron. Space
  Sci., 9, 1005672

\bibitem[{{Vecchio} {et~al.}(2021){Vecchio}, {Maksimovic}, {Krupar}, {Bonnin},
  {Zaslavsky}, {Astier}, {Dekkali}, {Cecconi}, {Bale}, {Chust}, {Guilhem},
  {Khotyaintsev}, {Krasnoselskikh}, {Kretzschmar}, {Lorf{\`e}vre},
  {Plettemeier}, {Sou{\v{c}}ek}, {Steller}, {{\v{S}}tver{\'a}k},
  {Tr{\'a}vn{\'\i}{\v{c}}ek}, \& {Vaivads}}]{Vecchio2021}
{Vecchio}, A., {Maksimovic}, M., {Krupar}, V., {et~al.} 2021, \aap, 656, A33

\bibitem[{Verbeke {et~al.}(2022)Verbeke, {Mays}, Kay, Riley, Palmerio,
  Dumbović, Mierla, Scolini, Temmer, Paouris, Balmaceda, Cremades, \&
  Hinterreiter}]{Verbeke2022}
Verbeke, C., {Mays}, M.~L., Kay, C., {et~al.} 2022, Adv. Space Res., in press

\bibitem[{{Veronig} {et~al.}(2002){Veronig}, {Temmer}, {Hanslmeier}, {Otruba},
  \& {Messerotti}}]{Veronig2002}
{Veronig}, A., {Temmer}, M., {Hanslmeier}, A., {Otruba}, W., \& {Messerotti},
  M. 2002, \aap, 382, 1070

\bibitem[{{Veronig} {et~al.}(2018){Veronig}, {Podladchikova}, {Dissauer},
  {Temmer}, {Seaton}, {Long}, {Guo}, {Vr{\v{s}}nak}, {Harra}, \&
  {Kliem}}]{Veronig2018}
{Veronig}, A.~M., {Podladchikova}, T., {Dissauer}, K., {et~al.} 2018, \apj,
  868, 107

\bibitem[{Veronig {et~al.}(2008)Veronig, Temmer, \& Vr\v{s}nak}]{Veronig2008}
Veronig, A.~M., Temmer, M., \& Vr\v{s}nak, B. 2008, \apj, 681, L113

\bibitem[{{von Rosenvinge} {et~al.}(2008){von Rosenvinge}, {Reames}, {Baker},
  {Hawk}, {Nolan}, {Ryan}, {Shuman}, {Wortman}, {Mewaldt}, {Cummings}, {Cook},
  {Labrador}, {Leske}, \& {Wiedenbeck}}]{vonRosenvinge2008}
{von Rosenvinge}, T.~T., {Reames}, D.~V., {Baker}, R., {et~al.} 2008, \ssr,
  136, 391

\bibitem[{{Vourlidas} {et~al.}(2013){Vourlidas}, {Lynch}, {Howard}, \&
  {Li}}]{Vourlidas2013}
{Vourlidas}, A., {Lynch}, B.~J., {Howard}, R.~A., \& {Li}, Y. 2013, \solphys,
  284, 179

\bibitem[{{Wallace} {et~al.}(2022){Wallace}, {Jones}, {Arge}, {Viall}, \&
  {Henney}}]{Wallace2022}
{Wallace}, S., {Jones}, S.~I., {Arge}, C.~N., {Viall}, N.~M., \& {Henney},
  C.~J. 2022, \apj, 935, 24

\bibitem[{{Wang} \& {Sheeley}(1992)}]{WangSheeley1992}
{Wang}, Y.-M. \& {Sheeley}, Jr., N.~R. 1992, \apj, 392, 310

\bibitem[{{Warmuth}(2015)}]{Warmuth2015}
{Warmuth}, A. 2015, Living Reviews in Solar Physics, 12, 3

\bibitem[{{Warmuth} {et~al.}(2020){Warmuth}, {{\"O}nel}, {Mann}, {Rendtel},
  {Strassmeier}, {Denker}, {Hurford}, {Krucker}, {Anderson}, {Bauer},
  {Bittner}, {Dionies}, {Paschke}, {Pl{\"u}schke}, {Sablowski}, {Schuller},
  {Senthamizh Pavai}, {Woche}, {Casadei}, {K{\"o}gl}, {Arnold},
  {Gr{\"o}belbauer}, {Schori}, {Wiehl}, {Csillaghy}, {Grimm}, {Orleanski},
  {Skup}, {Bujwan}, {Rutkowski}, \& {Ber}}]{Warmuth2020}
{Warmuth}, A., {{\"O}nel}, H., {Mann}, G., {et~al.} 2020, \solphys, 295, 90

\bibitem[{{Wentzel}(1984)}]{Wentzel94}
{Wentzel}, D.~G. 1984, \solphys, 90, 139

\bibitem[{{Wiedenbeck} {et~al.}(2017){Wiedenbeck}, {Angold}, {Birdwell},
  {Burnham}, {Christian}, {Cohen}, {Cook}, {Cummings}, {Davis}, {Dirks}, {Do},
  {Everett}, {Goodwin}, {Hanley}, {Hernandez}, {Kecman}, {Klemic}, {Labrador},
  {Leske}, {Lopez}, {Link}, {McComas}, {Mewaldt}, {Miyasaka}, {Nahory},
  {Rankin}, {Riggans}, {Rodriguez}, {Rusert}, {Shuman}, {Simms}, {Stone}, {von
  Rosenvinge}, {Weidner}, \& {White}}]{Wiedenbeck2017}
{Wiedenbeck}, M.~E., {Angold}, N.~G., {Birdwell}, B., {et~al.} 2017, in
  International Cosmic Ray Conference, Vol. 301, 35th International Cosmic Ray
  Conference (ICRC2017), 16

\bibitem[{{Wold} {et~al.}(2018){Wold}, {Mays}, {Taktakishvili}, {Jian},
  {Odstrcil}, \& {MacNeice}}]{Wold2018}
{Wold}, A.~M., {Mays}, M.~L., {Taktakishvili}, A.~r., {et~al.} 2018, Journal of
  Space Weather and Space Climate, 8, A17

\bibitem[{{Worden} \& {Harvey}(2000)}]{WordenHarvey2000}
{Worden}, J. \& {Harvey}, J. 2000, \solphys, 195, 247

\bibitem[{{Wuelser} {et~al.}(2004){Wuelser}, {Lemen}, {Tarbell}, {Wolfson},
  {Cannon}, {Carpenter}, {Duncan}, {Gradwohl}, {Meyer}, {Moore}, {Navarro},
  {Pearson}, {Rossi}, {Springer}, {Howard}, {Moses}, {Newmark},
  {Delaboudiniere}, {Artzner}, {Auchere}, {Bougnet}, {Bouyries}, {Bridou},
  {Clotaire}, {Colas}, {Delmotte}, {Jerome}, {Lamare}, {Mercier}, {Mullot},
  {Ravet}, {Song}, {Bothmer}, \& {Deutsch}}]{Wuelser2004}
{Wuelser}, J.-P., {Lemen}, J.~R., {Tarbell}, T.~D., {et~al.} 2004, in Society
  of Photo-Optical Instrumentation Engineers (SPIE) Conference Series, Vol.
  5171, Telescopes and Instrumentation for Solar Astrophysics, ed.
  S.~{Fineschi} \& M.~A. {Gummin}, 111--122

\bibitem[{{Zheleznyakov}(1965)}]{Zheleznyakov65}
{Zheleznyakov}, V.~V. 1965, \sovast, 9, 191

\bibitem[{{Zheleznyakov} \& {Zaitsev}(1970)}]{Zheleznyakov70a}
{Zheleznyakov}, V.~V. \& {Zaitsev}, V.~V. 1970, \sovast, 14, 47

\bibitem[{{Zheleznyakov} \& {Zlotnik}(1964)}]{Zheleznyakov64}
{Zheleznyakov}, V.~V. \& {Zlotnik}, E.~Y. 1964, \sovast, 7, 485

\bibitem[{{Zurbuchen} \& {Richardson}(2006)}]{Zurbuchen2006}
{Zurbuchen}, T.~H. \& {Richardson}, I.~G. 2006, \ssr, 123, 31

\end{thebibliography}

\begin{appendix}
\section{Spectral analysis of the solar flare}\label{flare_ap}

\begin{figure}[h!]  
   \centering
	\includegraphics[width=0.5\textwidth]{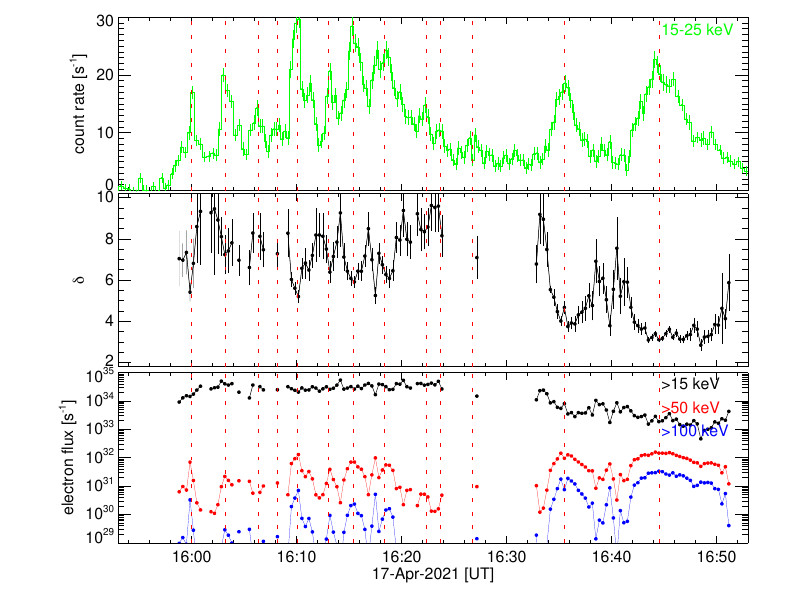}
	\caption{STIX fit results for the non-thermal electron component. From the top, the figure shows the background-subtracted STIX count rate in the 15-25~keV range, the spectral index of the injected electrons, $\delta$, the low-energy cutoff, $E_\mathrm{LC}$, and the injected electron fluxes above 15, 50 and 100~keV. The 13 non-thermal peaks are indicated by red dashed lines. Times refer to UT at the Sun.} 
	\label{fig_stix_spec}
\end{figure}

A quantitative characterization of the accelerated electrons is carried out by forward-fitting the combination of an isothermal and a non-thermal thick-target component to a time series of observed STIX count spectra using the OSPEX spectral analysis package\footnote{http://hesperia.gsfc.nasa.gov/ssw/packages/spex/doc/}. The non-thermal fit results are plotted in Fig.~\ref{fig_stix_spec}. From the top, the figure shows the background-subtracted STIX count rate in the energy range of 15-25~keV, the power-law index of the injected electron flux, $\delta$, and the electron flux. While the fit results for the low-energy cutoff were lying around 15~keV for most of the flare, the fit uncertainties of this parameter became very large during the last two peaks. We therefore adopt 15~keV as the effective cutoff for the whole event. It is important to point out that this is only an upper threshold, since the true cutoff energy could well be lower since it is masked by the thermal emission. Consequently, the total electron flux (bottom  panel) represents a lower estimate. For reference, we also plot here the fluxes above 50~keV and 100~keV, respectively. This assumes that the fitted power-law extends unbroken to high energies.

In Fig.~\ref{fig_stix_spec}, the spikes in the non-thermal count rate are indicated by red dashed lines. Note that the count rate in each spike is anticorrelated with the spectral index, consistent with the well-established soft-hard-soft spectral evolution \citep[e.g.][]{Grigis2004}. This is also reflected in the electron flux. The spectrum is rather soft during most spikes ($\delta=5.5$-9.5), but hardens significantly during the two major late peaks (with $\delta=3$ as a minimum). It is evident that during the two major late peaks more electrons have been accelerated to over 50~keV than in the main impulsive phase.

\section{Radio polarization and directivity analysis} \label{radio_ap}

Thanks to the fleet of spacecraft located at different vantage points, the evolution of the flare-CME event in the radio wavelengths provides a unique opportunity to understand its evolution. We use different capabilities of instrumentation, such as polarization of the radio waves, and cross-correlated flux measurements to locate and to understand the characteristics of the radio source.

\subsection{Polarization}

\begin{figure*}
    \centering
    \includegraphics[width=0.58\textwidth]{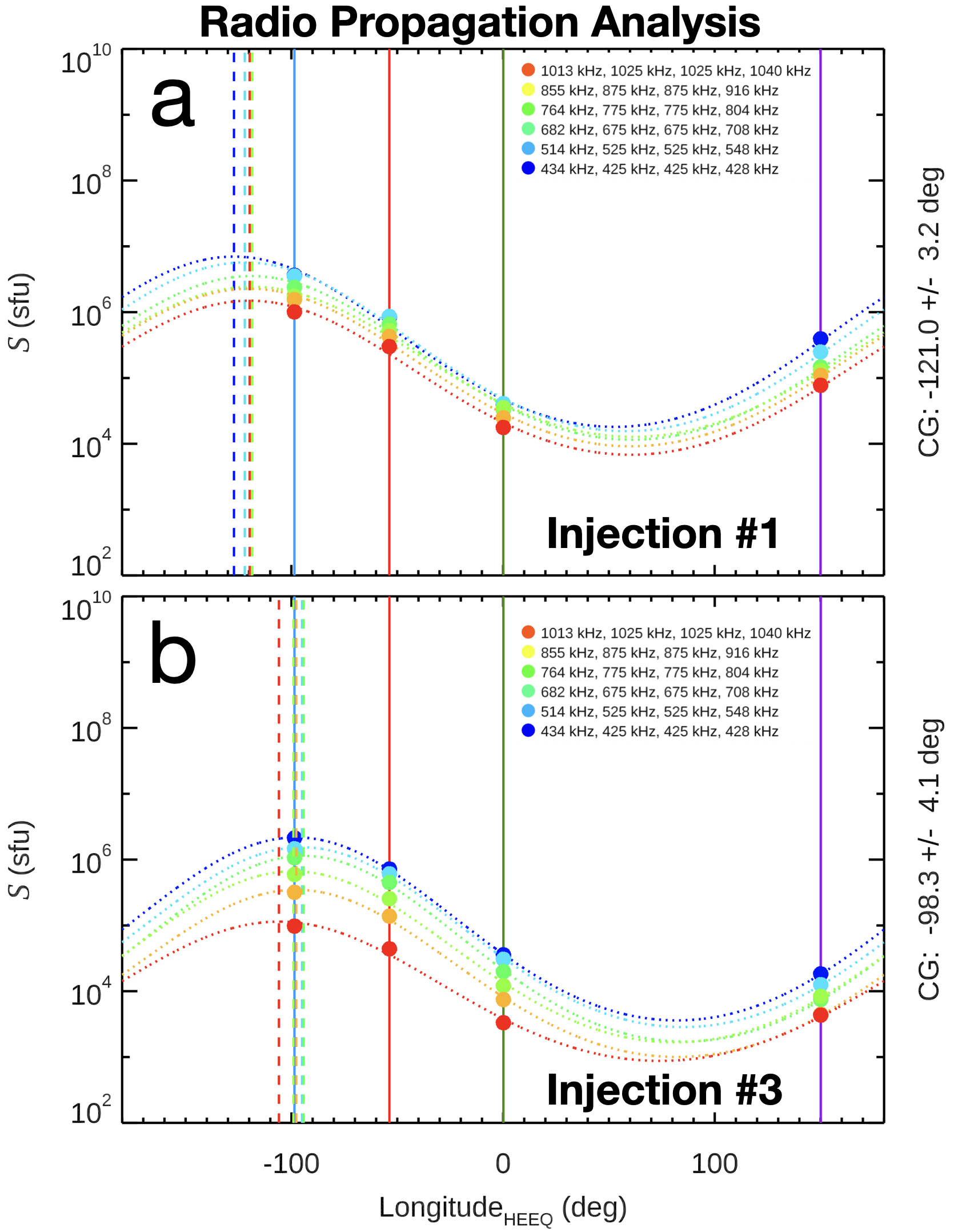}
    \includegraphics[width=0.41\textwidth]{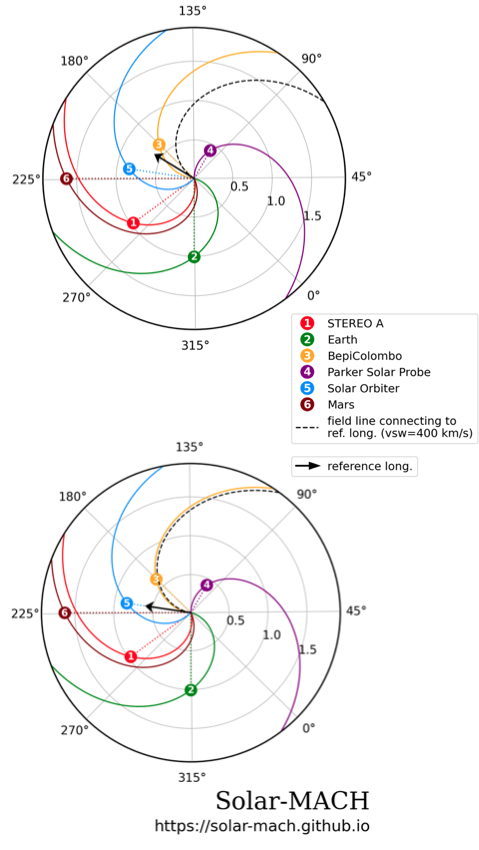}
      \caption{Radio propagation analysis of the TIII(1) and TIII (3). Left: Calibrated peak radio fluxes as a function of spacecraft locations $\lambda$ for six frequency channels denoted by the colored circles. Dotted lines are results of the equation (\ref{eq:lambda}) fitting. Vertical dashed lines indicate modelled direction of the peak fluxes. Right: Longitudinal spacecraft constellation plots like in Fig.~\ref{fig:solar-mach_and_multi_sc_SEP} (left) but with reference longitudes corresponding to the results of the radio propagation analysis for TIII(1) (top) and TIII(3) (bottom).  }
         \label{fig:radio_directions}
\end{figure*}
We calculate the net polarization (Stokes V/I) measurements of the radio waves by Parker Solar Probe explained in detail by \cite{Pulupa20}. Figure~\ref{fig:radio_multi_sc}, second panel from top, shows a relatively high degree of circular polarization for TIII(2) and TIII(4). A high degree of circular polarization generally suggests emission by a source that is propagating along magnetic field lines of unipolarity \citep[][]{Melrose78, Reiner07}. Additionally, it may also indicate higher beam velocities or efficient wave conversion \citep[electrostatic to electromagnetic; ][]{Wentzel94}. In the case of TIII(2) and TIII(4), this would mean that the emission was mostly directed in the viewing direction of Parker Solar Probe at the short-hectometer wavelengths. The sense of polarization of the bursts is left-handed, indicating that the electric field vector rotates counter-clockwise in the direction of propagation and likely originates from a region with negative magnetic field polarity. The decrease in polarization is expected at lower frequencies where the intensity (Stokes I) is greater than the circular polarization (Stokes V).

We also note that the circular polarization of TIII(4) is $\sim$70\% in a very narrowband at the vicinity of the TII with herringbones (TII(HB)), which may indicate that the emission was mostly fundamental \citep[higher harmonics are polarized significantly less;][]{Zheleznyakov64} and was generated in a region that was in the direct viewing direction of Parker Solar Probe. Furthermore, \citet{Melrose78} and \citet{Ledenev99} have shown that the degree of polarization is proportional to the ratio between the electron cyclotron frequency and the plasma frequency ($f_{ce}/f_{pe}$). This would indicate that a high degree of polarization is not only dependent on the viewing angle, but also on the magnitude of the magnetic field in the region of emission \citep{Reiner07}. Such a scenario also supports the shock acceleration of TIII(4) where the magnetic field compression is significantly large.

In the case of TIII(2), the degree of polarization is $\sim$25\%, which is still significantly larger than the nominal degree of polarization measured during type III storms \citep[$\sim$5\% ;][]{Reiner07, Pulupa20}. The emission in this case may be generated from plasma corresponding to the decametric wavelengths and the polarization fall-off corresponding to an \textit{f}ln(\textit{f}) rate may therefore result in a relatively lower polarization. We also note that the highest degree of polarization (or any polarization) is only observed during the initial rise phase of each type III within TIII(2). This indicates the presence of a polarized fundamental ($f_{pe}$) and a weakly polarized harmonic ($2f_{pe}$) component.

\subsection{Directivity analysis}

Figure~\ref{fig:radio_directions} shows radio source locations of TIII(1) and TIII(3). We compare calibrated radio fluxes measured by Parker Solar Probe, Solar Orbiter, STEREO~A and Wind at six frequency channels. We assume that the radio emission pattern $S$ as a function of heliocentric longitude $\lambda$ can be described by the von Mises distribution (also known as the circular normal distribution) as:
\begin{equation}
S(\lambda)=\frac{\exp(\kappa\cos(\lambda-\lambda_0))}{2\pi I_0(\kappa)}
\mbox{,}
\label{eq:lambda}
\end{equation}
where $\lambda_0$ is a direction corresponding to a peak radio flux, $\kappa$ is a measure of concentration, and $I_0$ is the modified Bessel function of the first kind of order 0, with this scaling constant chosen so that the distribution sums to unity. Dotted lines in Fig.~\ref{fig:radio_directions} (left) show results of this fit for both type III bursts based on peak fluxes, and Fig.~\ref{fig:radio_directions} (right) illustrates these injection directions (black arrows) with respect to the longitudinal spacecraft constellation. The results of the analysis suggest that the electron beam generating TIII(1) propagated in the longitude $-121.0^\circ\pm3.2^\circ$ (slightly east of the flare longitude), and the electron beam generating TIII(3) propagated in the longitude $-98.3^\circ\pm4.1^\circ$ (slightly west of the flare longitude).

\section{EUV wave kinematics}\label{EUV_wave_ap}

\begin{figure*}
   \centering
	\includegraphics[width=\textwidth] {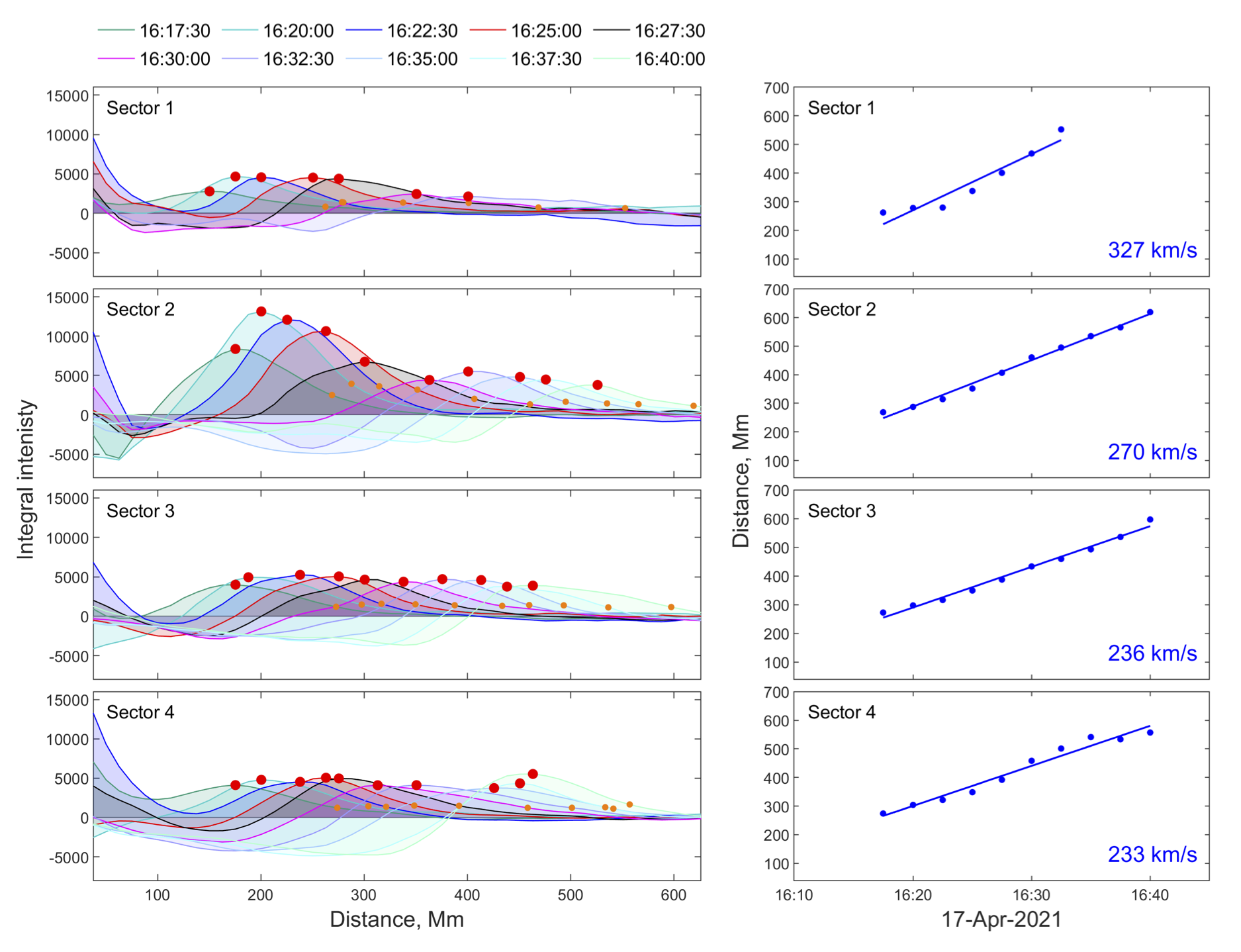}
	\caption{Kinematics of the EUV wave observed on the solar disk. Left panels: Perturbation profiles demonstrating the EUV wave propagation in the four angular sectors considered. Big red markers indicate the peak amplitude of each profile, while small orange markers show the estimated wavefront locations extracted at 30\% of the peak value, which we use for the assessment of the EUV wave kinematics presented in the right panels. A movie accompanying the figure is available online (movie2) and represents the dynamics of the intensity perturbation profiles.
	Right panels: Distance-time profiles of the EUV wave fronts for the four angular sectors. Dots indicate data points, and the solid lines show the linear fit. } 
	\label{fig_wave_on_disk}
\end{figure*}

To study the EUV wave kinematics and perturbation characteristics, we determine the location of the wavefronts and the perturbation amplitudes from intensity profiles from the given running-difference EUVI-A images using the ring analysis method \citep{Podladchikova2005, Podladchikova2019, Jebaraj20, Dumbovic2021}. We first design a spherical polar coordinate system centered at the source region and then split the solar sphere into rings of equal width of 12.5~Mm 
around the eruptive center. Additionally, we define four sectors with an angular width of 45$^\circ$, where the EUV wave propagation is most prominent (Fig.~\ref{fig_wave_on_disk}). For each sector, we calculate the integral intensity of all pixels in each ring and plot the derived intensity perturbation profiles smoothed with a forward-backward exponential smoothing method \citep{Brown1963} in the left panels of Fig.~\ref{fig_wave_on_disk}. The x-axis shows the distance (in Mm) from the eruptive center, which corresponds to the outer boundary of each ring, and the y-axis gives the resulting integral intensity in the considered angular sectors. The EUV wave is represented by the intensity increase in the perturbation profiles toward the peak amplitudes (big red markers), followed by a decay to background level. To obtain a robust estimate of the wavefront locations, we extract the outer distance at an intensity value corresponding to 30\% of the peak value above the background (small orange markers). The right panels of Fig.~\ref{fig_wave_on_disk} show the obtained evolution of the wavefront locations as function of time.
From the linear fits (solid lines) to the wavefront locations, which were followed up to about 600~Mm from the source region, we derive a mean velocity increasing gradually from sector 4 to sector 1 from 223 to 327~km~s$^{-1}$. In sector 2, the wave propagation could be further seen up to about 680~Mm from the source region; however, we do not include it in our analysis because of the relatively weak signal at these distances.   

\begin{figure}  
	\centering
	\includegraphics[width=0.75\textwidth]{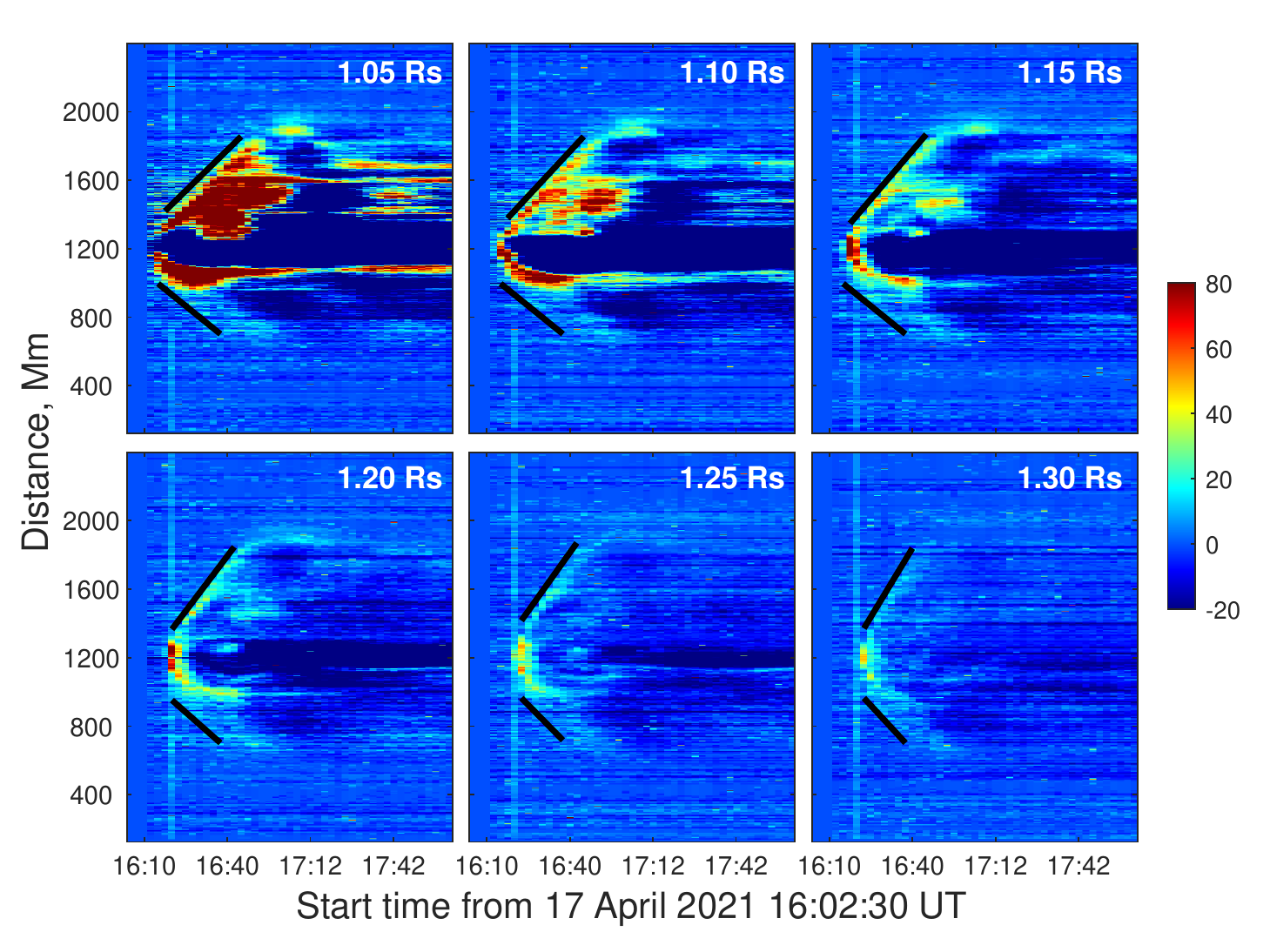}
	\caption{Stack plots derived from STEREO~A/EUVI 195~{\AA} base-difference images in vertical slits at heights from 1.05 to 1.30~$R_{\odot}$ above the solar surface from the Sun center. The EUV wave fronts are shown by black lines.}
	\label{fig_above_limb}
\end{figure}

To study the EUV wave kinematics above the solar limb and as function of height, we show in Fig.~\ref{fig_above_limb} stack plots generated from STEREO~A/EUVI 195~{\AA} base-difference images in vertical slits at different heights, from 1.05 to 1.30~$R_{\odot}$ as measured from the Sun center \cite[cf.,][]{Veronig2018}. To this aim, the emission is integrated over a vertical layer of a width of 18 EUVI pixels and stacked in time. In these stack plots, the EUV wave in the northern direction can be followed to a distance of about 740~Mm, propagating with velocities from 260 to 450~km s$^{-1}$ (for heights increasing from 1.05 to 1.15~$R_{\odot}$). As can be also seen in Fig.~\ref{fig_above_limb}, there is a wave reflection, which is best observed at a height of 1.15~$R_{\odot}$. In the southern direction, the velocities are smaller, increasing from 220--300~km~s$^{-1}$ with height, and the EUV wave propagation is seen only to about 350~Mm. This smaller extent of the wave propagation is most probably related to the southern coronal hole, which represents an obstacle to the wave propagation due to its high Alfv\'en speed \cite[e.g.,][]{Piantschitsch2018,Downs2021}. In the stack plots, one can also see that close to the eruption center, it is difficult to disentangle the CME flanks and the wave, but thereafter it is well seen that the CME flank expansion stops whereas the wave decouples from it and further propagates as a freely propagating fast MHD wave \cite[e.g.,][]{Warmuth2015}. Such behavior has been observed in previous events and led to the interpretation that the EUV wave is initiated by the fast initial lateral expansion of the CME flanks \citep[][]{Veronig2008,Veronig2018,Kienreich2009,Patsourakos2009}. 

\end{appendix}
\end{document}